\documentclass[english,prd,preprint,preprintnumbers,aps,eqsecnum,nofootinbib,superscriptaddress,notitlepage,showpacs,floatfix]{revtex4-1}
\usepackage[T1]{fontenc}
\usepackage[latin9]{inputenc}
\usepackage{dcolumn}
\usepackage{amsmath}
\usepackage{amssymb}
\usepackage{graphicx}
\usepackage{subfigure}
\usepackage{esint}
\usepackage[mathscr]{euscript}
\usepackage[pdftex]{color}
\usepackage[colorlinks=true,linkcolor=blue,filecolor=blue,urlcolor=blue,citecolor=blue,pdftex=true,plainpages=false]{hyperref}
\usepackage{ulem}
\usepackage[geometry]{ifsym}
\usepackage{babel}
\usepackage{bm}
\usepackage{marginnote}
\usepackage{tikz}
\usepackage{multirow}

\providecommand{\tabularnewline}{\\}

\graphicspath{{./figures/}}
\newcommand{\amlms}[2]{$a\approx #1 $~fm, $m'_l/m'_{h}= #2$}

\makeatother

\begin{document}

\title{\boldmath $|V_{ub}|$ from $B\to\pi\ell\nu$ decays and (2+1)-flavor lattice QCD}

\author{Jon. A. Bailey}
\affiliation{Department of Physics and Astronomy,
Seoul National University, Seoul, South Korea}

\author{A.~Bazavov}
\altaffiliation[Present address:~]{Department of Physics and Astronomy, University of Iowa, Iowa City, IA, USA}
\affiliation{Physics Department, Brookhaven National Laboratory, Upton, NY, USA}

\author{C.~Bernard}
\affiliation{Department of Physics, Washington University, St.~Louis, MO, USA}

\author{C.M.~Bouchard}
\affiliation{Department of Physics, The Ohio State University, Columbus, OH, USA}
\affiliation{Physics Department, College of William and Mary, Williamsburg, VA, USA}

\author{C.~DeTar}
%\affiliation{Physics Department, University of Utah, Salt Lake City, UT 84112, USA}
\affiliation{Department of Physics and Astronomy, University of Utah, Salt Lake City, UT, USA}

\author{D.~Du}
\email[]{dadu@syr.edu}
\affiliation{Department of Physics, University of Illinois, Urbana, IL, USA}
%\affiliation{Department of Physics, Syracuse University, Syracuse, NY 13244, USA}
\affiliation{Department of Physics, Syracuse University, Syracuse, NY, USA}

\author{A.X.~El-Khadra}
%\affiliation{Physics Department, University of Illinois, Urbana, IL 61801, USA}
\affiliation{Department of Physics, University of Illinois, Urbana, IL, USA}

\author{J.~Foley}
%\affiliation{Physics Department, University of Utah, Salt Lake City, UT 84112, USA}
\affiliation{Department of Physics and Astronomy, University of Utah, Salt Lake City, UT, USA}

\author{E.D.~Freeland}
%\affiliation{\textcolor{red}{UPDATE NEEDED?} Department of Physics, Benedictine University, Lisle, IL 60532, USA}
\affiliation{Liberal Arts Department, School of the Art Institute of Chicago, Chicago, IL, USA}

\author{E.~G\'amiz}
\affiliation{CAFPE and Departamento de Fisica Te\'orica y del Cosmos, Universidad de Granada, Granada, Spain}

\author{Steven~Gottlieb}
%\affiliation{Department of Physics, Indiana University, Bloomington, IN 47405 USA}
\affiliation{Department of Physics, Indiana University, Bloomington, IN, USA}

\author{U.M.~Heller}
%\affiliation{American Physical Society, Ridge, NY 11961, USA}
\affiliation{American Physical Society, Ridge, NY, USA}
% \affiliation{American Physical Society, One Research Road, Box 9000, Ridge, New York, USA}

\author{J.~Komijani}
\affiliation{Department of Physics, Washington University, St.~Louis, MO, USA}

\author{A.S.~Kronfeld}
%\affiliation{Fermi National Accelerator Laboratory, Batavia, IL 60510 USA}
\affiliation{Fermi National Accelerator Laboratory, Batavia, IL, USA}
\affiliation{Institute for Advanced Study, Technische Universit\"at M\"unchen, Garching, Germany}

\author{J.~Laiho}
%\affiliation{SUPA, School of Physics and Astronomy, University of Glasgow, Glasgow, UK}
%\affiliation{Department of Physics, Syracuse University, Syracuse, NY 13244, USA}
\affiliation{Department of Physics, Syracuse University, Syracuse, NY, USA}

\author{L.~Levkova}
%\affiliation{Physics Department, University of Utah, Salt Lake City, UT 84112, USA}
\affiliation{Department of Physics and Astronomy, University of Utah, Salt Lake City, UT, USA}

\author{Yuzhi Liu}
\affiliation{Department of Physics, University of Colorado, Boulder, CO, USA}

\author{P.B.~Mackenzie}
%\affiliation{Fermi National Accelerator Laboratory, Batavia, IL 60510 USA}
\affiliation{Fermi National Accelerator Laboratory, Batavia, IL, USA}

\author{Y. Meurice}
\affiliation{Department of Physics and Astronomy, University of Iowa, Iowa City, IA, USA}

\author{E.~Neil}
%\affiliation{Department of Physics, University of Colorado, Boulder, CO 80309, USA}
\affiliation{Department of Physics, University of Colorado, Boulder, CO, USA}
\affiliation{RIKEN-BNL Research Center, Brookhaven National Laboratory, Upton, NY, USA}

\author{Si-Wei Qiu}
\altaffiliation[Present address:~]{Laboratory of Biological Modeling, NIDDK, NIH, Bethesda, Maryland, USA}
\affiliation{Department of Physics and Astronomy, University of Utah, Salt Lake City, UT, USA}

\author{J.N.~Simone}
%\affiliation{Fermi National Accelerator Laboratory, Batavia, IL 60510 USA}
\affiliation{Fermi National Accelerator Laboratory, Batavia, IL, USA}

\author{R.~Sugar}
%\affiliation{Department of Physics, University of California, Santa Barbara, CA 93106, USA}
\affiliation{Department of Physics, University of California, Santa Barbara, CA, USA}

\author{D.~Toussaint}
%\affiliation{Physics Department, University of Arizona, Tucson, AZ 85721, USA}
\affiliation{Physics Department, University of Arizona, Tucson, AZ, USA}

\author{R.S.~Van~de~Water}
%\affiliation{Fermi National Accelerator Laboratory, Batavia, IL 60510 USA}
\affiliation{Fermi National Accelerator Laboratory, Batavia, IL, USA}

\author{R.~Zhou}
%\affiliation{Fermi National Accelerator Laboratory, Batavia, IL 60510 USA}
\affiliation{Fermi National Accelerator Laboratory, Batavia, IL, USA}

%\author{[Fermilab Lattice and MILC Collaborations]}
\collaboration{Fermilab Lattice and MILC Collaborations}
%\noaffiliation

\date{\today}

\begin{abstract}
	We present a lattice-QCD calculation of the $B\to\pi\ell\nu$ semileptonic form factors and a new determination of the CKM matrix element $|V_{ub}|$. We use the MILC asqtad 2+1-flavor lattice configurations at four lattice spacings and light-quark masses down to 1/20 of the physical strange-quark mass. We extrapolate the lattice form factors to the continuum using staggered chiral perturbation theory in the hard-pion and SU(2) limits. We employ a model-independent $z$ parameterization to extrapolate our lattice form factors from large-recoil momentum to the full kinematic range. We introduce a new functional method to propagate information from the chiral-continuum extrapolation to the $z$ expansion. We present our results together with a complete systematic error budget, including a covariance matrix to enable the combination of our form factors with other lattice-QCD and experimental results. To obtain $|V_{ub}|$, we simultaneously fit the experimental data for the $B\to\pi\ell\nu$ differential decay rate obtained by the BaBar and Belle collaborations together with our lattice form-factor results. We find $|V_{ub}|=(3.72\pm 0.16)\times 10^{-3}$ where the error is from the combined fit to lattice plus experiments and includes all sources of uncertainty. Our form-factor results bring the QCD error on $|V_{ub}|$ to the same level as the experimental error. We also provide results for the $B\to\pi\ell\nu$ vector and scalar form factors obtained from the combined lattice and experiment fit, which are more precisely-determined than from our lattice-QCD calculation alone. These results can be used in other phenomenological applications and to test other approaches to QCD.
\end{abstract}

 %Determination of CKM matrix elements

\maketitle

\section{introduction}

The Cabibbo-Kobayashi-Masakawa (CKM) matrix \cite{Cabibbo:1963yz,Kobayashi:1973fv} element $|V_{ub}|$ is one of the
fundamental parameters of the Standard Model and is an important input to searches for CP violation beyond the Standard Model. Constraints on new physics in the flavor sector are commonly cast in terms of over-constraining the apex of the CKM unitarity triangle.
In contrast to the well-determined angle $\beta$ of the unitarity triangle, the opposite
side $|V_{ub}/V_{cb}|$ is poorly determined, and the uncertainty
is currently dominated by $|V_{ub}|$. This is due to the fact that charmless decays of the $B$
meson have far smaller branching fractions than the
charmed decays, as well as the fact that the theoretical calculations are less precise than for $\sin2\beta$, $|V_{us}|$, or $|V_{cb}|$.
Currently the most precise determination of $|V_{ub}|$ is obtained
from charmless semileptonic $B$ decays, using exclusive or inclusive
methods that rely on the measurements of the branching fractions and the corresponding theoretical inputs. Exclusive determinations require knowledge of the form factors, while inclusive determinations rely on the operator product expansion, perturbative QCD, and non-perturbative input from experiments. 
There is a long standing discrepancy between $|V_{ub}|$ determined from inclusive and exclusive decays: the central values from these two approaches differ by about $3\sigma$. It was argued in Ref.~\cite{Crivellin:2014zpa} that this tension is unlikely to be due to new physics effects, and it is therefore important to examine the (theoretical and experimental) inputs to the $|V_{ub}|$ determinations. With the result obtained in this paper, the tension is reduced to $2.4\sigma$. 

In the limit of vanishing lepton mass, the Standard Model prediction for the differential decay rate of the exclusive semileptonic $B\to\pi\ell\nu$ decay is given by 
\begin{eqnarray}
\frac{d\Gamma(B\to\pi\ell\nu)}{dq^{2}} & = & \frac{G_{F}^{2}|V_{ub}|^{2}}{24\pi^{3}}|{\bm p}_{\pi}|^{3}|f_{+}(q^{2})|^{2}, \label{eq:partialdecayrate}
\end{eqnarray}
where $|\bm{p}_{\pi}|=\frac{1}{2M_{B}}\left[(M_{B}^{2}+M_{\pi}^{2}-q^{2})^{2}-4M_{B}^{2}M_{\pi}^{2}\right]^{1/2}$ 
is the pion momentum in the $B$-meson rest frame. To determine $|V_{ub}|$,
the form factor $|f_{+}(q^{2})|$ must be calculated with nonperturbative methods. The first unquenched lattice calculations of $|f_{+}(q^{2})|$ with 2+1 dynamical
sea quarks were performed by HPQCD \cite{hep-lat/0601021}
and the Fermilab/MILC collaborations \cite{0811.3640} several years ago. Here we extend and improve Ref.~\cite{0811.3640} in several ways.

The most recent exclusive determination of $|V_{ub}|$ from the Heavy Flavor Averaging Group (HFAG) \cite{1207.1158} is based on combined lattice plus experiment fits and yields $|V_{ub}|=(3.28\pm 0.29)\times 10^{-3}$, where the error includes both the experimental and theoretical uncertainties. The experimental data included in the average are the BaBar untagged six-$q^2$-bin data~\cite{1005.3288}, the BaBar untagged twelve-$q^2$-bin data~\cite{1208.1253}, the Belle untagged data~\cite{1012.0090}, and the Belle hadronic tagged~\cite{1306.2781} data. The theoretical errors on the form factors from lattice QCD \cite{0811.3640} are currently the dominant source of uncertainty in $|V_{ub}|$ \cite{1209.4674}. Hence a new lattice calculation of $f_+(q^2)$ with improved statistical and systematic errors is desirable \footnote{Note that there are several other efforts with 2 \cite{1211.6327} and 2+1 flavors of sea quarks \cite{Flynn:2015mha, 1310.3207}.}. To compare, the value of $|V_{ub}|$ from the inclusive method quoted by HFAG is about $(4.40\pm 0.20)\times 10^{-3}$ \cite{1207.1158} using the theory of Ref.~\cite{Bosch:2004bt}.

In this paper, we present a new lattice-QCD calculation of the $B\to\pi\ell\nu$
semileptonic form factors and a determination of $|V_{ub}|$. Our
calculation shares some features with the previous Fermilab/MILC calculation \cite{0811.3640} but makes several improvements. We quadruple the statistics on the previously used ensembles and improve our strategy for extracting the form factors by including excited states in our three-point correlator analysis. In addition, we include twice as many ensembles in this analysis. The new ensembles have smaller lattice spacings, with the smallest lattice spacing decreased by half. This analysis also includes ensembles with light sea-quark masses that are much closer to their physical values ($m_l/m_s=0.05$ versus $0.1$).  
The smaller lattice spacings and light-quark masses provide much better control over the dominant systematic error due to the chiral-continuum extrapolation. We find that heavy-meson rooted staggered chiral perturbation theory (HMrS$\chi$PT) in the SU(2) and hard-pion limits provides a satisfactory description of our data. All together, these improvements reduce the error on the form factors by a factor of about 3. Finally, we introduce a new functional method for the extrapolation over the full kinematic range.

The determination of $|V_{ub}|$ from a combined fit to our lattice form factors together with experimental measurements also yields a very precise determination of the vector and scalar form factors over the entire kinematic range. These form factors will be valuable input to other phenomenological applications in the Standard Model and beyond. An example is the rare decay $B\to\pi\ell^+\ell^-$, which we will discuss in a separate paper.

Because our primary goal in this work was a reliable and precise determination of $|V_{ub}|$, we employed a blinding procedure to minimize subjective bias.  At the stage of matching between the lattice and continuum vector currents, a slight multiplicative offset was applied to the data that was only known to two of the authors. The numerical value of the blinding factor was only disclosed after the analysis and error-estimation procedure, including the determination of $|V_{ub}|$, were essentially finalized.

This paper is organized as follows. In Sec.~\ref{secII}, we present our calculation of the form factors. We describe the lattice actions, currents, simulation parameters, correlation functions and fits to extract the matrix elements, renormalization of the currents, and adjustment of the form factors to correct for quark-mass mistunings. 
In Sec.~\ref{secIV}, we present the combined chiral-continuum extrapolation, followed by an itemized presentation of our complete error budget in Sec.~\ref{secV}. We then extrapolate
the form factors to the full $q^{2}$ range through the functional $z$ expansion
method in Sec.~\ref{secVI}. We also perform fits to lattice and experimental data simultaneously, to obtain $|V_{ub}|$. We conclude with a comparison to other results and discussion of the future outlook in Sec.~\ref{secVII}. Preliminary reports of this work can be found in Refs.~\cite{Du:2013kea,Bailey:2014fpx}.

% % %\input{section2}

\section{Lattice-QCD simulation} \label{secII}

In this section, we describe the details of the lattice simulation.
We briefly describe the calculation of the form factors in Sec.~\ref{sub:Methodology}. We also calculate the tensor form factor, which follows a analysis similar to that of the vector and scalar form factors. The tensor form factor enters the Standard-Model rate for $B\to\pi\ell^+\ell^-$ decay, and our final result for $f_T$ will be presented in a forthcoming paper. In Sec.~\ref{sub:actions-parameters}, we introduce
the actions and simulation parameters used in this analysis. This
is followed, in Sec.~\ref{sub:Currents-and-correlation}, by a brief discussion of the currents and lattice correlation
functions. The correlator fits to extract the lattice form factors are provided in Sec.~\ref{subsec:correlator_fits}. In Sec.~\ref{subsec:renormalization}, we discuss the renormalization of the lattice currents. In Sec.~\ref{subsec:tuning}, we correct the form factors {\it a posteriori} to account for the mistuning of the simulated heavy $b$-quark mass.

\subsection{Form-factor definitions\label{sub:Methodology}}\label{secIII}

The vector and tensor hadronic matrix elements relevant for $B\to\pi$ semileptonic
decays can be parameterized by the following three form factors:
\begin{eqnarray}
\langle\pi(p_\pi)|\mathcal{V^{\mu}}|B(p_B)\rangle & = & \left(p_B^{\mu}+p_\pi^{\mu}-\frac{M_{B}^{2}-M_{\pi}^{2}}{q^{2}}q^{\mu}\right)f_{+}(q^{2})+\frac{M_{B}^{2}-M_{\pi}^{2}}{q^{2}}\, q^{\mu}\, f_{0}(q^{2}),\label{eq:def_f+0}\\
\langle\pi(p_\pi)|\mathcal{T^{\mu\nu}}|B(p_B)\rangle & = & \frac{2}{M_{B}+M_{\pi}}\left(p_B^{\mu}p_\pi^{\nu}-p_B^{\nu}p_\pi^{\mu}\right)f_{T}(q^{2}),\label{eq:def_fT}
\end{eqnarray}
where $\mathcal{V}^{\mu}=\bar{q}\gamma^{\mu}b$, and $\mathcal{T}^{\mu\nu}=i\bar{q}\sigma^{\mu\nu}b$.
In lattice gauge theory and in chiral perturbation theory, it is convenient to parameterize the vector-current matrix elements by \cite{hep-ph/0101023}
\begin{eqnarray}
\langle\pi(p_\pi)|\mathcal{V^{\mu}}|B(p_B)\rangle & = & \sqrt{2M_{B}}\left[v^{\mu}f_{\parallel}(E_{\pi})+p_{\pi,\perp}^{\mu}f_{\perp}(E_{\pi})\right],\label{eq:def_fpv}
\end{eqnarray}
where $v^{\mu}=p_B^{\mu}/M_{B}$ is the four velocity of the $B$ meson
and $p_{\pi,\perp}^{\mu}=p_\pi^{\mu}-(p_\pi\cdot v)v^{\mu}$ is the projection
of the pion momentum in the direction perpendicular to $v^{\mu}$.
The pion energy is related to the lepton momentum transfer $q^{2}$ by $E_{\pi}=p_\pi\cdot v=(M_{B}^{2}+M_{\pi}^{2}-q^{2})/(2M_{B})$.
With this setup, we have 
\begin{eqnarray}
f_{\parallel}(E_{\pi}) & = & \frac{\langle\pi(p_\pi)|\mathcal{V}^{4}|B(p_B)\rangle}{\sqrt{2M_{B}}},\label{eq:def_fv}\\
f_{\perp}(E_{\pi}) & = & \frac{\langle\pi(p_\pi)|\mathcal{V}^{i}|B(p_B)\rangle}{\sqrt{2M_{B}}}\,\frac{1}{p_\pi^{i}},\label{eq:def_fp}\\
f_{T}(q^{2}) & = & \frac{M_{B}+M_{\pi}}{\sqrt{2M_{B}}}\,\frac{\langle\pi(p_\pi)|\mathcal{T}^{4i}|B(p_B)\rangle}{\sqrt{2M_{B}}}\,\frac{1}{p_\pi^{i}},\label{eq:fT}
\end{eqnarray}
where no summation is implied by the repeated indices here. The form factors $f_{+}$ and $f_0$ are
\begin{eqnarray}
f_{+}(q^{2}) & = & \frac{1}{\sqrt{2M_{B}}}\left[f_{\parallel}+(M_{B}-E_{\pi})f_{\perp}\right] ,\label{eq:fvp_f+} \\
f_{0}(q^{2}) & = & \frac{\sqrt{2M_{B}}}{M_{B}^{2}-M_{\pi}^{2}}\left[(M_{B}-E_{\pi})f_{\parallel}+(E_{\pi}^{2}-M_{\pi}^{2})f_{\perp}\right].\label{eq:fvp_f0}
\end{eqnarray}

\subsection{Actions and parameters \label{sub:actions-parameters}}

\begin{figure}
	\centering \includegraphics[scale=1]{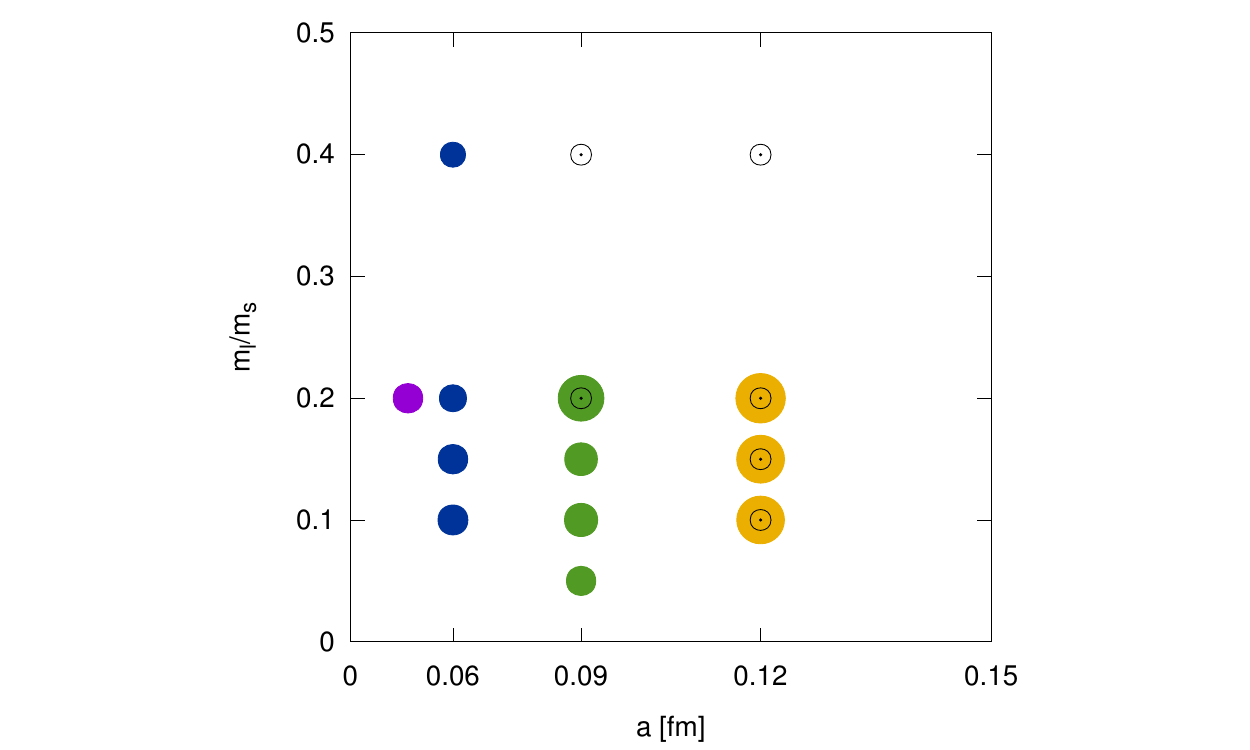}
	
	\caption{ Lattice spacings and light-quark masses used in this analysis. The area of each filled disk is proportional to the number of configurations in the ensemble. Open black circles indicate ensembles use in the analysis of Ref.~\cite{0811.3640}; those with ($a$, $m'_l/m'_h$)=(0.12 fm, 0.4) and (0.09 fm, 0.4) used in Ref.~\cite{0811.3640} (open circles without disks) are not used in this analysis.  
		\label{fig:ensembles}}
\end{figure}

The lattice gauge-field configurations we use have been generated by the
MILC Collaboration \cite{0903.3598,Bernard:2001av, Aubin:2004wf}, and some of their properties are listed in Table~\ref{tab:ensembles}. These twelve
ensembles have four different lattice spacings ranging from $a\approx 0.12$~fm to $a\approx 0.045$~fm with several light sea-quark masses at most lattice spacings in the range $0.05 \le a m'_l/a m'_h \le 0.4$. The parameter range is shown in Fig.~\ref{fig:ensembles}. We use
the Symanzik-improved %($O(\alpha_{s}a^{2})$)
gauge action \cite{Weisz:1982zw,Weisz:1983bn,Luscher:1984xn} for the gluons and the tadpole-improved (asqtad) staggered action \cite{hep-lat/9609036,Lepage:1998vj,hep-lat/9806014,hep-lat/9805009,hep-lat/9903032,hep-lat/9912018} for the 2+1 flavors of dynamical sea quarks and for the light valence quarks. Both Table~\ref{tab:ensembles} and Fig.~\ref{fig:ensembles} also indicate the ensembles used
in the previous Fermilab/MILC calculation
\cite{0811.3640}. The current analysis benefits from an almost quadrupled increase in the statistics over that of Ref.~\cite{0811.3640}, as well as finer lattice spacings and lighter sea-quark masses. All
ensembles have large enough spatial volume, $M_{\pi}L \geq 3.8$, such
that the systematic error due to finite-size effects is
negligible compared to other uncertainties.

\begin{table}[h]
	\centering \caption{Parameters of the MILC asqtad gauge-field ensembles used in this
		analysis. From left to right: approximate lattice
		spacing $a$ in fm, the (light/strange)-quark mass ratio $am'_{l}/am'_{h}$, the coupling constant $\beta$, the tadpole parameter $u_{0}$ determined from the plaquette, lattice volume, the number of configurations $N_{\text{cfg}}$, $M_{\pi}L$ ($L$
		is the spatial length of the lattice), and the number of configurations of the four ensembles that were used in Ref.~\cite{0811.3640}. \label{tab:ensembles}
	}

	\begin{tabular}{cccccccc}
		\hline
		\hline  
		$\approx$$a$(fm)   &  $am'_{l}/am'_{h}$   &  $\beta$   &  $u_{0}$  &  volume   & \;\;  $N_{\text{cfg}}$ \;  &\;  $M_{\pi}L$ \;\;  &  $N_\text{cfg}$(Ref.~\cite{0811.3640})) \tabularnewline
		\hline
		0.12   &  0.01/0.05   &  6.760   &  0.8677  &  $20^{3}\times64$   &  2259   &  4.5   &  592 \tabularnewline
		&  0.007/0.05   &  6.760   &  0.8678  &  $20^{3}\times64$   &  2110   &  3.8   &  836\tabularnewline
		&  0.005/0.05   &  6.760   &  0.8678  &  $24^{3}\times64$   &  2099   &  3.8   &  529\tabularnewline
		\hline
		0.09   &  0.0062/0.031   &  7.090   &  0.8782  &  $28^{3}\times96$   & 1931   &  4.1   &  557 \tabularnewline
		&  0.00465/0.031   &  7.085   &  0.8781  &  $32^{3}\times96$   &  984   &  4.1   & \tabularnewline
		&  0.0031/0.031   &  7.080   &  0.8779  &  $40^{3}\times96$   & 1015   &  4.2   & \tabularnewline
		&  0.00155/0.031   &  7.075   &  0.877805  &  $64^{3}\times96$   &  791   &  4.8   & \tabularnewline
		\hline
		0.06   &  0.0072/0.018   &  7.480   &  0.8881  &  $48^{3}\times144$   &  593   &  6.3   & \tabularnewline
		&  0.0036/0.018   &  7.470   &  0.88788  &  $48^{3}\times144$   & 673   &  4.5   & \tabularnewline
		&  0.0025/0.018   &  7.465   &  0.88776  &  $56^{3}\times144$   &  801   &  4.4   & \tabularnewline
		&  0.0018/0.018   &  7.460   &  0.88764  &  $64^{3}\times144$   &  827   &  4.3   & \tabularnewline
		\hline
		0.045   &  0.0028/0.014   &  7.810   &  0.89511 &  $64^{3}\times192$   &  801   &  4.6   & \tabularnewline
		\hline 
		\hline 
	\end{tabular}
\end{table}

In this calculation, we work in the full-QCD limit, so that the light valence-quark masses $am_{l}$ are the same as the
light sea-quark masses $am'_l$, which are degenerate. For the bottom quarks, we use
the Fermilab interpretation \cite{hep-lat/9604004} of the Sheikholeslami-Wohlert
clover action \cite{Sheikholeslami:1985ij}. In Table~\ref{tab:simulation_params}, we list parameters for the valence quarks.

\begin{table}[h]
	\centering \caption{Heavy-quark masses and other parameters used in the simulation.
		Starting in the third column: the clover
		parameter $c_\text{SW}$, the simulation $b$-quark mass parameter $\kappa'_b$, the current rotation parameter $d'_1$, the number of sources $N_\text{src}$ and the
		two source-sink separations $T$. Note that we use the same valence light-quark mass as $m'_l$ in the sea except the $a=0.09$fm, $m'_l/m'_h=0.00465/0.031$ ensemble where a slightly different valence mass $am_l = 0.0047$ is used. \label{tab:simulation_params}}
	
	\begin{tabular}{ccccccc}
		\hline 
		\hline 
		$\approx a$(fm)  & $am'_{l}/am'_{h}$  &\;\; $c_{SW}$  \;\; & $\kappa'_{b}$ & $d'_{1}$&  $N_\text{src}$  & $T$  \tabularnewline
		\hline 
		0.12  & 0.01/0.05   & 1.531  & 0.0901 & 0.09334 &4 & 18,19 \tabularnewline
		& 0.007/0.05   & 1.530  & 0.0901 & 0.09332 &4 & 18,19 \tabularnewline
		& 0.005/0.05    & 1.530  & 0.0901 & 0.09332 &4 & 18,19 \tabularnewline
		\hline 
		0.09  & 0.0062/0.031   & 1.476  & 0.0979 & 0.09677 &4 & 25,26 \tabularnewline
		& 0.00465/0.031    & 1.477  & 0.0977 & 0.09671&4 & 25,26 \tabularnewline
		& 0.0031/0.031   & 1.478  & 0.0976 & 0.09669 &4 & 25,26 \tabularnewline
		& 0.00155/0.031    & 1.4784  & 0.0976 & 0.09669 &4 & 25,26 \tabularnewline
		\hline 
		0.06  & 0.0072/0.018   & 1.4276  & 0.1048 & 0.09636 &4 & 36,37 \tabularnewline
		& 0.0036/0.018    & 1.4287  & 0.1052 & 0.09631 &4 & 36,37 \tabularnewline
		& 0.0025/0.018    & 1.4293  & 0.1052 & 0.09633 &4 & 36,37 \tabularnewline
		& 0.0018/0.018    & 1.4298  & 0.1052 & 0.09635 &4 & 36,37 \tabularnewline
		\hline 
		0.045  & 0.0028/0.014   & 1.3943  & 0.1143 & 0.08864 &4 & 48,49 \tabularnewline
		\hline 
		\hline 
	\end{tabular}
\end{table}

Table~\ref{tab:derived_params} lists the values of $r_1/a$ on each ensemble, along with other derived parameters, where $r_{1}$
is the characteristic distance between two static quarks such that the force between them satisfies $r_{1}^{2}F(r_{1})=1.0$ \cite{hep-lat/0002028,hep-lat/9310022}.
The absolute lattice scale $r_1$ is obtained by comparing the Particle Data Group value of $f_\pi$ with lattice calculations of $r_1f_\pi$ from MILC~\cite{0910.2966} and HPQCD \cite{0910.1229}, yielding the absolute scale $r_1 = 0.3117(22)$~fm \cite{1112.3051}. This value is consistent with the independent, but less precise, determination $r_1 = 0.323(9)$ from RBC/UKQCD using domain-wall fermions~\cite{Arthur:2012opa}.

\begin{table}[h]
	\centering \caption{Derived parameters from the simulation. Starting in the third column: relative scale $r_1/a$, the
		Goldstone pion mass $M_{\pi}$, root-mean-square (RMS) pion mass $M_{\pi}^\text{RMS}$, and the critical hopping parameter $\kappa_{\text{crit}}$ which enters our definition of the heavy-quark mass. \label{tab:derived_params} }

	\begin{tabular}{cccccc}
		\hline 
		\hline 
		$\approx a$(fm)  & $am'_{l}/am'_{h}$  & $r_{1}/a$  & $M_{\pi}$(MeV)  & $M_{\pi}^{\text{RMS}}$(MeV)   & $\kappa_{\text{crit}}$ \tabularnewline
		\hline 
		0.12  & 0.01/0.05   	& 2.7386  & 389  & 532  & 0.14091 \tabularnewline
		& 0.007/0.05  	& 2.7386  & 327  & 488  & 0.14095 \tabularnewline
		& 0.005/0.05  	& 2.7386  & 277  & 456  & 0.14096 \tabularnewline
		\hline                                         
		0.09  & 0.0062/0.031  	& 3.7887  & 354  & 413  & 0.139119 \tabularnewline
		& 0.00465/0.031  & 3.7716  & 307  & 374  & 0.139134 \tabularnewline
		& 0.0031/0.031  	& 3.7546  & 249  & 329  & 0.139173 \tabularnewline
		& 0.00155/0.031  & 3.7376  & 177  & 277  & 0.139190 \tabularnewline
		\hline                                         
		0.06  & 0.0072/0.018  	& 5.3991  & 450  & 466  & 0.137582 \tabularnewline
		& 0.0036/0.018  	& 5.3531  & 316  & 340  & 0.137632 \tabularnewline
		& 0.0025/0.018  	& 5.3302  & 264  & 291  & 0.137667 \tabularnewline
		& 0.0018/0.018  	& 5.3073  & 224  & 255  & 0.137678 \tabularnewline
		\hline                                         
		0.045 & 0.0028/0.014  	& 7.2082  & 324  & 331  & 0.136640 \tabularnewline
		\hline 
		\hline 
	\end{tabular}
\end{table}

\subsection{Currents and correlation functions \label{sub:Currents-and-correlation}}

We calculate the two-point and three-point functions  
\begin{eqnarray}
C_{P}(t;\bm{p}) & = & \sum_{\bm{x}}e^{i\bm{p\cdot}\bm{x}}\langle O_{P}(0,\bm{0})O_{P}^{\dagger}(t,\bm{x})\rangle,\label{eq:def_C_2pt}\;\;\text{and}\\
C_{J}(t,T;\bm{p}) & = & \sum_{\bm{x,y}}e^{i\bm{p\cdot}\bm{y}}\langle O_{\pi}(0,\bm{0})J(t,\bm{y})O_{B}^{\dagger}(T,\bm{x})\rangle,\label{eq:def_C_3pt}
\end{eqnarray}
where $P=B$, $\pi$ labels the pseudoscalar meson, the
operators $O_{P}$ ($O_{P}^{\dagger}$) annihilate
(create) the states with the quantum numbers of the pseudoscalar
meson $P$ on the lattice, and $J=V^{\mu},T^{\text{\ensuremath{\mu\nu}}}$
are the lattice currents.

For the $B$ meson, we use a mixed-action interpolating operator $O_{B}$
which is a combination of a Wilson clover bottom quark and a staggered
light quark \cite{0811.3640}:
\begin{eqnarray}
O_{B}(x) & =\sum_{y} & \bar{\psi}(y)S(y,x)\gamma_{5}\Omega(x)\chi(x),\label{eq:O_B}
\end{eqnarray}
where $\Omega(x)=\gamma_{1}^{x_{1}}\gamma_{2}^{x_{2}}\gamma_{3}^{x_{3}}\gamma_{4}^{x_{4}}$,
$x=(\bm{x},t)$, and $S(x,y)$ is a smearing function. For the
pion, we use the operator 
\begin{eqnarray}
O_{\pi}(x) & = & (-1)^{\sum_{i=1}^{4}x_{i}}\bar{\chi}(x)\chi(x),\label{eq:O_pi}
\end{eqnarray}
which is constructed from two 1-component staggered quarks.

The current operators are constructed in a similar way: 
\begin{eqnarray}
V^{\mu}(x) & = & \bar{\Psi}(x)\gamma^{\mu}\Omega(x)\chi(x),\label{eq:V_lat}\;\;\text{and}\\
T^{\mu\nu}(x) & = & \bar{\Psi}(x)\sigma^{\mu\nu}\Omega(x)\chi(x),\label{eq:T_lat}
\end{eqnarray}
where the heavy quark field spinor $\Psi$ is rotated to remove tree-level $O(a)$ discretization effects, via \cite{hep-lat/9604004} 
\begin{eqnarray}
\Psi(x) & = & (1+a\, d_{1}\bm{\bm{\gamma}\cdot}\bm{D}_\text{lat})\psi.\label{eq:Psi_b}
\end{eqnarray}

\begin{figure}
	\centering { 
	
	\includegraphics[scale=1]{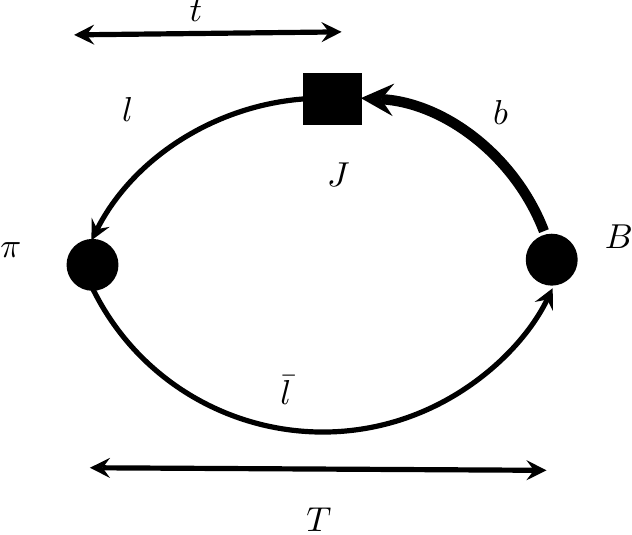}
	
}
	\caption{  Illustrative diagram for the three-point correlation functions. 
		\label{fig:3pt}}
\end{figure}

Figure~\ref{fig:3pt} illustrates the three-point correlation function used to obtain the lattice form factors. The current operator $J$ is inserted between the $b$- and $l$-quark lines. The three-point functions depend on both the current insertion time
$t$ and the temporal separation $T$ between the $\pi$ and $B$ mesons. The signal to noise ratio is largely
determined by $T$. A convenient approach is to fix the source-sink separation $T$ in the simulations and then insert the current operators at every time slice in between. The source-sink separations $T$
at different lattice spacings, sea-quark masses, and recoil momenta are chosen to be approximately the same in physical units. To minimize statistical uncertainties and reduce excited-state contamination, we tested data with different source-sink separations before choosing those shown in Table~\ref{tab:simulation_params}. The $B$ meson is at rest in our simulation, while the
daughter pion is either at rest or has a small three-momentum.
The light-quark propagator is computed from a point source so that one inversion of the Dirac operator can be used to obtain multiple momenta. The spatial source location is varied randomly from one configuration to the next to minimize autocorrelations. The $b$-quark source is always implemented with smearing based on a Richardson 1S wave function~\cite{menscher2005} after fixing to Coulomb gauge. 
We compute both the two-point function $C_{\pi}(t;\bm{p})$
and three-point function $C_{J}(t,T;\bm{p})$
at several of the lowest possible pion momenta in a finite box: $\bm{p}=(2\pi/L)(0,0,0),$
$(2\pi/L)(1,0,0)$, $(2\pi/L)(1,1,0)$, $(2\pi/L)(1,1,1)$, and $(2\pi/L)(2,0,0)$,
where contributions from each momentum are averaged over permutations of components. We find the correlation functions with momentum $(2\pi/L)(2,0,0)$ too noisy to be useful, so we exclude these data from our analysis.

\subsection{Two-point and three-point correlator fits} \label{subsec:correlator_fits}

In this subsection, we describe how to extract the desired matrix element from two- and three-point correlation functions. 
With our choice for the valence-quark actions and for the interpolating operators, the two- and three-point functions take the form \cite{hep-lat/0211014}
\begin{eqnarray}
C_{P}(t;\bm{p}) & = & \sum_{n=0}^{\infty}(-1)^{n(t+1)}|Z_{P}^{(n)}(\bm{p})|^{2}\,\left[e^{-E_{P}^{(n)}(\bm{p})t}+e^{-E_{P}^{(n)}(\bm{p})(N_{t}-t)}\right],\label{eq:C_2pt}\\
C_{J}(t,T;\bm{p}) & = & \sum_{m,n=0}^{\infty}(-1)^{m(t+1)}(-1)^{n(T-t-1)}Z_{\pi}^{(m)}(\bm{p})\,\mathcal{M}_{J}^{(mn)}\, Z_{B}^{(n)}(\bm{0})\: e^{-E_{\pi}^{(m)}(\bm{p})t-M_{B}^{(n)}(T-t)},\nonumber\\
\label{eq:C_3pt}
\end{eqnarray}
where $N_t$ is the temporal length of the lattice and 
\begin{eqnarray}
Z_{P}^{(n)}(\bm{p}) & = & \frac{|\langle0|O_{P}|P^{(n)}(\bm{p})\rangle|}{\sqrt{2E_{p}^{(n)}(\bm{p})}},\label{eq:Z_P}\\
\mathcal{M}{}_{J}^{(mn)} & = & \frac{\langle\pi^{(m)}(\bm{p})|J|B^{(n)}\rangle}{\sqrt{2E_{\pi}^{(m)}(\bm{p})}\sqrt{2M_{B}^{(n)}}}.\label{eq:M_Gamma}
\end{eqnarray}
Note that due to the staggered action used for the light quarks, the meson interpolating operators also couple to the positive parity (scalar) states which oscillate in Euclidean times $t$ and $T$ with the factors $(-1)^{n(t+1)}$ and $(-1)^{n(T-t)}$.  

Our goal is to extract $\mathcal{M}_{J}^{(00)}$,
the ground state matrix element from these correlation functions. To suppress the contributions from the positive parity states to the ratio, we follow the averaging procedure of Ref.~\cite{0811.3640}, which exploits the oscillating sign in their Euclidean time dependence. The time averages can be thought of as a smearing over neighboring time slices
$\{t,t+1,t+2\}\times\{T,T+1\}$ to significantly reduce the overlap with opposite-parity states. Denoting the
averaged correlators by $\overline{C_{P}}$ and $\overline{C_{J}}$, we then use the ratio~\cite{0811.3640} 
\begin{eqnarray}
R_{J}(t,T;\bm{p}) & = & \frac{\overline{C_{J}}(t,T;\bm{p})}{\sqrt{\overline{C_{\pi}}(t;\bm{p})\overline{C_{B}}(T-t;\bm{0})}}\sqrt{\frac{2E_{\pi}^{(0)}(\bm{p})}{e^{-E_{\pi}^{(0)}(\bm{p})t-M_{B}^{(0)}(T-t)}}},\label{eq:R_Gamma}
\end{eqnarray}
where $E_{\pi}^{(0)}(\bm{p})$ and $M_{B}^{(0)}$ are the ground-state
pion energy and $B$-meson rest mass, respectively. The uncertainty in the $B$-meson rest mass has significant impact on the ratio $R_J$, so we follow a two-step procedure. We first determine the pion and $B$-meson ground-state energy as precisely as possible using the corresponding two-point functions. We then feed these ground-state energies into the ratio $R_J$, preserving the correlations with jackknife resampling.

\begin{figure}
	\centering 
	\includegraphics[width=0.48\textwidth]{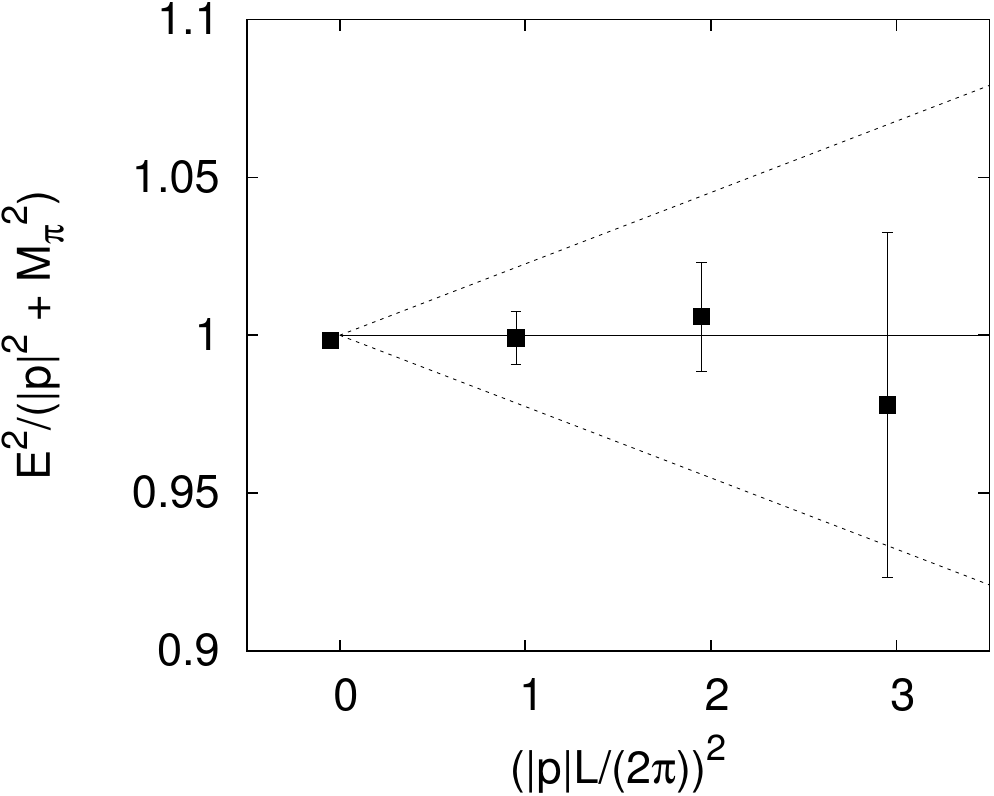}\hfill
	\includegraphics[width=0.48\textwidth]{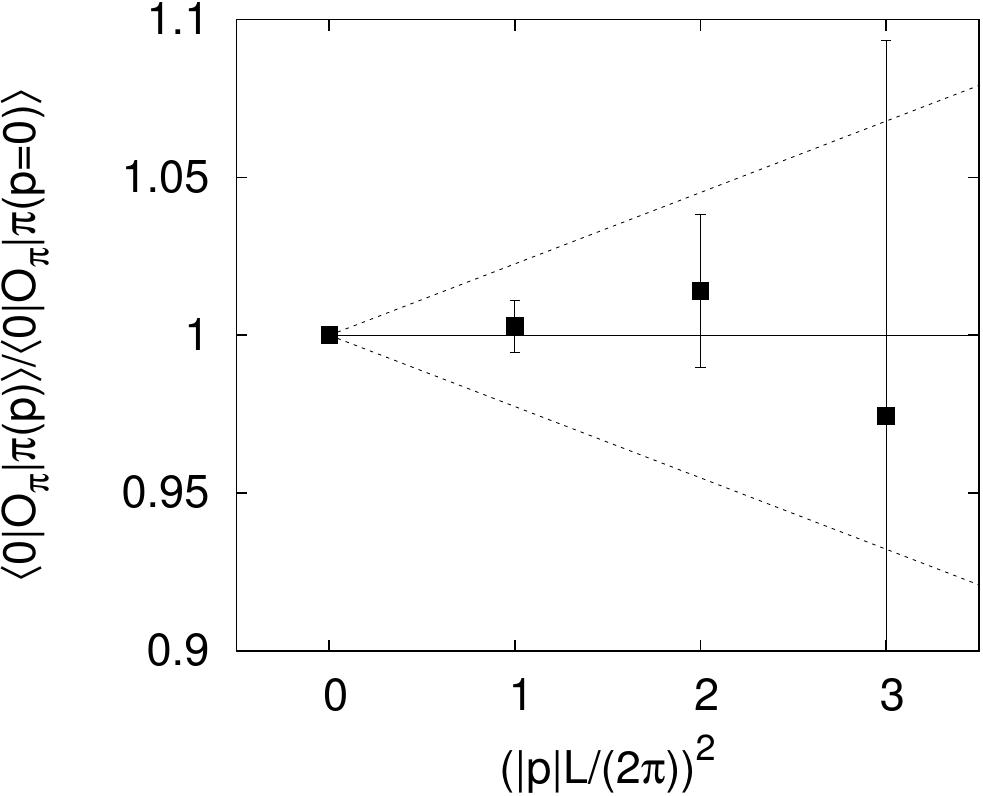} 
	
	\caption{Dispersion-relation tests on the ensemble \amlms{0.12}{0.1}.
		Left: fit results of $E^2_{\pi}(\bm{p})$ from two-point functions
		are compared with the continuum dispersion relation $|\bm{p}|^{2}+M_{\pi}^{2}$. Right: wave-function overlaps $\langle 0|O_\pi|\pi(\bm{p})\rangle$ at different momenta are compared.
		The dotted lines show the expected size of the leading momentum-dependent discretization
		errors based on power counting, which are of ${\mathcal O}( \alpha_s^2 |{\bm{p}}|^2 a^2)$.
		\label{fig:dispersion_relation}}
\end{figure}

\begin{figure}
	\center{\includegraphics[width=0.50\textwidth]{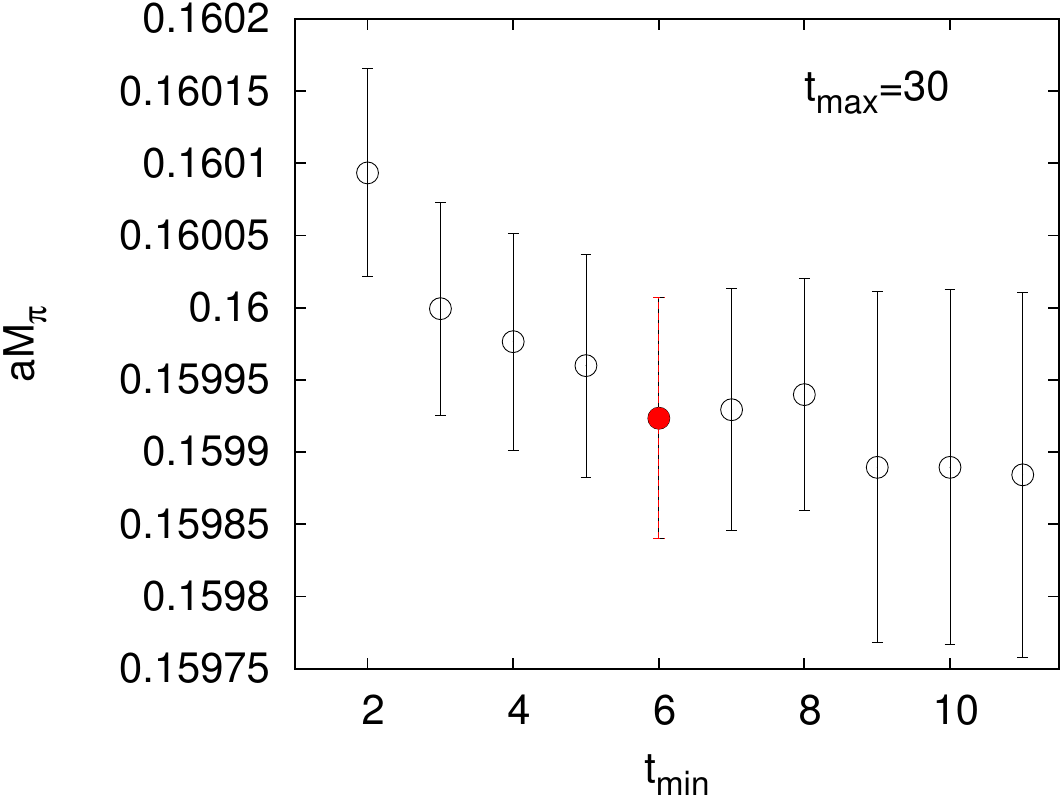}\hfill\includegraphics[width=0.46\textwidth]{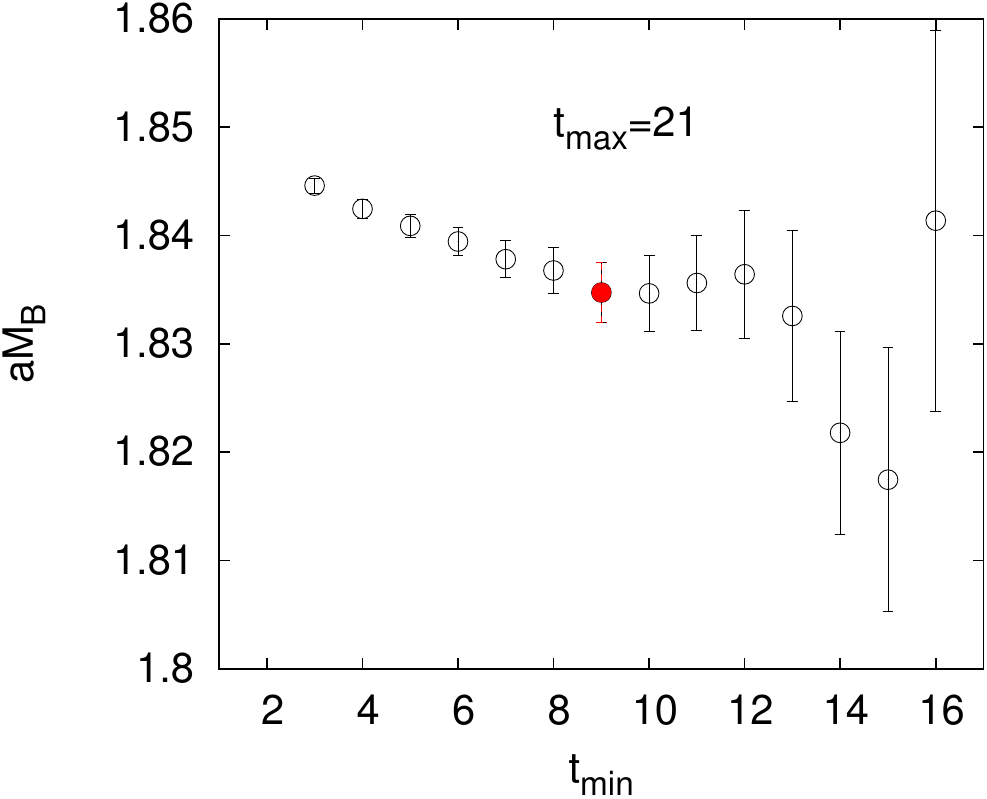}} 
	
	\caption{Pion (left) and $B$ meson (right) mass in lattice units from the two-point
		correlator fits versus different choices of $t_\text{min}$ for the \amlms{0.12}{0.1} ensemble. The $t_\text{min}$ marked in red (filled) is that of
		our preferred fit. The pion and $B$ meson correlators are fit to 2+0 and 1+1 states
		with $t_\text{max}=30,21$, respectively.\label{fig:Stability-pion-B}}
\end{figure}

For the pion two-point functions at zero momentum, the oscillating
states --- the terms in Eq.~(\ref{eq:C_2pt}) with odd powers of $(-1)$ --- do not appear. Thus, we fit the pion two-point functions using Eq.~(\ref{eq:C_2pt})
with the lowest two non-oscillating states ($n=0,2$). For the two-point
functions with nonzero momentum, the contribution from oscillating
states is small but noticeable. We find that we only need to include the lowest
three states ($n=0,1,2)$ in the fits. Because the momenta
we consider are typically small compared to $2\pi/a$, the continuum dispersion
relation is satisfied within statistical errors, as shown in Fig.~\ref{fig:dispersion_relation}.
In the main analysis, we therefore use the mass $M_{\pi}$ from the zero-momentum fit and the continuum dispersion relation to set $E_{\pi}^{(0)}(\bm{p})=\sqrt{|\bm{p}|^{2}+M_{\pi}^{2}}$\, 
for non-zero momentum. Because the zero-momentum energy 
has significantly smaller statistical error than that of nonzero momentum, using this choice and the dispersion relation for nonzero-momentum energy leads
to a more stable and precise determination of $\mathcal{M}_{J}^{(00)}$.
Table~\ref{tab:fit_range_2pt} lists the relevant fit ranges for the
two-point fits. In the two-point correlators (except the zero-momentum pion two-point correlators), the noise grows rapidly 
with increasing $t$, the distance away from the pion source in the temporal direction.
The data points at large $t$ are not useful, and including them would lead to a larger covariance matrix which would be difficult to resolve given the limited number of configurations. We choose the upper end of the
fit ranges $t_\text{max}$ such that the relative error does not exceed 20\%.
The lower end $t_\text{min}$ is chosen such that the excited state contamination is sufficiently small, {\it i.e.},
the resulting central values of the ground state energy are stable against variations in $t_\text{min}$ as shown
in Fig.~\ref{fig:Stability-pion-B} (left).

In our analysis, there are two places where quantities from the $B$-meson two-point functions are needed. The first is for $M_B^{(0)}$ in Eq.~(\ref{eq:R_Gamma}). The second is for the $B$-meson excited state energy in Eq.~(\ref{eq:R_gamma_exp}) below.  
For the determination of $M_B^{(0)}$ in Eq.~(\ref{eq:R_Gamma}) we use two-point functions constructed with a 1S-smearing function in the interpolating operators for the source and sink.  The 1S-smeared operator has good overlap with the ground state and a much smaller overlap with the excited states than the local source operator, thus reducing excited-state contributions to the corresponding correlators. We fit the 1S-1S smeared $B$-meson two-point correlators with relatively large $t_\text{min}$
to only two states ($n<2$ in Eq.~(\ref{eq:C_2pt})). To choose $t_\text{max}$,
we again apply the 20\%-rule on the relative error. The lower bound $t_\text{min}$
is chosen in a manner similar to the pion two-point fits and the stability plot
is shown in Fig.~\ref{fig:Stability-pion-B} (right). The chosen fit ranges
are shown in Table~\ref{tab:fit_range_2pt}.

\begin{table}
	\caption{Fit ranges $[t_\text{min},t_\text{max}]$ of the two-point correlator fits used to obtain the rest masses of the pion and $B$ mesons. \label{tab:fit_range_2pt} }

	\begin{tabular}{cccc}
		\hline
		\hline  
		$\approx a$(fm)  & $am'_{l}/am'_{h}$  & $C_\pi$ (2+0)  \;\;& $C_B^{(1S)}$\textbf{ }(1+1)\textbf{ }\tabularnewline
		\hline 
		0.12  & 0.01/0.05  & [6, 30]  & [9, 21] \tabularnewline
		& 0.007/0.05  & [6, 30]  & [9, 21] \tabularnewline
		& 0.005/0.05  & [6, 30]  & [9, 21] \tabularnewline
		\hline 
		0.09  & 0.0062/0.031  & [9, 47]  & [12, 32] \tabularnewline
		& 0.00465/0.031  & [9, 47]  & [12, 29] \tabularnewline
		& 0.0031/0.031  & [9, 47]  & [13, 29] \tabularnewline
		& 0.00155/0.031  & [9, 47]  & [14, 29] \tabularnewline
		\hline 
		0.06  & 0.0072/0.018  & [13, 71]  & [14, 41] \tabularnewline
		& 0.0036/0.018  & [13, 71]  & [14, 41] \tabularnewline
		& 0.0025/0.018  & [13, 71]  & [14, 41] \tabularnewline
		& 0.0018/0.018  & [13, 71]  & [15, 41] \tabularnewline
		\hline 
		0.045  & 0.0028/0.014  & [17, 74]  & [17, 61] \tabularnewline
		\hline 
		\hline 
	\end{tabular}
\end{table}

\begin{figure}
	\includegraphics[width=0.46\textwidth]{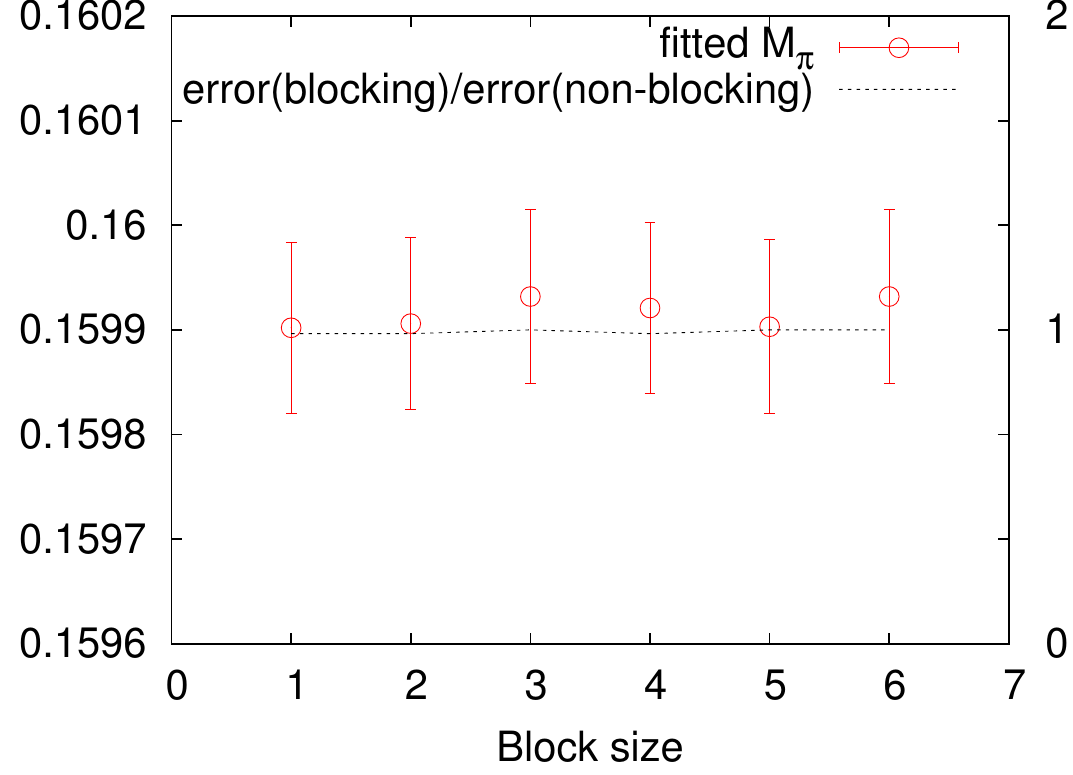}\hfill\includegraphics[width=0.46\textwidth]{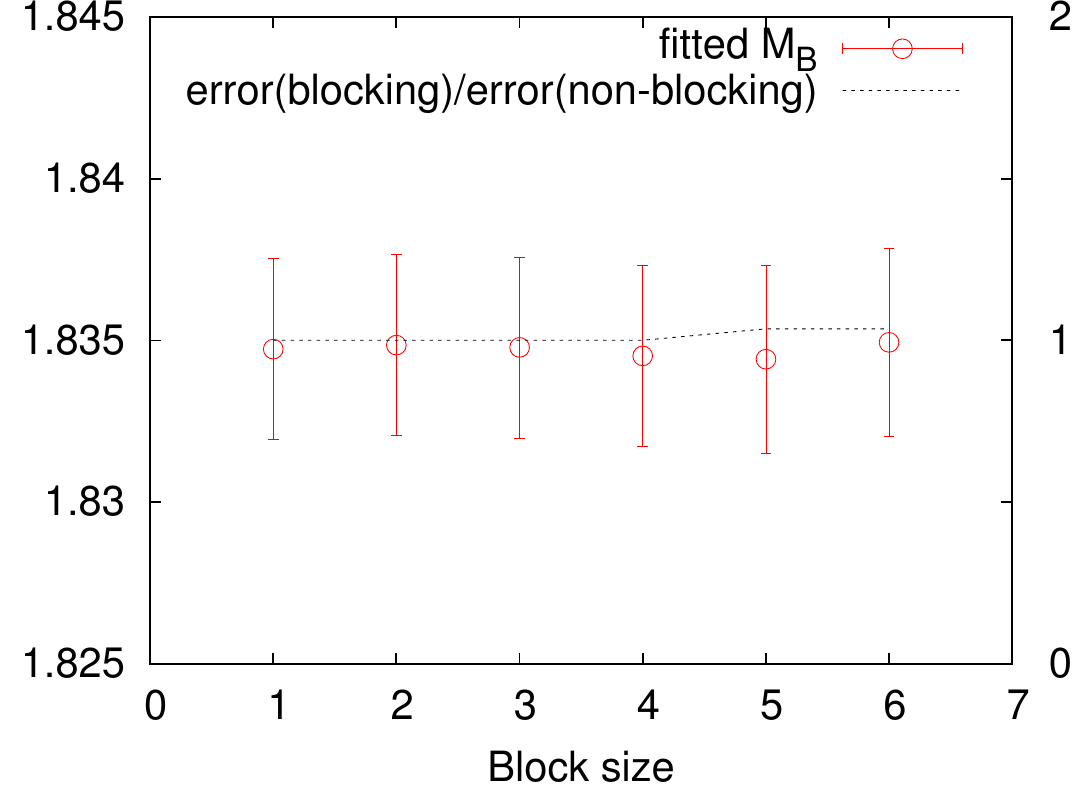} \\
	\includegraphics[width=0.46\textwidth]{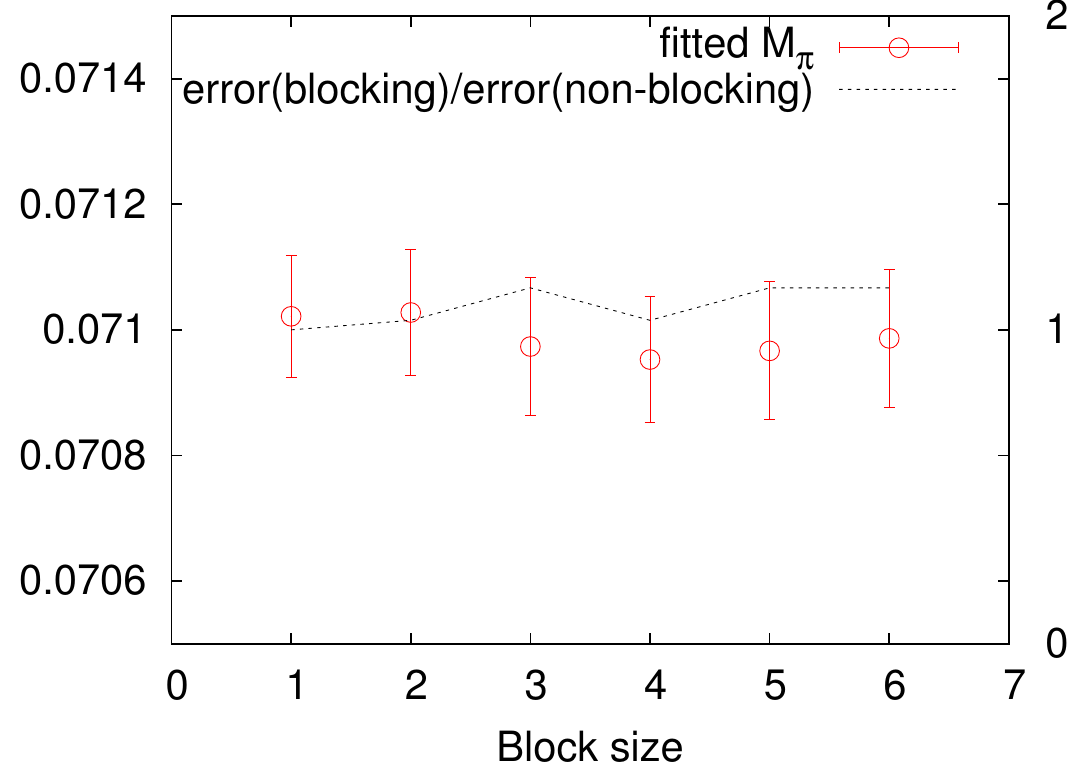}\hfill\includegraphics[width=0.46\textwidth]{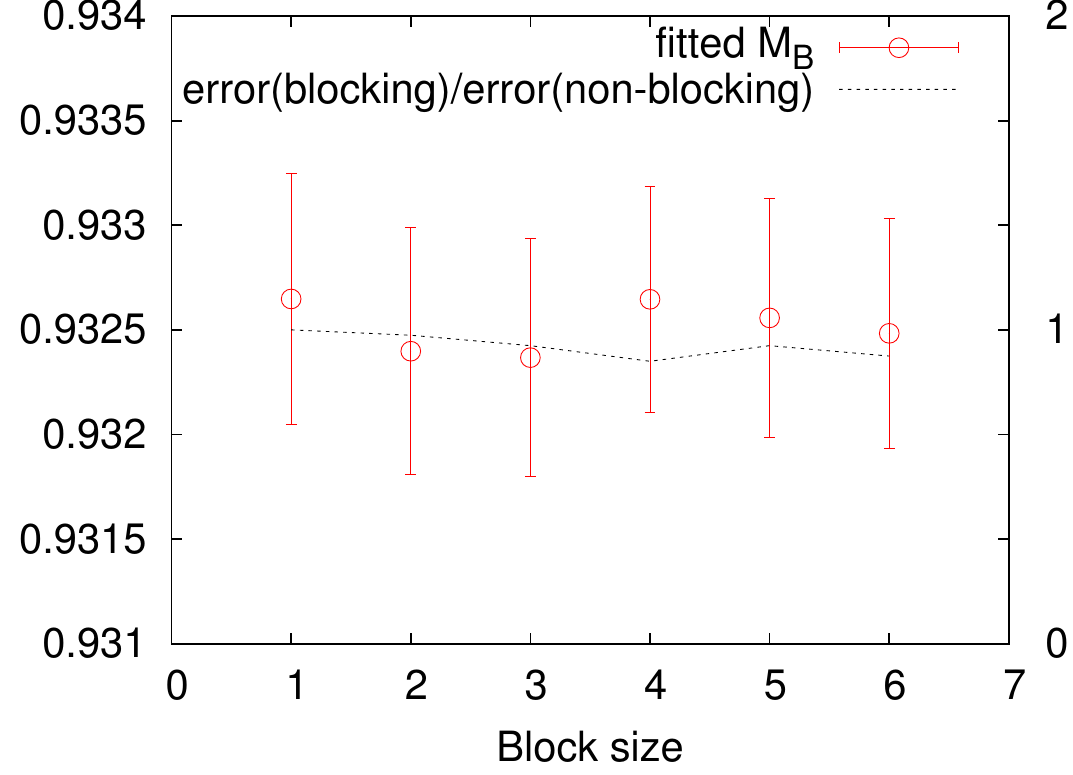}
	
	\caption{Fitted $M_{\pi}$ and $M_{B}$ in lattice units versus the block
		size for the \amlms{0.12}{0.1} ensemble (top) and for the \amlms{0.045}{0.2} ensemble (bottom). The dashed line shows the ratio of the fit errors from the blocking and non-blocking data, which can be read off from the right $y$-axis. \label{fig:MB_Mpi_blocking}}
\end{figure}

We test for autocorrelations by blocking the configurations on each ensemble with different block sizes, and then using a single-elimination jackknife procedure to propagate the
statistical error to the two-point correlator fits for $M_{\pi}$
and $M_{B}^{(0)}$. We do not observe any noticeable autocorrelations in all the ensembles we use, as illustrated in Fig.~\ref{fig:MB_Mpi_blocking} for the coarsest and finest ensembles, and choose not to block the data.

The ratios in Eq.~(\ref{eq:R_Gamma}) have the advantage that the wavefunction overlap factors $Z_{P}$ cancel,
but the trade-off is that we need an additional factor --- the
square root term on the right-hand side --- to remove the leading $t$ dependence
in the ratio. If the lowest lying states dominated
the ratio $R_J$, then it would be constant in $t$
and proportional to the lattice form factor $f_{J}$. The subscript $J$ now runs over $\perp$, $\parallel$, and $T$, corresponding
to the operators $V^i$, $V^4$, and $T^{4i}$, respectively. Our previous analysis employed a simple plateau fit constant in time. With our improved statistics, the small excited-state contributions to the ratio are significant and cannot be neglected. On the other hand, even with our improved statistics, we find that contributions to $R_J$ from wrong-parity states are still negligible. We use two different fit strategies to remove excited state contributions and use the consistency between them as an added check that any remaining excited state contamination is negligibly small.

\begin{figure}
	\centering{\includegraphics[width=0.6\linewidth]{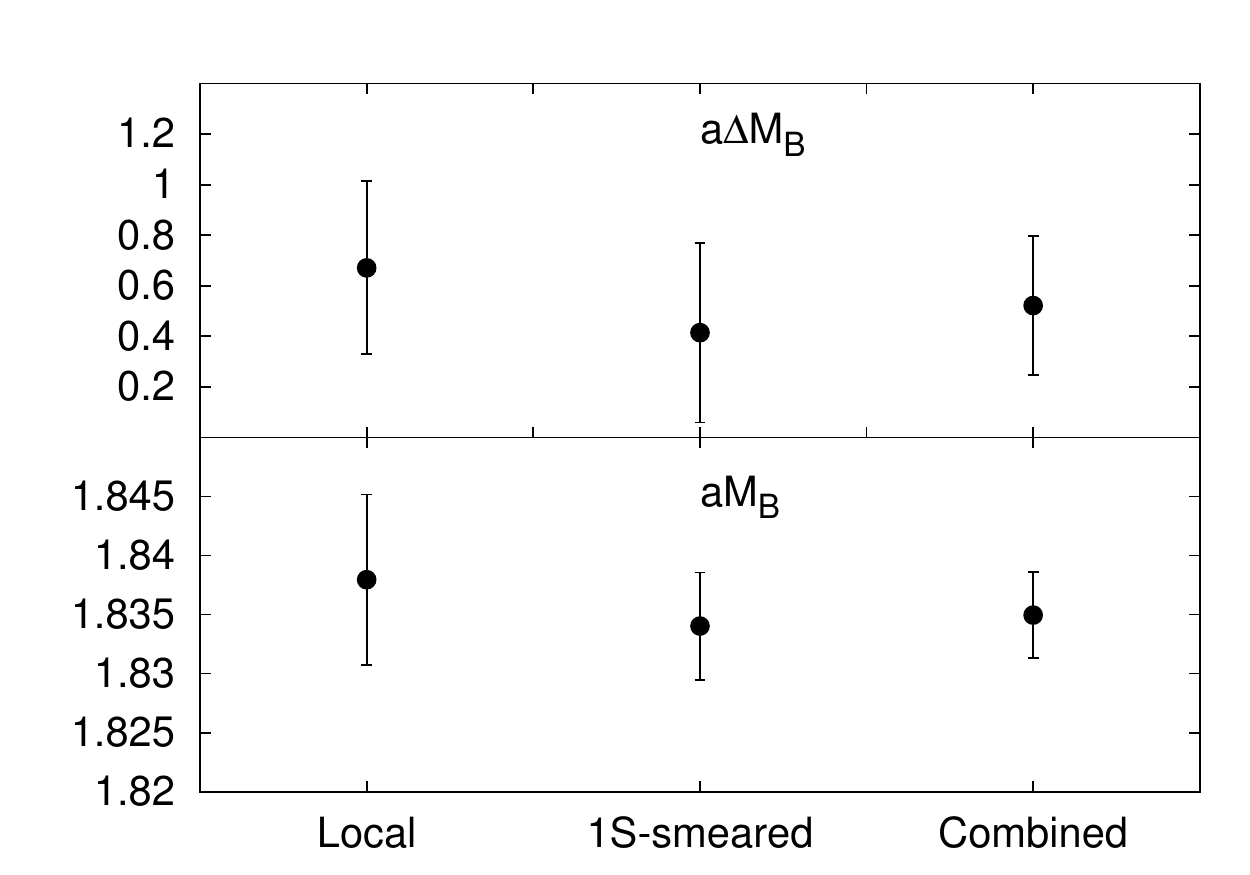}}
	
	\caption{Ground-state energy (bottom) and the first excited-state splitting (top)
		of the $B$ meson on the \amlms{0.12}{0.1} ensemble, from different fits to local, 1S-smeared
		and both two-point correlators. \label{fig:MB_1S_d}}
\end{figure}

The first strategy starts with the ratio in Eq.~(\ref{eq:R_Gamma}) and minimally extends the plateau fitting scheme by including the first excited state of the $B$ meson in the following form:
\begin{eqnarray}
\frac{R_{J}(t)}{h_{J}} & = & f_{J}^{\rm lat}\left[1+\mathcal{A}_{J}e^{-\Delta M_{B}(T-t)}\right],\label{eq:R_gamma_exp}
\end{eqnarray}
where $\mathcal{A}_J$ and $f_J^{\rm lat}$ are unconstrained fit parameters, $\Delta M_{B} \equiv M_B^{(2)}-M_B^{(0)}$ is the lowest energy splitting of the pseudoscalar $B$ meson, and the prefactors are $h_{\parallel}=1,\, h_{\perp}=p_{\pi}^{i}$ and $h_{T}=(\sqrt{2M_{B}}p_{\pi}^{i})/(M_{B}+M_{\pi})$. 
We choose the fit ranges for $R_J$ such that contributions from pion excited states to $R_J$ can be neglected.
The fit parameter $\Delta M_{B}$ is determined by the $B$-meson two-point correlators. In practice, we fit the ratio in Eq.
(\ref{eq:R_gamma_exp}) along with the $B$-meson two-point correlation functions with $\Delta M_B$ as a common parameter. We find it beneficial in the combined fit to include both the local and
smeared two-point correlation functions. We use 2+2 states for both correlators, but use a different set of fit ranges (listed in Table~\ref{tab:fit_range_ratio}). The results of these two-point fits are shown in Fig.~\ref{fig:MB_1S_d}. The agreement in the $B$-meson energies between the separate and combined fits is very good, but the combined fit leads to smaller errors.
\begin{table}
	\caption{The fit ranges $[t_\text{min},t_\text{max}]$ of the combined two-point and
		three-point ratio fits to obtain the lattice form factors. \label{tab:fit_range_ratio} }
	\begin{tabular}{ccccccc}
		\hline 
		\hline 
		$\approx a$(fm)  & $am'_{l}/am'_{h}$  & $C_B^{(d)}$  & $C_B^{(1S)}$  & $R_{\parallel}$  & $R_{\perp}$  & $R_{T}$  \tabularnewline
		\hline 
		0.12  & 0.01/0.05  & [9, 23]  & [7, 21]  & [6,12]  & [6,12]  & [6,12] \tabularnewline
		& 0.007/0.05  & [9, 23]  & [7, 21]  & [6,12]  & [6,12]  & [6,12] \tabularnewline
		& 0.005/0.05  & [9, 23]  & [7, 21]  & [6,12]  & [6,12]  & [6,12] \tabularnewline
		\hline 
		0.09  & 0.0062/0.031  & [12, 32]  & [9, 32]  & [9,16]  & [9,16]  & [9,16] \tabularnewline
		& 0.00465/0.031  & [12, 32]  & [9, 29]  & [9,16]  & [9,16]  & [9,16] \tabularnewline
		& 0.0031/0.031  & [12, 32]  & [9, 29]  & [9,16]  & [9,16]  & [9,16] \tabularnewline
		& 0.00155/0.031  & [12, 29]  & [9, 29]  & [9,15]  & [9,15]  & [9,15] \tabularnewline
		\hline 
		0.06  & 0.0072/0.018  & [13, 41]  & [9, 41]  & [12,22]  & [12,22]  & [12,22] \tabularnewline
		& 0.0036/0.018  & [13, 41]  & [9, 41]  & [12,22]  & [12,22]  & [12,22] \tabularnewline
		& 0.0025/0.018  & [13, 41]  & [9, 41]  & [12,22]  & [12,22]  & [12,22] \tabularnewline
		& 0.0018/0.018  & [13, 41]  & [9, 41]  & [12,21]  & [12,21]  & [12,21] \tabularnewline
		\hline 
		0.045  & 0.0028/0.014  & [16, 61]  & [10, 61]  & [16,26]  & [16,26]  & [16,26] \tabularnewline
		\hline
		\hline  
	\end{tabular}
\end{table}

\begin{figure}
	\center{\includegraphics[width=0.48\textwidth]{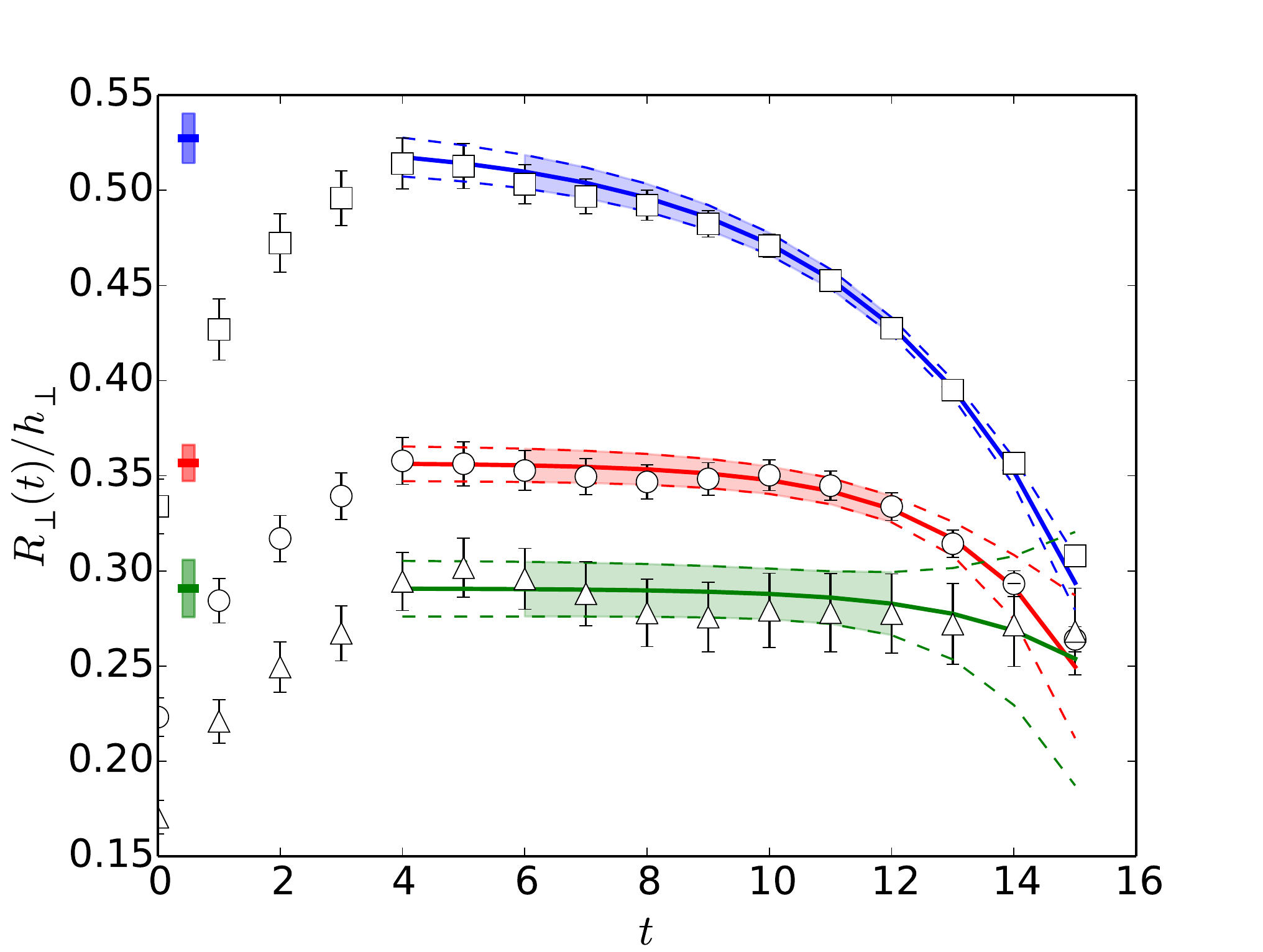} \hfill \includegraphics[width=0.48\textwidth, , trim = 60 60 80 60]{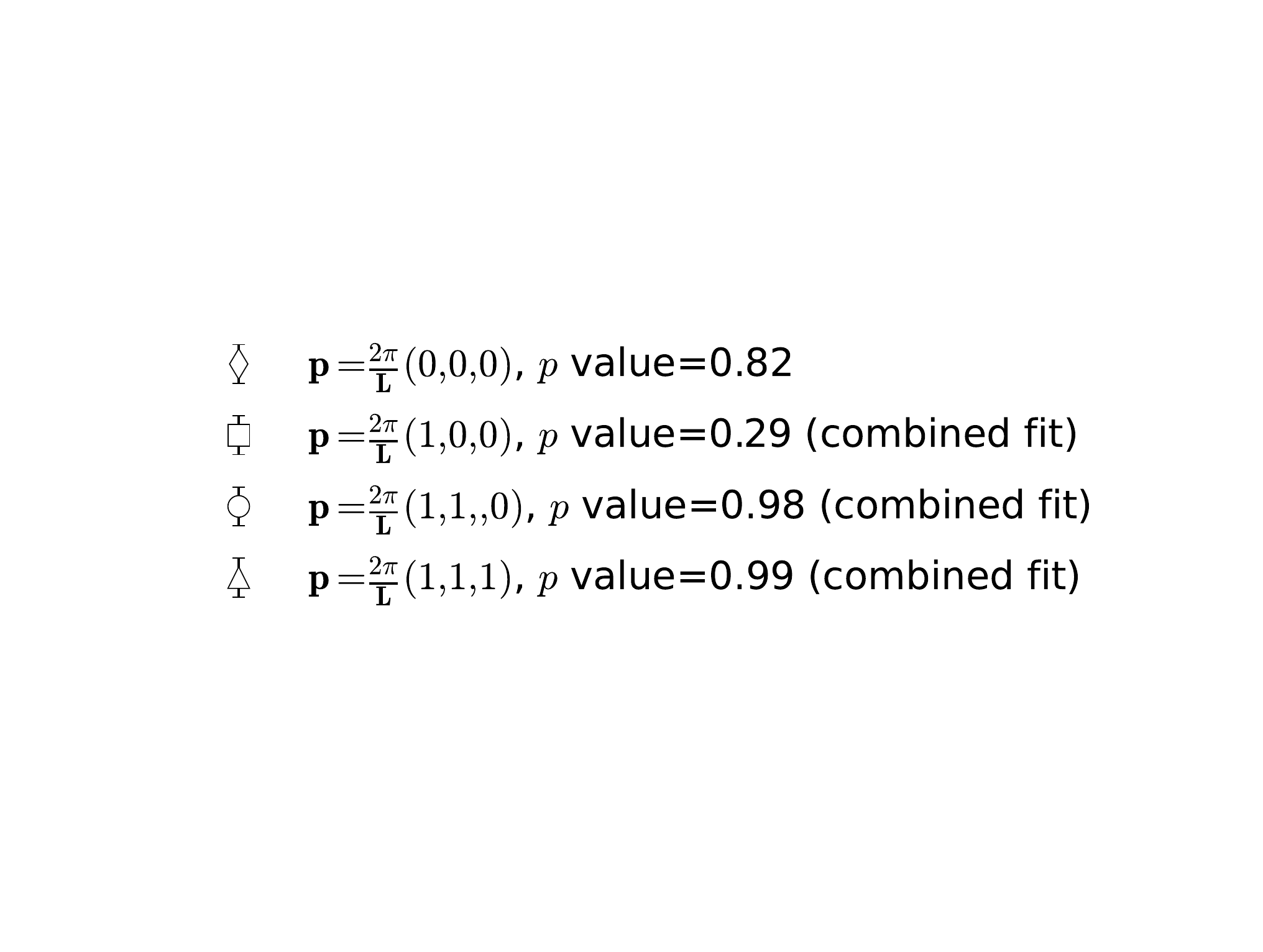}}
	\center{\includegraphics[width=0.48\textwidth]{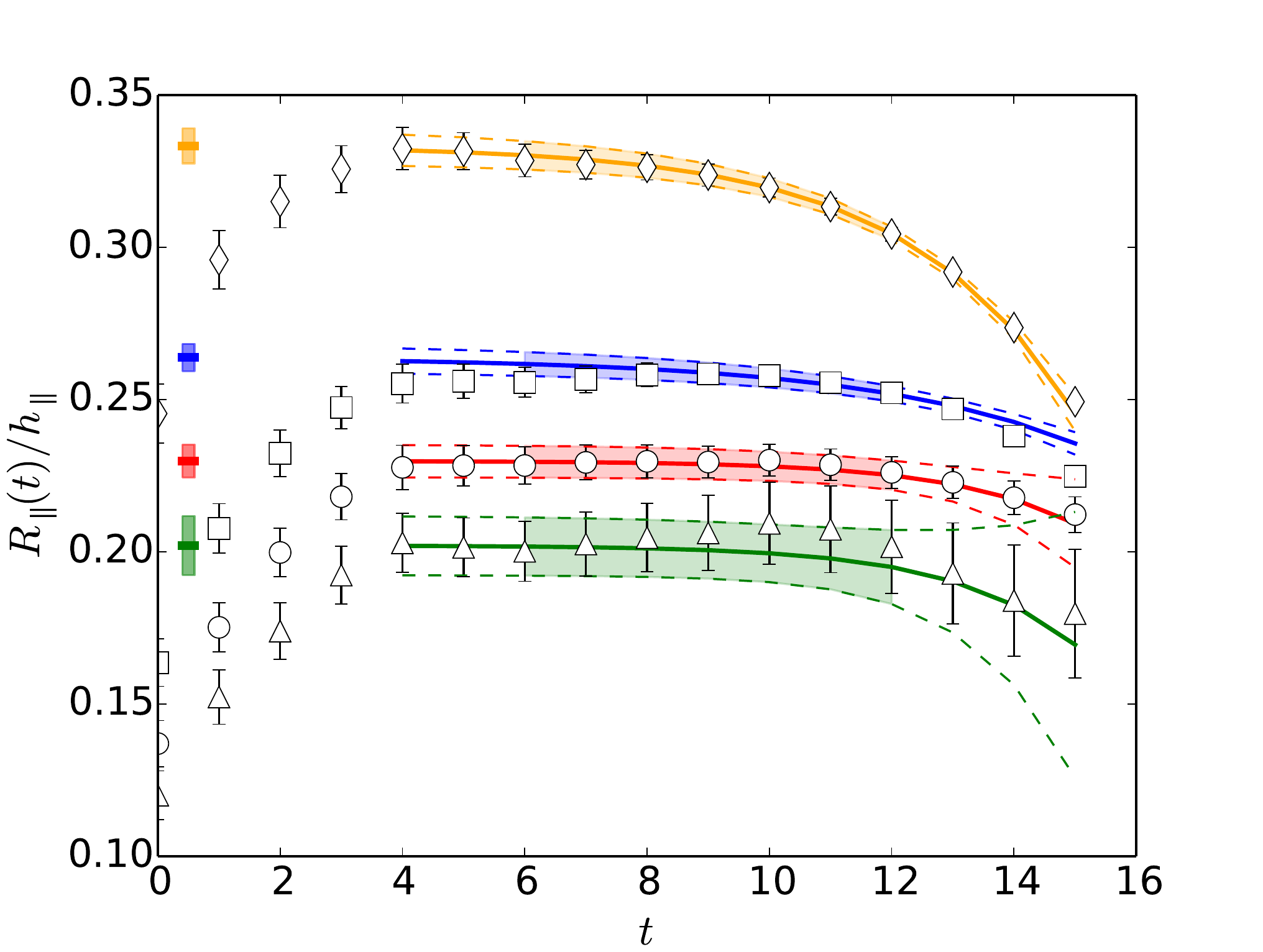}\hfill \includegraphics[width=0.48\textwidth]{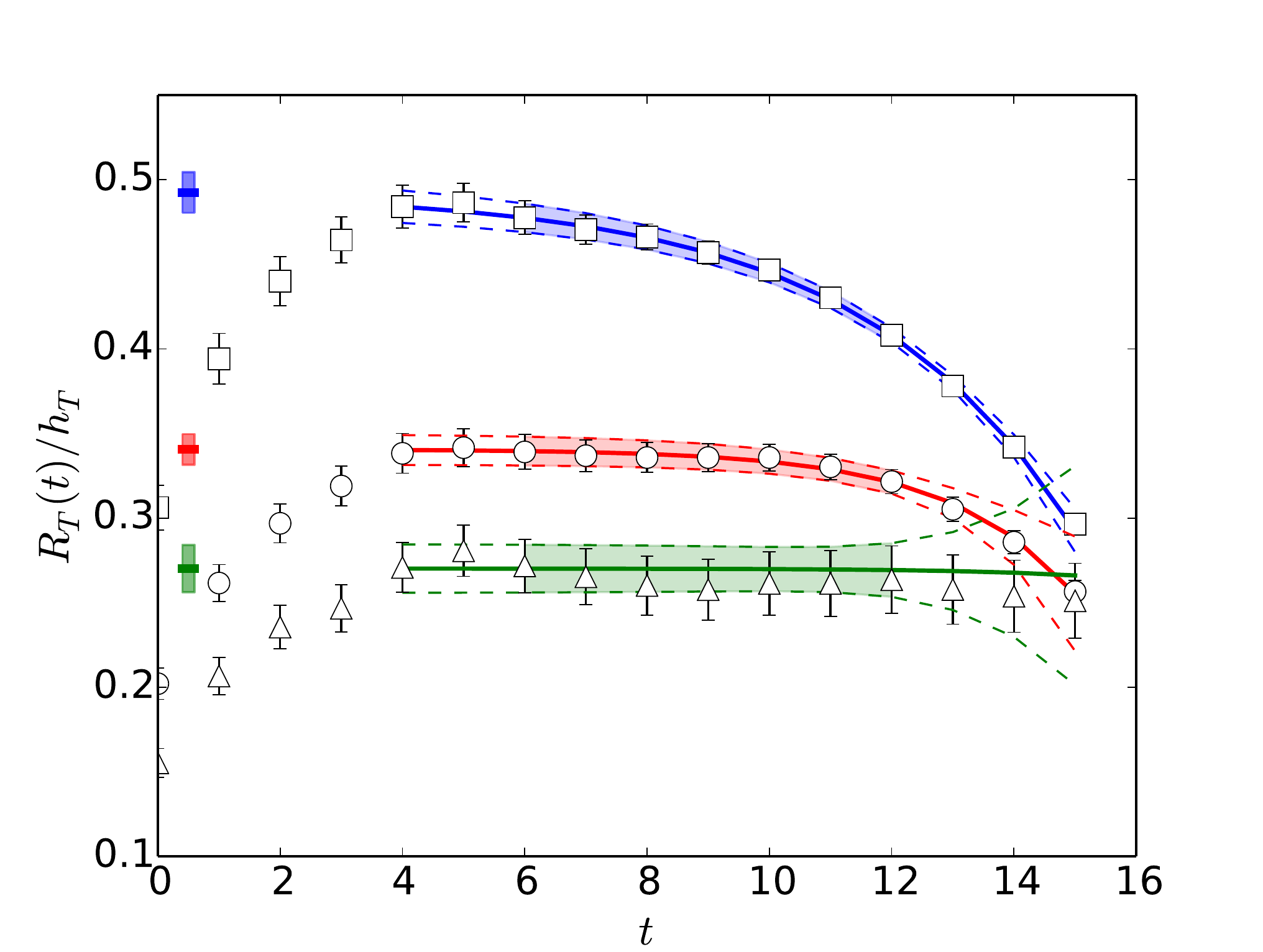}}

	\caption{ Fits of ratios $R_{\perp}/h_{\perp}$ (top left), $R_{\parallel}/h_{\parallel}$ (bottom left)
		and $R_{T}/h_{T}$ (bottom right) for the \amlms{0.12}{0.1} ensemble. The open data points with error bars
		are the ratios constructed from two- and three-point correlators with various momenta. The fit results are shown as solid lines with dashed error widths. The shaded bands indicate the range of data employed in the fit. The resulting lattice form factor $f_J$ are shown by
		the colored bars on the left of the plots. \label{fig:corr_fits}}
\end{figure}

\begin{figure}
	\includegraphics[scale=0.65]{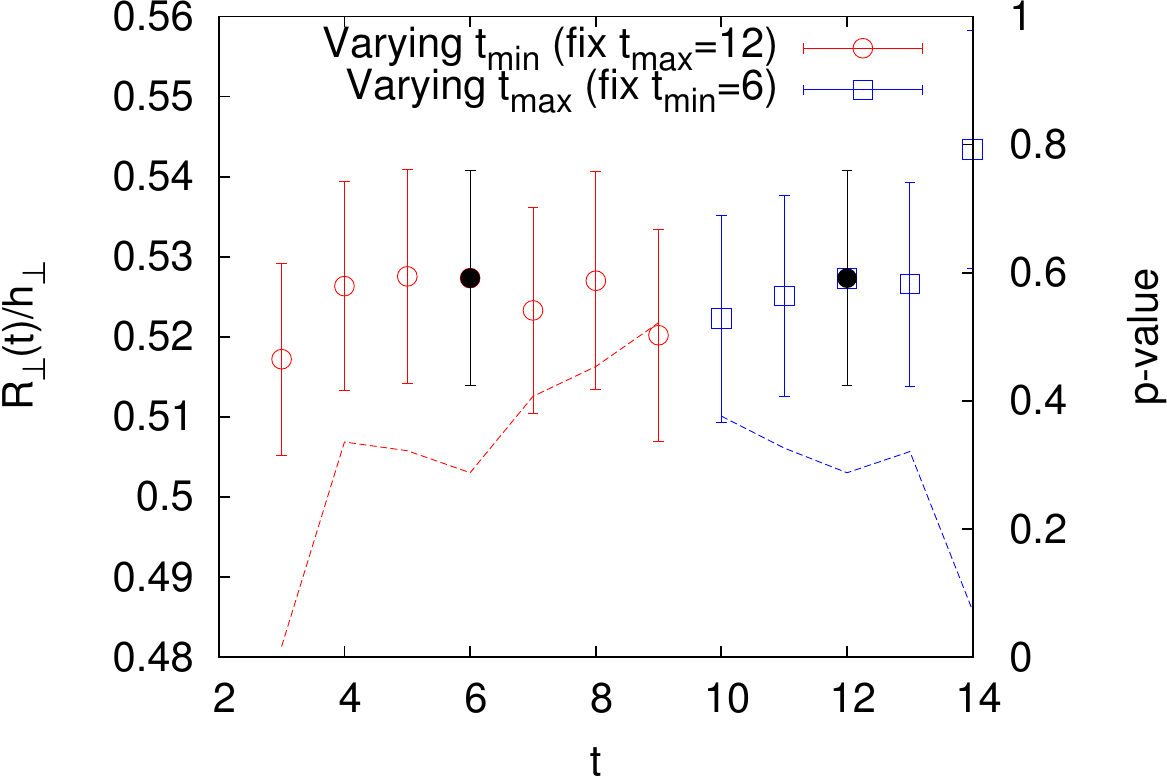}\;
	\includegraphics[scale=0.65]{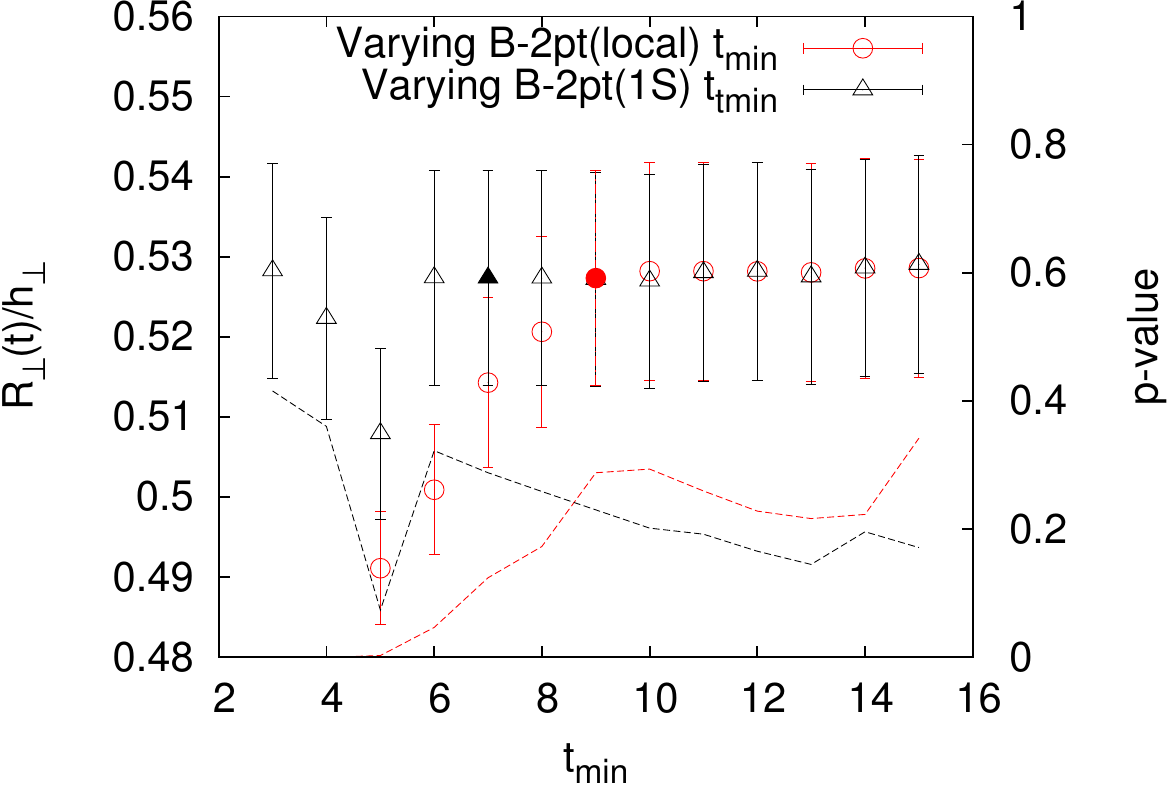}
	
	\caption{ Left: fitted ratio $R_{\perp}/h_{\perp}$ on the \amlms{0.12}{0.1} ensemble with momentum $\bm{p}=(2\pi/L)(1,0,0)$ for different combinations of $t_\text{min},t_\text{max}$ of the three-point
		correlator in the combined fit, where circles are results of varying $t_\text{min}$ with fixed $t_\text{max}=12$ and squares are that of varying $t_\text{max}$ with fixed $t_\text{min}=6$. Right: the same $R_{\perp}/h_{\perp}$ upon variations of $t_\text{min}$ of
		the two $B$-meson two-point correlators (local and $1S$-smeared) in the combined fit, where $t_\text{max}$ is given in Table~\ref{tab:fit_range_ratio}. The filled
		points in both plots show the preferred fit ranges and the dashed lines indicate
		the $p$ values of the fits. \label{fig:Stability-ratios}}
\end{figure}

To summarize our strategy, for the case of zero momentum, we fit the ratio $R_{\parallel}(t)$
together with the local and smeared $B$-meson two-point correlators $C_{B}^\text{(d)}$, $C_{B}^{\text{(1S)}}$
simultaneously. For non-zero momentum $\bm{p}$, $\Delta M_{B}$
is common to all three ratios,
$R_{\parallel},R_{\perp},R_{T}$. Thus, we perform a combined fit to the five quantities: $R_{\parallel},\; R_{\perp},\; R_{T},\; C_{B}^\text{(d)}$ and $C_{B}^{(\text{1S})}$. Figure~\ref{fig:corr_fits} shows an example
of these fits. Figure~\ref{fig:Stability-ratios} shows the stability plots of $R_\perp$ against the variations in the fit ranges of the ratio fits, and the variations in the fit ranges of both two-point correlators. The preferred fit ranges are set to be in the stable region upon these variations. 

\begin{figure}
	\center{\includegraphics[width=0.5\textwidth]{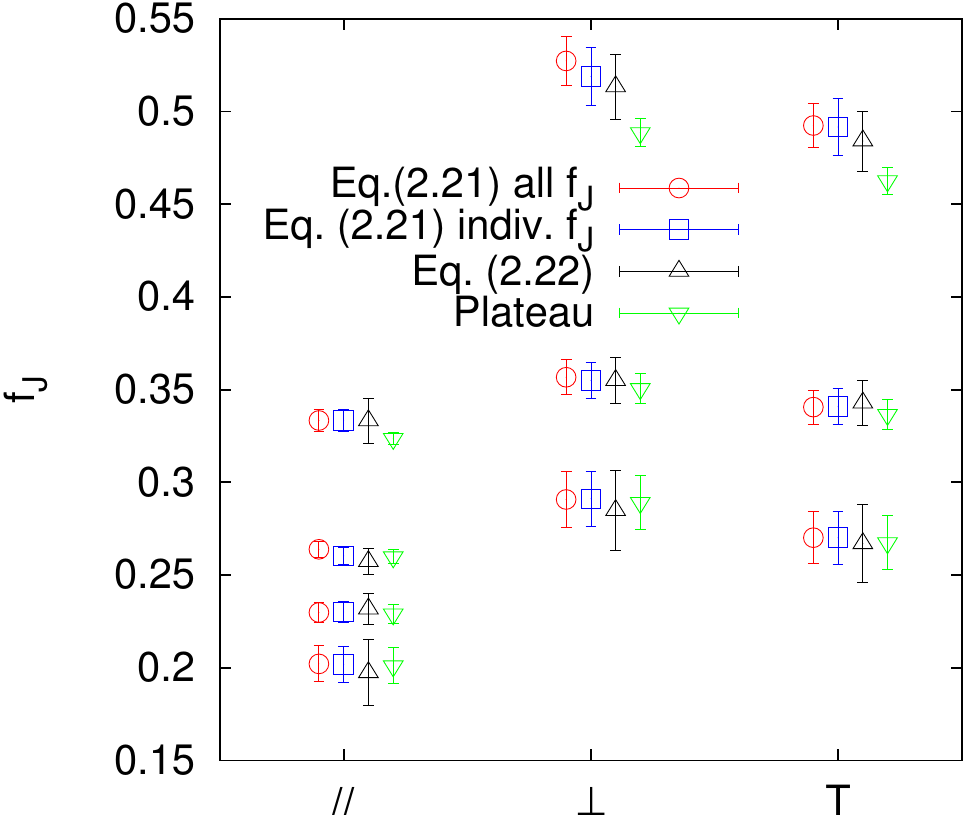}}
	
	\caption{Fitted lattice form factors $f_J$ at different momenta: from top to bottom, $(2\pi/L)(0,0,0)$ ($f_\parallel$ only), $(2\pi/L)(1,0,0)$, $(2\pi/L)(1,1,0)$ and $(2\pi/L)(1,1,1)$. Each compares the following different fit schemes: separate (squares) or combined (circles) fits of the three lattice form factors $f_J$ using Eq.~(\ref{eq:R_gamma_exp}), fit including the full
		first excited states as defined in Eq.~(\ref{eq:R_Gamma_nonavg})
		(upward-pointing triangles) and the simple plateau fit (downward-pointing triangles).
		\label{fig:corr_fits_schemes}}
\end{figure}

Our second fit strategy includes excited-state contributions from both the pion and the $B$ meson. It starts with a different ratio, without time averages, which ensures that there are enough data points to constrain all the parameters:
\begin{eqnarray}
\tilde{R}_{J}(t) & = & \frac{C_{J}(t,T;\bm{p})}{C_{\pi}(t;\bm{p})C_{B}(T-t;\bm{0})},\label{eq:R_Gamma_nonavg}
\end{eqnarray}
where the two- and three-point correlators are defined in Eqs.~(\ref{eq:def_C_2pt}) and (\ref{eq:def_C_3pt}). We fit $\tilde{R}_{J}(t)$ with all the possible states with $m,n\leq 2$ in Eqs.~(\ref{eq:C_2pt}) and (\ref{eq:C_3pt}), combining the fits to the pion and $B$-meson two-point correlators. We compare the fit results of the two different
fit schemes in Fig.~\ref{fig:corr_fits_schemes}. The first (simple) fit model described in Eq.~(\ref{eq:R_gamma_exp}) gives, fitting either simultaneously or individually to the three lattice form factors $f_J$, results that are consistent with the second
fit model that includes the full set of first excited states in Eq.~(\ref{eq:R_Gamma_nonavg}). In contrast, the plateau
fits to $R_J$ defined in Eq.~(\ref{eq:R_Gamma}) yield results that are as much as one statistical $\sigma$ smaller. 
%about one $\sigma$ smaller inthe case of zero momentum for $R_\parallel$ and p100 momentum for $R_{\perp, T}$.
In summary, we find that the first fit strategy described by Eq.~(\ref{eq:R_gamma_exp}) is sufficient to remove contributions from excited states, and we therefore adopt this method for the main analysis. 

\subsection{Matching} \label{subsec:renormalization}

We match the lattice currents to continuum QCD with the relation, 
\begin{eqnarray}
\mathcal{J} \doteq Z_{J_{bl}} J, \label{eq:current_matching}
\end{eqnarray}
where $\mathcal{J}$ and $J$ denote the vector or tensor currents in the continuum and lattice theories, respectively, and ``$\doteq$'' means ``has the same matrix elements'' \cite{Kronfeld:2000ck}. We calculate the current renormalization with the mostly nonperturbative renormalization method \cite{hep-ph/0101023, Harada:2001fi}, 
\begin{eqnarray}
Z_{J_{bl}} & = & \rho_{J_{bl}}\sqrt{Z_{V^4_{bb}}Z_{V^4_{ll}}},\label{eq:Z_bl}
\end{eqnarray}
where $Z_{V^4_{bb}}$ and $Z_{V^4_{ll}}$ are the matching factors for
the corresponding flavor-conserving vector currents. These factors capture most of the current renormalization. The remaining flavor off-diagonal contribution to the matching factor, $\rho_{J_{bl}}$, is close to unity.

We calculate the factors $Z_{V^4_{bb}}$ and $Z_{V^4_{ll}}$ nonperturbatively
for each ensemble by computing the matrix elements of the flavor-conserving vector currents and using the relations 
\begin{eqnarray}
1 & = & Z_{V^4_{ll}}\langle\pi|V^4_{ll}|\pi\rangle,\label{eq:ZV_ll}\\
1 & = & Z_{V^4_{bb}}\langle B_{s}|V^4_{bb}|B_{s}\rangle , \label{eq:ZV_bb}
\end{eqnarray}
where the lattice current $V^4_{ll}$ is a bilinear of light staggered
quark fields and $V^{4}_{bb}$ is a bilinear of clover heavy quark
fields. The
factors $Z_{V^4_{bb}}$ and $Z_{V^4_{ll}}$ are listed in Table~\ref{tab:renorm_factors}.
Because there is very little $m_{l}$ dependence in the factor $Z_{V^4_{ll}}$, we use the same $Z_{V^4_{ll}}$ for
ensembles with different light quark masses but the same lattice spacing. The factor $Z_{V^4_{bb}}$ depends crucially on the heavy $b$ quark mass, though it has negligible light quark mass dependence.

We use lattice perturbation theory \cite{hep-lat/9209022} to compute the remaining renormalization
factors $\rho_{J}$ at one-loop. Due to the cancellation of the tadpole
contributions in the radiative corrections to the left and right side of Eq.~(\ref{eq:Z_bl}), the factors $\rho_{J}$ are
very close to one. They have the perturbative expansion 
\begin{eqnarray}
\rho_{J} & = & 1+\alpha_{V}(q^{*})\rho_{J}^{[1]}+O(\alpha_{V}^{2}),\label{eq:rho_Gamma}
\end{eqnarray}
where we take the strong coupling in the $V$-scheme \cite{hep-lat/9209022} at a scale $q^*$ that corresponds to the typical gluon loop momentum. In practice, we choose $q^*=2/a$. The details of the calculation of the one-loop coefficients $\rho_{J}^{[1]}$ will be presented elsewhere. The values used in this work are shown in Table~\ref{tab:renorm_factors}.

\begin{table}
	\caption{The parameters for the renormalization of the form factors. The first two columns label the ensemble with its approximate lattice spacing and the sea light- and strange-quark mass ratio. The third column is the simulation $\kappa'_b$. The fourth and fifth columns are the nonperturbative heavy-heavy and light-light renormalization factors $Z_{V^4_{bb}}$, $Z_{V^4_{ll}}$. The sixth, seventh, and eighth columns are the one-loop estimates of $\rho_{V^4}$, $\rho_{V^i}$ and $\rho_T$, respectively. The tensor current has a nonzero anomalous dimension; the numbers reported here match to the $\overline{\rm MS}$ scheme at renormalization scale $\mu=m_2$, which corresponds to the pole mass. 
	Note that with our convention $Z_{V^4_{ll}}$ and $Z_{V^4_{bb}}$ are normalized so that at tree-level
	$Z^{[0]}_{V^4_{ll}} = 2$ and $Z^{[0]}_{V^4_{bb}} = 1 - 6 u_0 \kappa_b'$~\cite{1112.3051}. As a result, $Z_{V_4^{bb}}$ only approaches 1 in the limit $m_b \to \infty$.
		\label{tab:renorm_factors}}
	\begin{tabular}{ccccccccc}
		\hline 
		\hline 
		$\approx a$(fm)  & $am'_{l}/am'_{h}$  & $\kappa'_b$  & $Z_{V^4_{bb}}$  & $Z_{V^4_{ll}}$  & $\rho_{V^{4}_{bl}}$  & $\rho_{V^{i}_{bl}}$ & $\rho_{T_{bl}}$ \tabularnewline
		\hline 
		0.12   & 0.01/0.05  & 0.0901  & 0.5015(8)  & 1.741(3)       & 1.006214 & 0.973023 & 1.033350    \tabularnewline
		& 0.007/0.05  & 0.0901  & 0.5015(8)  & 1.741(3)     & 1.006252 & 0.973109 & 1.033280   \tabularnewline
		& 0.005/0.05  & 0.0901  & 0.5015(8)  & 1.741(3)     & 1.006197 & 0.973082 & 1.033270   \tabularnewline
		& 0.01/0.05  & 0.0860  & 0.5246(9)  & 1.741(3)      & 1.012999 & 0.977290 & 1.030650   \tabularnewline
		& 0.01/0.05  & 0.0820  & 0.5469(10)  & 1.741(3)     & 1.018261 & 0.980129 & 1.028960   \tabularnewline
		\hline 
		0.09   & 0.0062/0.031  & 0.0979  & 0.4519(15)  & 1.776(5)  & 0.999308 & 0.975822 & 1.036590   \tabularnewline
		& 0.00465/0.031  & 0.0977  & 0.4530(15)  & 1.776(5) & 0.999405 & 0.975775 & 1.036390   \tabularnewline
		& 0.0031/0.031  & 0.0976  & 0.4536(15)  & 1.776(5)  & 0.999441 & 0.975744 & 1.036350   \tabularnewline
		& 0.00155/0.031  & 0.0976  & 0.4536(15)  & 1.776(5) & 0.999416 & 0.975703 & 1.036390   \tabularnewline
		\hline
		0.06   & 0.0072/0.018  & 0.1048  & 0.4089(21)  & 1.807(7)  & 0.995605 & 0.979279 & 1.042390   \tabularnewline
		& 0.0036/0.018  & 0.1052  & 0.4065(21)  & 1.807(7)  & 0.995371 & 0.979260 & 1.043160   \tabularnewline
		& 0.0025/0.018  & 0.1052  & 0.4065(21)  & 1.807(7)  & 0.995350 & 0.979217 & 1.043190   \tabularnewline
		& 0.0018/0.018  & 0.1052  & 0.4065(21)  & 1.807(7)  & 0.995327 & 0.979176 & 1.043250   \tabularnewline
		\hline
		0.045  & 0.0028/0.014  & 0.1143  & 0.3564(65)  & 1.841(6)  & 0.994195 & 0.984351 & 1.058790   \tabularnewline
		\hline
		\hline  
	\end{tabular}

\end{table}

\subsection{Heavy-quark mass correction} \label{subsec:tuning}

In the clover action, the hopping parameter $\kappa_b$ corresponds to the bare $b$-quark mass. When we started generating data for this analysis, we had a good estimate for the bottom-quark $\kappa'_b$ on each ensemble, but not the final tuned values, which were obtained as described in Appendix~C of Ref.~\cite{Bailey:2014tva}. We therefore need to adjust the form factors {\it a posteriori} to account for the slightly mistuned values of $\kappa_b$.

The $\kappa_b$ parameters are adjusted so that the corresponding $B_s$ kinetic masses match the experimentally-measured value \cite{Bailey:2014tva}. Table~\ref{tab:tuned_kappa} shows both the simulation and final tuned $\kappa_b$ values. For some ensembles, the difference between the two is as large as 7$\sigma$ of the statistical uncertainty associated with the tuning procedure. We study the $\kappa_b$-dependence of the lattice form factors by generating data on the \amlms{0.12}{0.2} ensemble, with two additional $\kappa'_b$ values, $0.0860$ and $0.0820$, and all other simulation parameters unchanged. 
\begin{table}
	\caption{Parameters needed to apply heavy-quark mass corrections. The third column contains the value $\kappa'_b$ used for the calculation, the fourth column contains the tuned value $\kappa_b$ with its statistical error. Subsequent columns contain the percentage shift in $m_2$ and each of the form factors. \label{tab:tuned_kappa}}
	\begin{tabular}{cccccccc}
		\hline
		\hline  
		$a$($\approx$fm)  & $am'_{l}/am'_{h}$  & $\kappa_b'$  & $\kappa_b$ \;&\; $\frac{\Delta m_2}{m_2}$(\%) \;&\; $\frac{\Delta f_\perp}{f_\perp}$(\%) \;&\; $\frac{\Delta f_\parallel}{f_\parallel}$(\%) \;&\; $\frac{\Delta f_T}{f_T}$(\%) \tabularnewline
		\hline 
		0.12  & 0.01/0.05  & 0.0901  & 0.0868(9)  	& 10.9 	& 1.79	& 1.55	& 1.60	\tabularnewline
		& 0.007/0.05  & 0.0901  & 0.0868(9)  		& 10.9	& 1.80	& 1.57	& 1.58	\tabularnewline
		& 0.005/0.05  & 0.0901  & 0.0868(9)  		& 10.9	& 1.81	& 1.58	& 1.56	\tabularnewline
		\hline 
		0.09  & 0.0062/0.031  & 0.0979  & 0.0967(8) & 4.3	& 0.69	& 0.60	& 0.62	\tabularnewline
		& 0.00465/0.031  & 0.0977  & 0.0966(8)  	& 3.9	& 0.63	& 0.55	& 0.56	\tabularnewline
		& 0.0031/0.031  & 0.0976  & 0.0965(8)  	& 3.9	& 0.64	& 0.56	& 0.55	\tabularnewline
		& 0.00155/0.031  & 0.0976  & 0.0964(8)  	& 4.2	& 0.70	& 0.61	& 0.59	\tabularnewline
		\hline 
		0.06  & 0.0072/0.018  & 0.1048  & 0.1054(5) &$-$2.4	& $-$0.37	& $-$0.36	& $-$0.44	\tabularnewline
		& 0.0036/0.018  & 0.1052  & 0.1052(5)  	& 0.0	& 0.0	& 0.0	& 0.0	\tabularnewline
		& 0.0025/0.018  & 0.1052  & 0.1051(5)  	& 0.4	& 0.06	& 0.06	& 0.06	\tabularnewline
		& 0.0018/0.018  & 0.1052  & 0.1050(5)  	& 0.8	& 0.13	& 0.11	& 0.11	\tabularnewline
		\hline 
		0.045  & 0.0028/0.014  & 0.1143  & 0.1116(4)& 14.3	& 2.34	& 2.03	& 2.10	\tabularnewline
		\hline 
		\hline 
	\end{tabular}
\end{table}
To generalize the $\kappa_b$ dependence from this ensemble
to others, we work with the quark kinetic mass $m_2$ instead of $\kappa_b$ itself. We expand the form factor $f$ ($f=f_{\perp,\parallel,T})$
in $m_{2}^{-1}$ about a reference point $\bar{m}_{2}^{-1}$ (which corresponds to the tuned $\kappa_b$) as follows \begin{eqnarray}
f(m_{2}^{-1},a^{2},m_{l},E_{\pi}) & \approx & f(\bar{m}_{2}^{-1},a^{2},m_{l},E_{\pi})+\frac{\partial f(\bar{m}_{2}^{-1},a^{2},m_{l},E_{\pi})}{\partial\bar{m}_{2}^{-1}}\left(m_{2}^{-1}-\bar{m}_{2}^{-1}\right)\nonumber \\
& = & f(\bar{m}_{2}^{-1},a^{2},m_{l},E_{\pi})\left[1+\frac{1}{f}\frac{\partial f}{\partial\bar{m}_{2}^{-1}}\left(m_{2}^{-1}-\bar{m}_{2}^{-1}\right)\right]\nonumber\\
& = & f(\bar{m}_{2}^{-1},a^{2},m_{l},E_{\pi})\left[1-\frac{\partial\ln f}{\partial\ln\bar{m}_{2}}(\frac{\bar{m}_{2}}{m_{2}}-1)\right], \label{eq:kappa_tune_f}
\end{eqnarray}
where the masses and $E_\pi$ are all in $r_1$ units. To obtain $f$ at the reference point, we need to
find the dimensionless normalized slope $-(\partial \ln f/\partial \ln \bar{m}_{2})$.

\begin{figure}
	\includegraphics[width=0.48\linewidth, trim = 50 20 50 20]{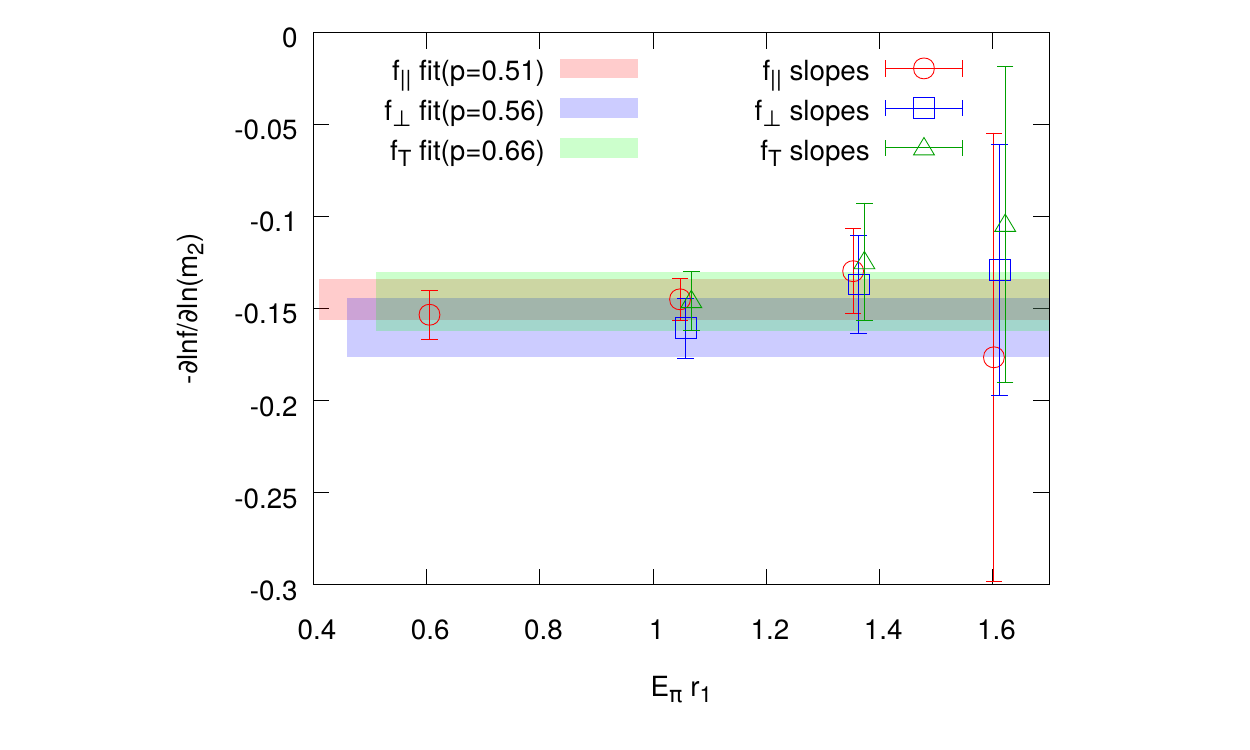}
	\hfill
	\includegraphics[width=0.48\linewidth, trim = 50 20 50 20]{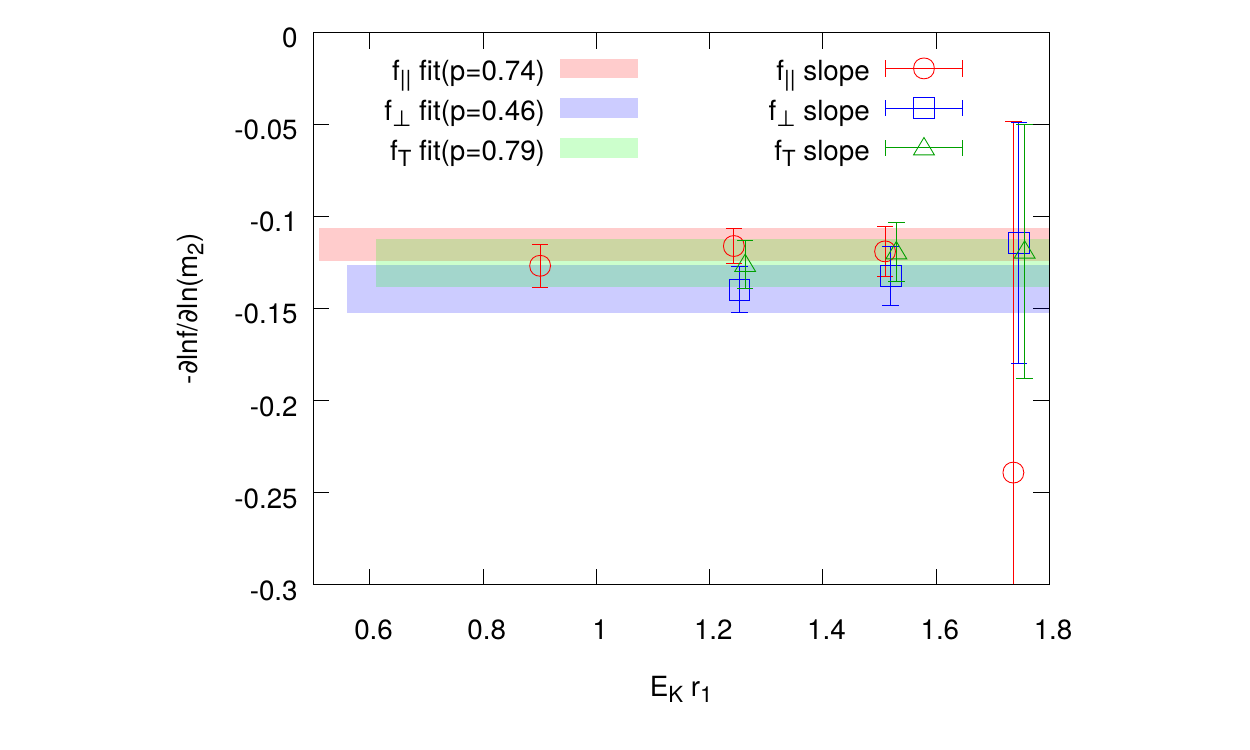}
	
	\caption{The normalized slopes $-(\partial \ln f/\partial \ln \bar{m}_{2})$ of the three different form factors $f_{\parallel,\perp,T}$
		at several momenta for $B\to\pi$ (left) and $B\to K$ (right) semileptonic decays. The horizontal shaded error
		bands in each plot are the results of correlated fits to all momenta for each form factor. \label{fig:normalized_slope}}
\end{figure}

We use exactly the same procedure as described in Sec.~\ref{subsec:correlator_fits} for $\kappa'_b=0.0901$ to obtain the $B\to\pi\ell\nu$
form factors $f_{\parallel,\perp,T}$ for the additional values $\kappa'_b=0.0860$ and $0.0820$. We apply the matching factors given in Table~\ref{tab:renorm_factors}. Finally, we take $\bar{m}_{2}$ to be the kinetic mass corresponding to $\kappa_b=0.0868$ (the tuned kappa given in Table \ref{tab:tuned_kappa}) and use it as the reference point. We fit each form factor at each momentum for the three data points to the linear form given in Eq.~(\ref{eq:kappa_tune_f}), taking $f(\bar{m}_{2}^{-1})$ and $-(\partial \ln f/\partial \ln \bar{m}_{2})$ as fit parameters. The result is shown in Fig.~\ref{fig:normalized_slope} (left). As shown in
the plot, the normalized slope $-(\partial \ln f/\partial \ln m_{2})$ has a very mild $E_{\pi}$
dependence. Therefore, for each form factor we perform a correlated fit to all momenta to obtain a single common normalized slope. The result is shown in Table \ref{tab:normalized_slope}. Fitting the data to a linear form in $E_{\pi}$ results in a slope statistically consistent with zero.  

To examine the light-quark mass dependence of the
normalized slopes, we repeat the same procedure for the $B\to K$ semileptonic form factors with a heavier daughter valence quark $am_s = 0.0349$, which is close to the physical strange-quark mass.
The results are plotted in Fig.~\ref{fig:normalized_slope} (right). We fit the points of each form factor to a constant and tabulate the results in Table \ref{tab:normalized_slope}. Comparing the normalized slopes for $f^{B\to\pi}$ and $f^{B\to K}$, taking into account statistical correlations, we observe a mild but statistically-significant light daughter-quark mass dependence. So we fit the slopes for $f^{B\to\pi}$ and $f^{B\to K}$ simultaneously to a linear form,
\begin{equation}
-\frac{\partial \ln f}{\partial \ln \bar{m}_{2}} = c + d\; \frac{m_l}{m_s},
\end{equation}
where $m_l/m_s = 0.2$ and $1.0$ for $f^{B\to\pi}$ and $f^{B\to K}$, respectively. The results for the parameters $c$ and $d$ are given in Table \ref{tab:normalized_slope}. Note that the results in Table \ref{tab:normalized_slope} are also used in Ref.~\cite{ranzhou}.

\begin{table}[b]
	\caption{Fitted normalized slopes to all momenta for $f^{B\to\pi}$, $f^{B\to K}$ and their combined fit with daughter-quark mass dependence.  \label{tab:normalized_slope}}
	\begin{tabular}{cccccccc}
		\hline 
		\multirow{2}{*}{} & \multicolumn{2}{c}{$B\to\pi$} & \multicolumn{2}{c}{$B\to K$} & \multicolumn{3}{c}{Combined fit with $m_l/m_s$ dependence}\tabularnewline
		\cline{2-8} 
		\;\;&\;\; $p$ value \;\;&\;\; $\frac{\partial\ln f}{\partial\ln\bar{m}_{2}}$ \;\;&\;\; $p$ value \;\;&\;\; $\frac{\partial\ln f}{\partial\ln\bar{m}_{2}}$ \;\;\;&\;\;\; $p$ value & $c$ & $d$\tabularnewline
		\hline 
		\hline 
		$f_{\perp}$ & 0.56 & 0.145(11) & 0.46 & 0.115(9) & 0.4 & $-$0.146(11) & 0.032(3)\tabularnewline
		$f_{\parallel}$ & 0.51 & 0.160(16) & 0.74 & 0.139(13) & 0.84 & $-$0.165(17) & 0.025(8)\tabularnewline
		$f_{T}$ & 0.66 & 0.146(16) & 0.79 & 0.126(13) & 0.88 & $-$0.137(17) & 0.034(8)\tabularnewline
		\hline 
	\end{tabular}
\end{table}

We use the parameters $c$ and $d$ in Table \ref{tab:normalized_slope} to determine the normalized slope $-(\partial \ln f/\partial \ln \bar{m}_{2})$ for each ensemble. Although the dependence of the normalized slopes on the light daughter-quark mass is resolvable, the effects are small for the ensembles we use in the analysis (with light daughter-quark masses ranging from $0.05m_s$ to $0.4m_s$). We expect similarly small effects from the spectator-quark masses. We also expect that the lattice-spacing dependence of the normalized slopes is small, because it is a dimensionless ratio. We therefore correct each lattice form factor in each ensemble by a factor 
\begin{eqnarray}
1+\frac{\Delta f}{f} & = & \left[1-\frac{\partial\ln f}{\partial\ln\bar{m}_{2}}\left(\frac{\bar{m}_{2}}{m_{2}}-1\right )\right]^{-1}, \label{eq:df_f}
\end{eqnarray}
where $m_2$ and $\bar{m}_2$ are the kinetic masses corresponding to the simulation $\kappa'_b$ and tuned $\kappa_b$, respectively. The resulting relative shift for each ensemble is shown in Table \ref{tab:tuned_kappa}. Although the corrections to $\kappa_b$ itself are significant for some ensembles, the corresponding corrections to the form factors are much smaller ($\lesssim 2.3\%$), as a consequence of the small normalized slopes.

\section{Chiral-continuum extrapolation}\label{secIV}

Here we extrapolate the
form factors at four lattice spacings
with several unphysical light-quark masses to the continuum limit and physical light-quark mass. We use heavy-meson rooted staggered chiral perturbation
theory (HMrS$\chi$PT) \cite{0704.0795,Aubin:2005aq}, in the hard-pion
and SU(2) limits. We also incorporate heavy-quark discretization
effects into the chiral-continuum extrapolation.

\subsection{SU(2) staggered chiral perturbation theory in the hard-pion limit}

The full-QCD next-to leading order (NLO) HMrS$\chi$PT expression for the semileptonic
form factors can be written \cite{0704.0795}
\begin{eqnarray}
f_{J}^\text{NLO} & = & f_{J}^{(0)}\left[ c_{0}^{J}(1+\delta f_{J,\text{logs}})+c_{1}^{J}\chi_\text{val}+c_{2}^{J}\chi_\text{sea}+c_{3}^{J}\chi_{E}+c_{4}^{J}\chi_{E}^{2}+c_{5}^{J}\chi_{a^{2}}\right],\label{eq:chiral_f}
\end{eqnarray}
where $J=\perp,\parallel,T$. Note that the expressions are in units of the mass-independent scale $r_1$ and the coefficients $c_i^J$ have the dimension of $r_1^{-3/2}$. The leading-order terms are 
\begin{eqnarray}
f_{\perp,T}^{(0)} & = & \frac{1}{f_{\pi}}\frac{g_{B^{*}B\pi}}{E_{\pi}+\Delta_{B^{*}}+\delta D_\text{logs}},\label{eq:chiral_fp_LO}\\
f_{\parallel}^{(0)} & = & \frac{1}{f_{\pi}},\label{eq:chiral_fv_LO}
\end{eqnarray}
with $g_{B^{*}B\pi}$ the $B^*\text{-}B\text{-}\pi$ coupling constant and $\Delta_{B^{*}}\equiv M_{B^*}-M_B$
the $B^*\text{-}B$ mass splitting. The terms $\delta f_{J,\text{logs}}$
and $\delta D_\text{logs}$ are the one-loop nonanalytic contributions
in the chiral expansion, and depend upon the light pseudoscalar meson mass and energy \cite{0704.0795}. Note that in the heavy-quark expansion $f_T$ is proportional to $f_\perp$ up to $\mathcal{O}(1/m_b)$. We therefore use the same pole location and nonanalytic corrections for $f_T$ as $f_\perp$. The terms analytic in $\chi_{i}$ are introduced to cancel the scale dependence
arising from the nonanalytic contribution in Eq.~(\ref{eq:chiral_f}).
The dimensionless variables $\chi_i$ are proportional to the quark mass, pion energy, and lattice spacing. We define 
\begin{eqnarray}
\chi_\text{val} & = & \frac{2\mu m_{l}}{8\pi^{2}f_{\pi}^{2}},\label{eq:xval}\\
\chi_\text{sea} & = & \frac{\mu(2m'_{l}+m'_{h})}{8\pi^{2}f_{\pi}^{2}},\label{eq:xsea}\\
\chi_{E} & = & \frac{\sqrt{2}E_{\pi}}{4\pi f_{\pi}},\;\;\text{and}\label{eq:xE}\\
\chi_{a^{2}} & = & \frac{a^{2}\bar{\Delta}}{8\pi^{2}f_{\pi}^{2}}.\label{eq:x_a2}
\end{eqnarray}
Note that the valence mass $m_l$ is equal to the sea mass $m'_l$ in our data. The low-energy constant $\mu$ relates the pseudoscalar meson masses to the quark masses,
\begin{equation}
M^2_{\xi,\text{PS}} = (m_1+m_2)\mu + a^2\Delta_\xi,
\end{equation}
and $\Delta_\xi$ is the mass splitting for staggered taste $\xi$. The average taste splitting in Eq.~(\ref{eq:x_a2}) is $\bar{\Delta} \equiv \frac{1}{16} \sum_\xi \Delta_\xi$. The quantities
$\mu$ and $\Delta_\xi$ are obtained from the MILC Collaboration's analysis of light pseudoscalar mesons and are shown in Table~\ref{tab:taste_splits}.

We constrain the parameter $g_{B^*B\pi}$ with a prior. The value of $g_{B^*B\pi}$ has been calculated with lattice QCD in the static limit \cite{1109.2480,1203.3378} or with a relativistic $b$ quark \cite{Flynn:2013kwa} on gauge fields generated with domain-wall or Wilson sea quarks \cite{1011.4393}. We set the prior, based on these lattice-QCD calculations, to be $g_{B^*B\pi}=0.45\pm0.08$, where the error covers the differences among different determinations of the coupling. The LO and NLO coefficients, $\{c_{i},0\leq i\leq5\}$, 
are well determined by the data. Note that the formula given in Eq.~(\ref{eq:chiral_f}) is slightly different from that in Ref.~\cite{0811.3640} where the NLO coefficients therein are our $|c_i^J/c_0^J|$ ($i\neq 0$). With the introduction of variables $\chi_i$ defined in Eqs.~(\ref{eq:xval})-(\ref{eq:x_a2}), we should expect that $|c^J_i/c^J_0|\lesssim 1$ ($i\neq 0$), or $|c_i^J|\lesssim |c_0^J|$. In the actual fits, $|c_0^{\perp,\parallel} r_1^{3/2}| \lesssim 0.6$ and $|c_0^{T} r_1^{3/2}| \lesssim 1.0$. Note that the coefficients $c_i^J$ are dimensionful, and they are evaluated here in $r_1$ units. We constrain them with loose priors: $c_i^{\perp,\parallel} r_1^{3/2} = 0\pm 1 $ and $c_i^{T} r_1^{3/2}=0\pm 2$.

Standard HMrS$\chi$PT uses the assumption that the external and loop pions are soft, {\it i.e.}, $E_{\pi}\sim M_{\pi}$ \cite{hep-lat/0305001,0710.3496}. In our work, however, the external pion energies can be quite large, in some cases as much as 7 times the physical pion mass, and standard HMrS$\chi$PT may not converge well enough in this range. Indeed, the fit of the lattice form factor $f_{\parallel}$ to Eq.~(\ref{eq:chiral_f}) gives a poor
confidence level ($p\sim 0$), which is not improved by including higher-order contributions in the chiral expansion. Bijnens and Jemos \cite{1006.1197} proposed an approach called hard-pion $\chi$PT, in which the
internal energetic pions are integrated out and the $E_\pi$ dependence is absorbed into the low
energy constants.~\footnote{The factorization of hard-pion $\chi$PT breaks down starting at three loops~\cite{Colangelo:2012ew}, but we only use the one-loop non-analytic terms.} Since hard-pion $\chi$PT provides a more appropriate description of our data, we adopt it in this analysis. The explicit
expressions for the hard-pion nonanalytic terms $\delta f_{J,\text{logs}}^\text{hard}$ using SU(3) chiral perturbation theory as well as its SU(2) limit are given in the
appendix of Ref.~\cite{ranzhou}. We take the SU(2) limit by integrating out the strange quark. The resulting expression has no explicit strange-quark mass dependence, which has been absorbed into the value of the low energy constants. The SU(2) hard-pion $\chi$PT provides a better description of our $f_\parallel$ data than the SU(3) hard-pion $\chi$PT ($p$ value $0.29$ versus $0.09$ from the NLO $\chi$PT fit with priors). We also find that the chiral expansion converges faster using SU(2) $\chi$PT when including higher-order chiral corrections in the fit to our data, which results in smaller $\chi$PT truncation errors than from using SU(3) $\chi$PT. Finally, Ref.~\cite{0710.3496} provides phenomenological arguments to prefer the application of SU(2) HM$\chi$PT over SU(3) to lattice-QCD data.
We therefore use the SU(2) formula for our central value fit, but also check the consistency with the SU(3) fits in Sec.~\ref{secV}. 

Based on the above discussion, we use the following conditions for $f_\perp, f_\parallel$ and $f_T$ in Eq.~(\ref{eq:chiral_f}):
\begin{eqnarray}
\delta f_{J,\text{logs}} &=& \delta f_{J,\text{logs}}^\text{ hard, SU(2)}, \label{eq:df_log_hard}\\
\delta D_\text{logs} & = & 0,\label{eq:dD_log_hard} \\
c_2^J & = & 0, \label{eq:c_2_hard}\\
\chi_\text{val} & = & \frac{2(2\mu m_{l})}{8\pi^{2}f_{\pi}^{2}}-\frac{a^{2}\Delta_{I}/3}{8\pi^{2}f_{\pi}^{2}},\label{eq:X_val_hard}
\end{eqnarray}
where Eq.~(\ref{eq:dD_log_hard}) is a consequence of the hard-pion limit, Eq.~(\ref{eq:c_2_hard}) and the factor 2 in the first term of Eq.~(\ref{eq:X_val_hard}) follow from the fact that we take $m_l=m'_l$ and $m'_h$ has been integrated out, Eq.~(\ref{eq:X_val_hard}) preserves the chiral scale independence of the SU(2) hard-pion NLO expression, and $a^{2}\Delta_{I}$ is the taste
splitting of the taste-singlet pseudoscalar meson mass.

\begin{table}
	\caption{Fixed parameters that enter the chiral-continuum extrapolation fit function. The taste splittings $r_{1}^{2}a^{2}\Delta_{\xi}$, $\xi=P,A,T,V,I$
		are for the pseudoscalar, axial-vector, tensor,
		vector and scalar tastes, respectively. The pseudoscalar taste splittings are zero by virtue of the remnant chiral symmetry of staggered fermions. The hairpin parameters $r_{1}^{2}a^{2}\delta'_{V(A)}$ were determined in a combined fit to light-pseudoscalar quantities at multiple lattice spacings. We take the result for these couplings at the $a\approx0.12$~fm lattice and scale them to other lattice spacings by the ratio $\Delta_\text{rms}(a)/\Delta_\text{rms}(0.12\text{fm})$
		where $\Delta_\text{rms}$ is the rooted mean square of the taste splittings.
		The continuum value of the low-energy constant $\mu$ is evaluated at the same scale as the $0.09$ fm lattice in our mass-independent scheme. \label{tab:taste_splits}}

	\begin{tabular}{ccccccc}
		\hline 
		\hline 
		$a$(fm)  & 0.12  & 0.09  & 0.06  & 0.045  & 0  & \tabularnewline
		\hline 
		$r_{1}^{2}a^{2}\Delta_{P}$  & 0  & 0  & 0  & 0  & 0  & \tabularnewline
		$r_{1}^{2}a^{2}\Delta_{A}$  & 0.22705  & 0.07469  & 0.02635  & 0.01041  & 0  & \tabularnewline
		$r_{1}^{2}a^{2}\Delta_{T}$  & 0.36616  & 0.12378  & 0.04298  & 0.01698  & 0  & \tabularnewline
		$r_{1}^{2}a^{2}\Delta_{V}$  & 0.48026  & 0.15932  & 0.05744  & 0.22692  & 0  & \tabularnewline
		$r_{1}^{2}a^{2}\Delta_{I}$  & 0.60082  & 0.22065  & 0.07039  & 0.02781  & 0  & \tabularnewline
		$r_{1}^{2}a^{2}\delta'_{V}$  & 0.0  & 0.0  & 0.0  & 0.0  & 0  & \tabularnewline
		$r_{1}^{2}a^{2}\delta'_{A}$  & $-0.28$  & $-0.09$  & $-0.03$  & $-0.01$  & 0  & \tabularnewline
		$r_1\mu$  & 6.83190  & 6.63856  & 6.48665  & 6.41743  & 6.015349  & \tabularnewline
		\hline
		\hline  
	\end{tabular}
\end{table}

The fits of the lattice form factors using NLO SU(2) hard-pion HMrS$\chi$PT have acceptable
confidence levels. We find, however, that there is a sizable shift in the fit result when including higher-order terms in the $\chi$PT expansion. We therefore need to study the effects of higher-order contributions in the chiral expansion.

\subsection{Next-to-next-to-leading order (NNLO) corrections}

We supplement the NLO SU(2) hard-pion $\chi$PT expression with the following NNLO analytic terms
\begin{eqnarray}
\delta f_{J,\text{analytic}}^\text{NNLO} & = & c_{6}^{J}\chi_\text{val}\chi_{E}+c_{7}^{J}\chi_{a^{2}}\chi_{E}+c_{8}^{J}\chi_{E}^{3}+c_{9}^{J}\chi_\text{val}^{2}\nonumber \\
&  & +c_{10}^{J}\chi_\text{val}\chi_{E}^{2}+c_{11}^{J}\chi_{a^{2}}\chi_\text{val}+c_{12}^{J}\chi_{a^{2}}\chi_{E}^{2}+c_{13}^{J}\chi_{a^{2}}^{2}+c_{14}^{J}\chi_{E}^{4},\label{eq:df_NNLO}
\end{eqnarray}
such that the complete NNLO $\chi$PT expression is,
\begin{eqnarray}
f_{J}^\text{NNLO} & = & f_{J}^\text{NLO} + f_J^{(0)}\delta f_{J,\text{analytic}}^\text{NNLO}.\label{eq:chiral_f_NNLO}
\end{eqnarray}
Note that $f_J^\text{NLO}$ here uses the hard-pion and SU(2) $\chi$PT, as manifested in Eqs.~(\ref{eq:df_log_hard})-(\ref{eq:X_val_hard}). All light-quark discretization errors that arise from taste violations are included here; generic errors from light-quark and gluon action, which are $\mathcal{O}(\alpha_s a^2 \Lambda^2)$, are discussed in Sec.~\ref{sec:chiral:HQ}.

Again, the expectation from chiral perturbation theory is that the coefficients of these analytic terms should satisfy $|c_i^J/c_0^J|\sim\mathcal{O}(1)$ when written in terms of the dimensionless variables $\chi$ given in Eqs.~(\ref{eq:xval})--(\ref{eq:x_a2}). We set the priors for the NNLO coefficients for $f_{\perp, \parallel}$,  $\{c^{\perp,\parallel}_{i}r_1^{3/2},6\leq i\leq 14\}$, to be $0\pm 0.6$, since the role of these terms is simply to absorb the effects of higher-order contributions in the chiral expansion. This width 0.6 corresponds to the size of $|c^{\perp,\parallel}_0r_1^{3/2}|$. For the same reason, we set the priors for the coefficients for $f_T$, $\{c^{T}_{i}r_1^{3/2},6\leq i\leq 14\}$, to be $0\pm 1.0$. 
Doubling the prior widths leads to negligible shifts on the central values of the form factors and less than 20\% increases in the fit errors.

\subsection{Heavy-quark discretization effects}\label{sec:chiral:HQ}

The chiral-continuum extrapolation implemented in Eq.~(\ref{eq:chiral_f_NNLO}) accounts for
the discretization effects from the gluons and the light staggered quarks. Discretization effects from the
heavy $b$ quark need a separate treatment. Heavy-quark discretization errors arise from
the short-distance mismatch of higher-dimension Lagrangian and current
operators \cite{Kronfeld:2000ck, Harada:2001fi}. By power counting, such mismatches are of
$\mathcal{O}(a^{2}\Lambda^{2})$ or $\mathcal{O}(\alpha_{s}a\Lambda)$
where $\Lambda$ is a QCD scale appropriate for the heavy-quark expansion. We follow the
same method for incorporating the heavy-quark discretization effects described in Ref.~\cite{1112.3051} and include the following error
function in Eq.~(\ref{eq:chiral_f}),
\begin{eqnarray}
\delta f^\text{HQ}_{J} & = & \left(z_{E}^{J}f_{E}+z_{X}^{J}f_{X}+z_{Y}^{J}f_{Y}\right)(a\Lambda)^{2}+\left(z_{B}^{J}f_{B}+z_{3}^{J}f_3\right)(\alpha_{s}a\Lambda)+z_{0}^{J}\alpha_{s}(a\Lambda)^{2},\label{eq:HQ_discretization}
\end{eqnarray}
where the mismatch functions $f_{E,X,Y,B,3}$ are given in the Appendix of Ref.~\cite{1112.3051}. The error functions $f_{B,E}$ arise from mismatches of operators
in the Lagrangian, while functions $f_{X,Y,3}$ arise from those of
the vector current. The last term in Eq.~(\ref{eq:HQ_discretization}) accounts for higher order heavy-quark and generic light-quark and gluon errors not included in Eq.~(\ref{eq:chiral_f_NNLO}), which is of the order $\alpha_{s}(a\Lambda)^{2}$. The fit parameters are constrained with priors: $0\pm1$ for $z_{Y},z_{B},z_{0}$ and $0\pm\sqrt{2}$ for $z_{X}\,,z_{3}$; the latter two are wider because the functions $f_{X}$ and $f_{3}$
both appear twice \cite{Harada:2001fi}.

\begin{figure}
	%\center{\includegraphics[scale=0.65]{continuum_physics}}
	\center{\includegraphics[width=0.48\textwidth]{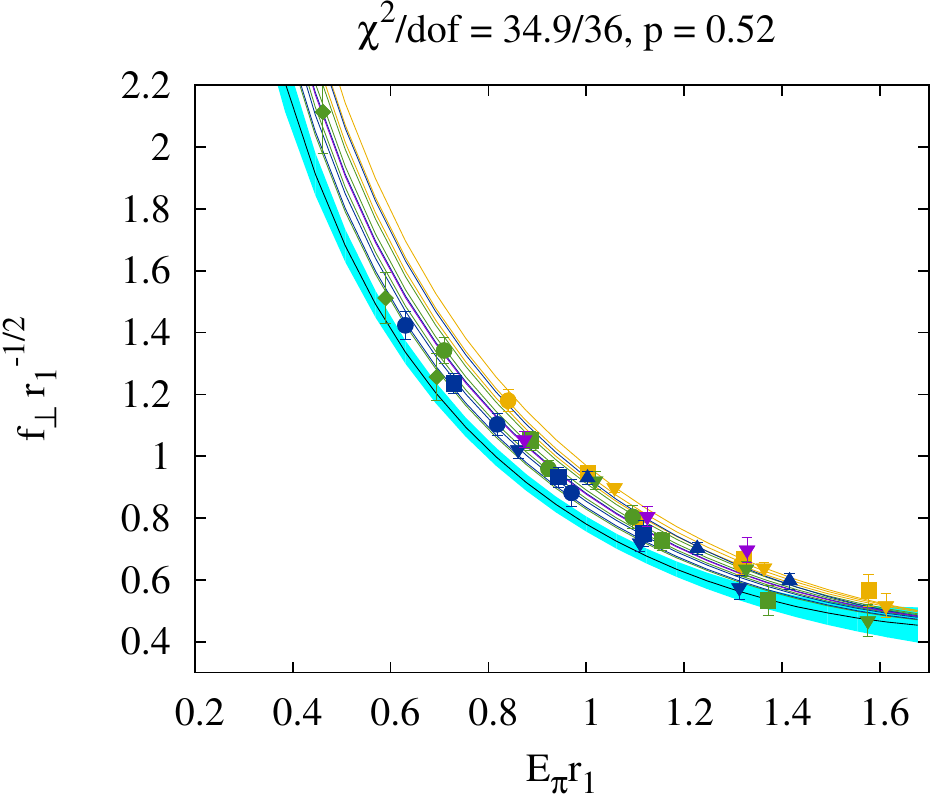} \hfill \includegraphics[width=0.48\textwidth, trim = 60 20 60 60]{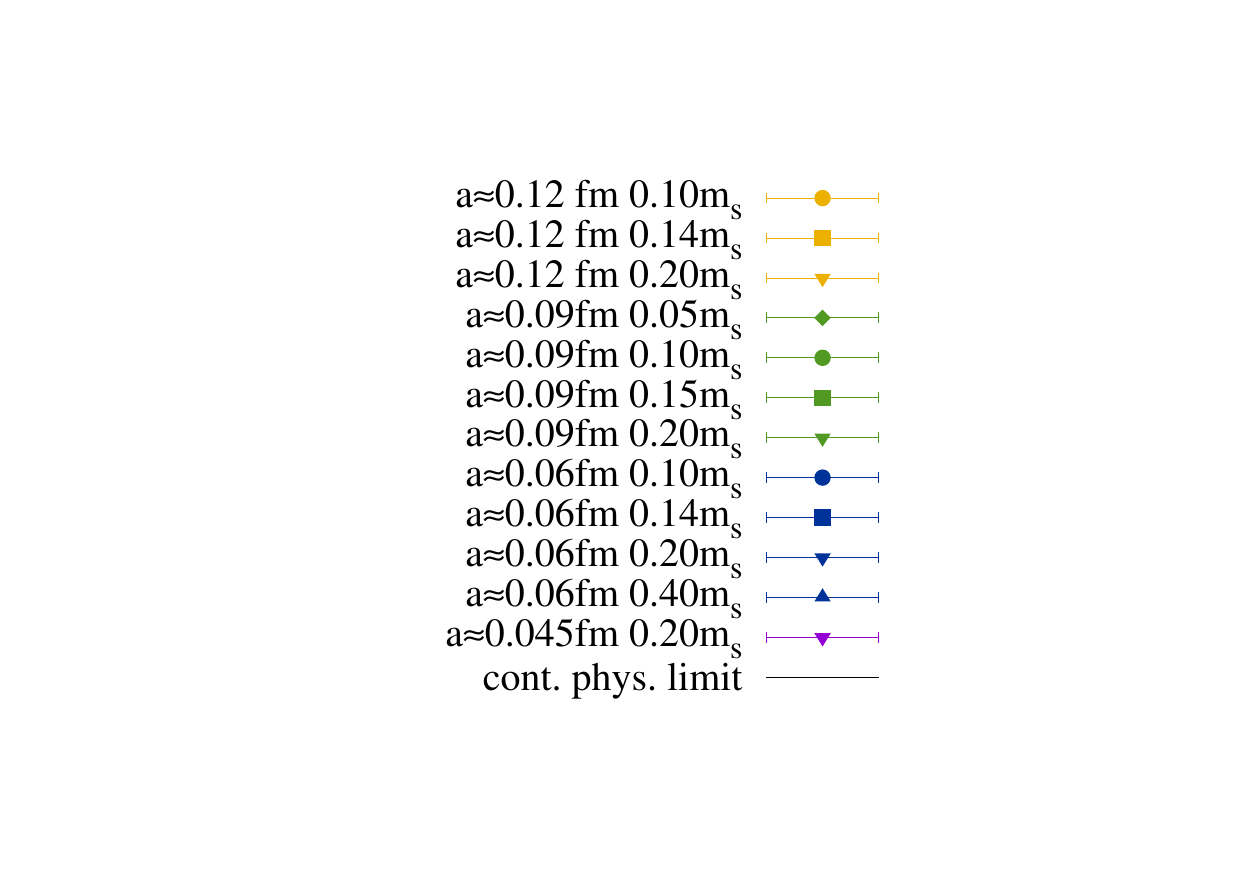}}
	\center{\includegraphics[width=0.48\textwidth]{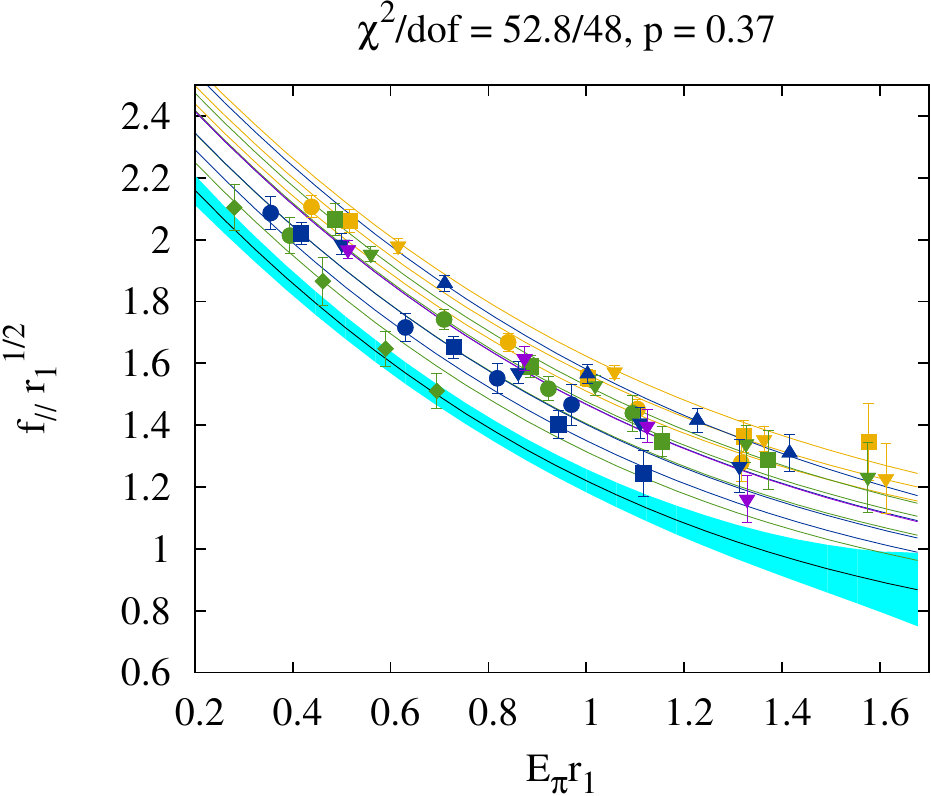}\hfill \includegraphics[width=0.48\textwidth]{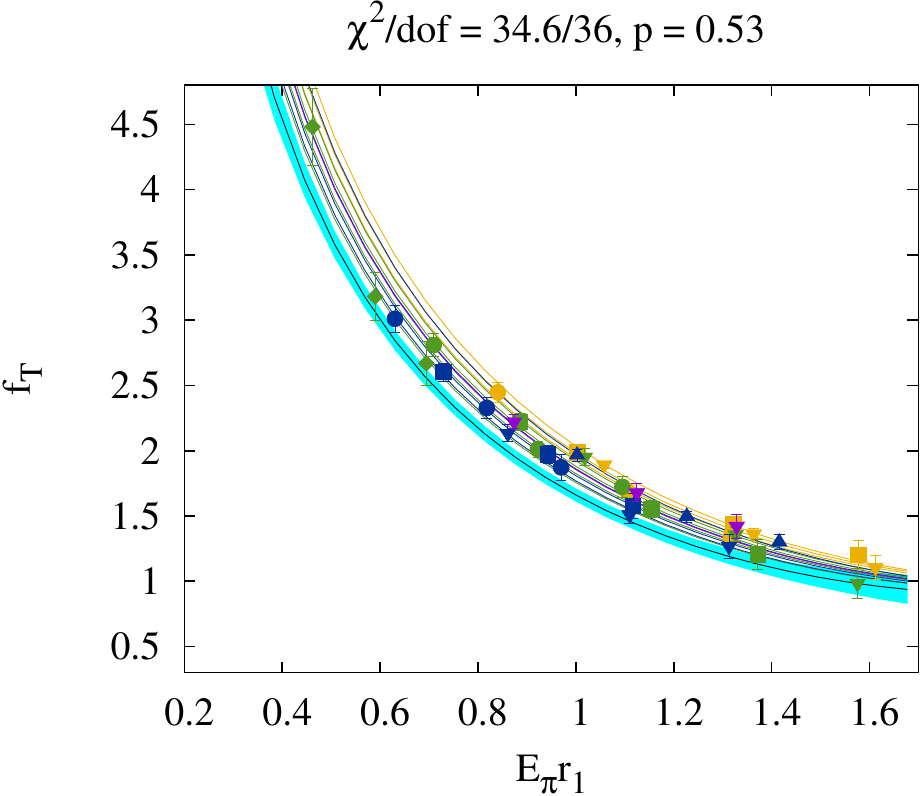}}
	
	\caption{Chiral-continuum extrapolation of lattice form
		factors $f_{\perp}$ (upper left), $f_{\parallel}$ (lower left) and $f_T$ (lower right) as functions of $E_{\pi}$, where all quantities
		are in $r_{1}$ units. The colors denote the lattice spacings: 0.12
		fm (gold), 0.09 fm (green), 0.06 fm (blue) and 0.045 fm (violet).
		The symbols denote the light-quark masses $m'_l/m'_h$: 0.05~(diamond), 0.1~(circle), 0.15~(square), 0.2~(downward-pointing triangle), and 0.4~(upward-pointing triangle). The colored lines correspond to the fit results
		evaluated at the parameters of the ensembles. The physical-mass continuum-limit curve is shown as a black curve with cyan error band. \label{fig:Hard-pion-SU(2)-NNLO}}
\end{figure}

To summarize, after incorporating the heavy-quark discretization effects, the complete NNLO SU(2) hard-pion HMrS$\chi$PT expression is
\begin{eqnarray}
f_{J}^\text{NNLO+HQ} & = & f_{J}^\text{NNLO}\times(1+\delta f_{J}^\text{HQ}),\label{eq:chiral_f_NNLO_HQ}
\end{eqnarray}
where $f_{J}^\text{NNLO}$ is defined in Eq.~(\ref{eq:chiral_f_NNLO}). With this treatment, the uncertainty due to truncating the chiral expansion at NNLO (cf. Sec.~\ref{secV} below), NNLO light-quark and gluon discretization effects, and LO heavy-quark discretization
effects are incorporated in the fit error of the chiral-continuum extrapolation. The fits
for $f_{\perp}$, $f_\parallel$, and $f_{T}$ to Eq.~(\ref{eq:chiral_f_NNLO_HQ}) are
shown in Fig.~\ref{fig:Hard-pion-SU(2)-NNLO}.

\begin{figure}
	\includegraphics[width=0.5\linewidth]{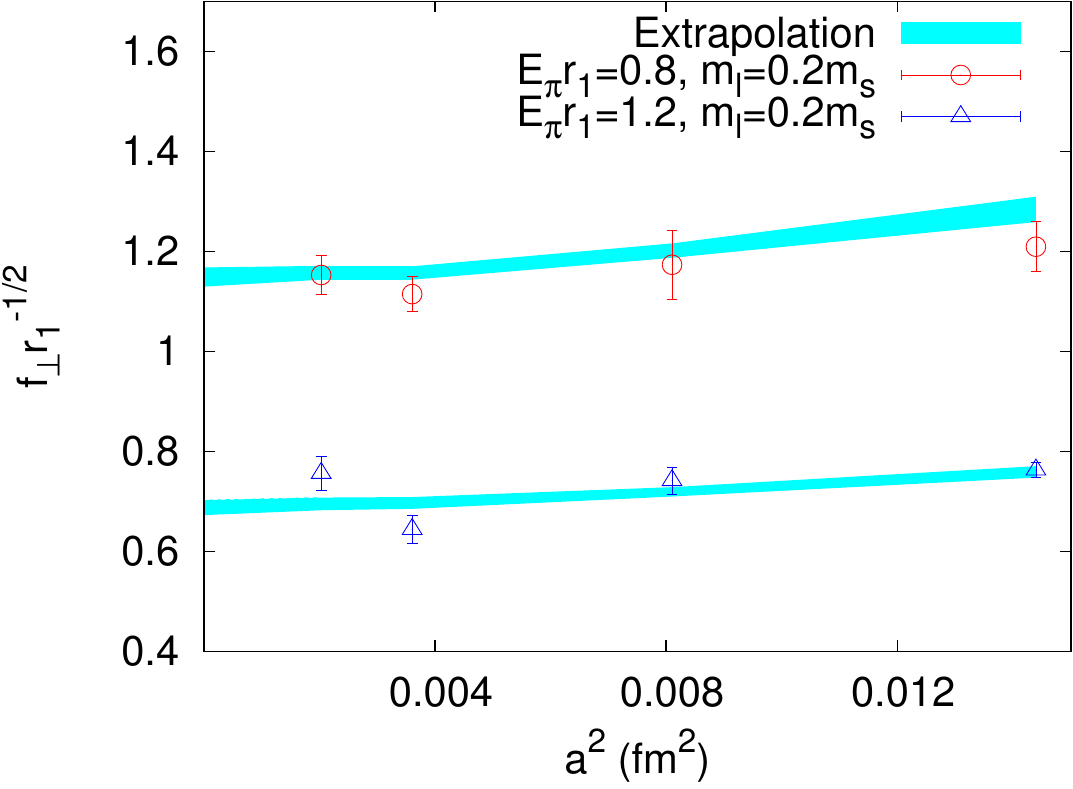}\hfill\includegraphics[width=0.5\linewidth]{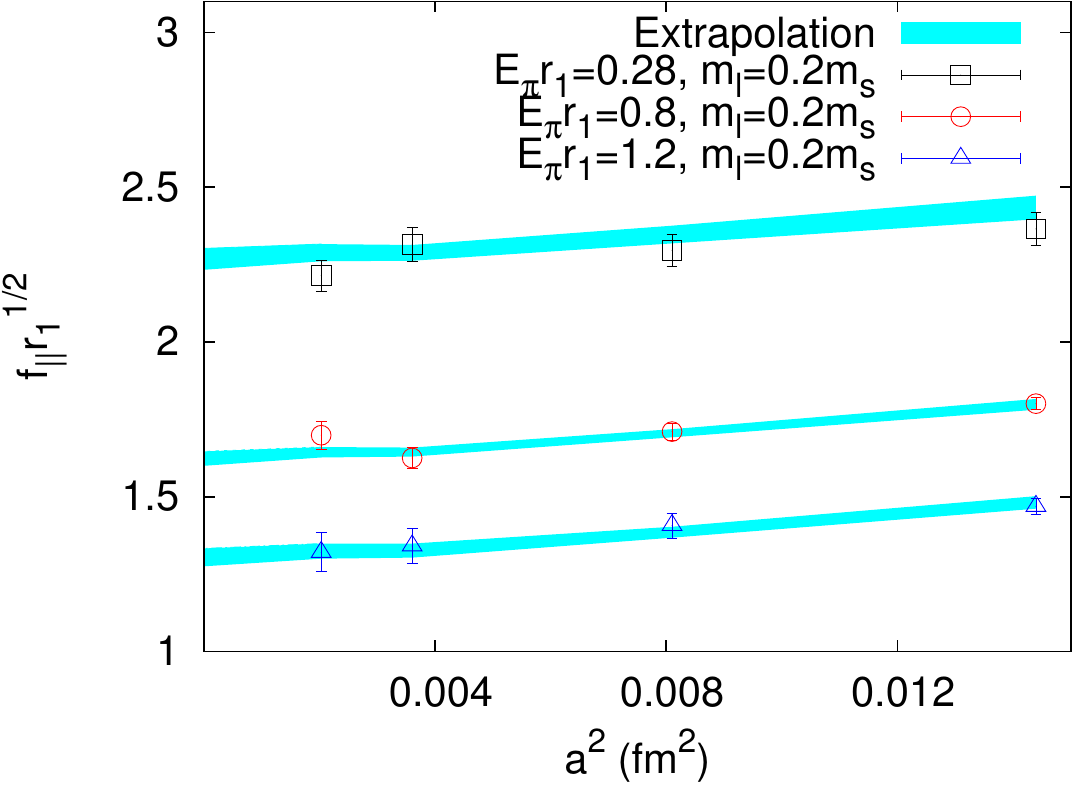}
	
	\caption{Discretization effects in the form factor $f_{\perp}$ (left) and $f_{\parallel}$
		(right) at a few kinematic points.
		The plots show the form factors on the $m'_l=0.2m'_h$
		ensembles at each lattice spacing vs.\ $a^{2}$ for various pion momenta (a slight extrapolation/interpolation
		is applied to adjust the raw data to the same $E_{\pi}r_{1}$). The range $E_{\pi}r_{1}\in [0.28,1.2]$
		is used in the $q^2$ extrapolation to the full kinematic range. \label{fig:discretization_effect}}
\end{figure}

To examine the size of discretization effects, we plot the form
factors $f_{\perp}$ and $f_{\parallel}$ with light-quark mass $m'_l=0.2m'_{h}$
at each lattice spacing versus $a^{2}$ in Fig.~\ref{fig:discretization_effect}. As we can see from the plots, the observed lattice-spacing dependence is very mild, with the data points at the largest lattice spacing ($a\approx 0.12$~fm) only about two statistical sigma away from the continuum limit. 

\section{Systematic error budget} \label{secV}

The error output from the central-value fit described in Sec.~\ref{sec:chiral:HQ} already includes the systematic errors due to the light- and heavy-quark discretization effects and the uncertainty on $g_{B^{*}B\pi}$. We now discuss other sources
of systematic uncertainty. We tabulate systematic error budgets for $f_+$ and $f_0$ at a representative kinematic point $q^2=20$~GeV$^2$ within the range of lattice data in Table~\ref{tab:Error-budgets-20}. We also present the error budget for the full simulated lattice momentum range in Fig.~\ref{fig:errors_q2}.

\subsection{Chiral-continuum extrapolation}

As discussed above, our central fit uses NNLO\footnote{NLO + NNLO analytic terms.} SU(2) hard-pion HMrS$\chi$PT including contributions
from heavy-quark discretization effects and the uncertainty in $g_{B^*B\pi}$. Here we consider variations of the fit function and the data included to estimate truncation and other systematic effects. 

\begin{figure}
	\center{\includegraphics[scale=0.85]{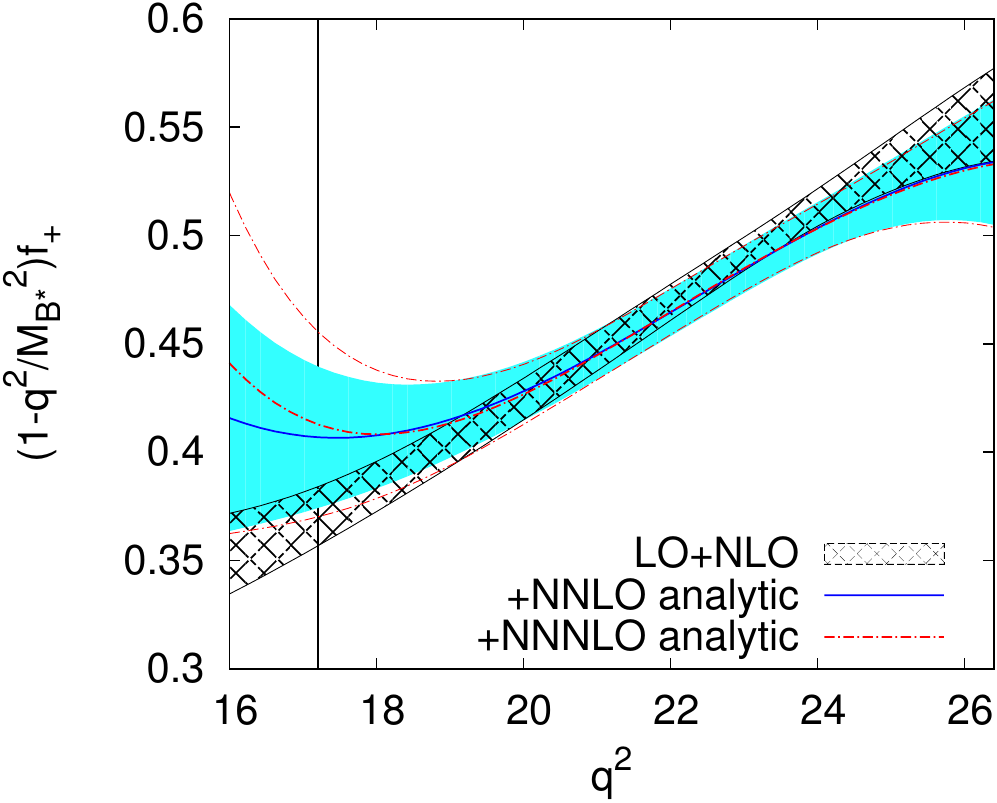}}
	
	\caption{The function $(1-q^{2}/M_{B^{*}}^{2})f_{+}$ obtained from the chiral-continuum
		fits with NLO logarithms and analytic terms through NLO (hatched black band), NNLO (solid cyan band) and NNNLO (dashed red lines). Errors shown are fit statistical errors only. The vertical line indicates the lowest $q^2$ of the lattice data.
		\label{fig:sys_NLO_NNLO_NNNLO}}
\end{figure}

\begin{figure}
	\center{\includegraphics[scale=0.75]{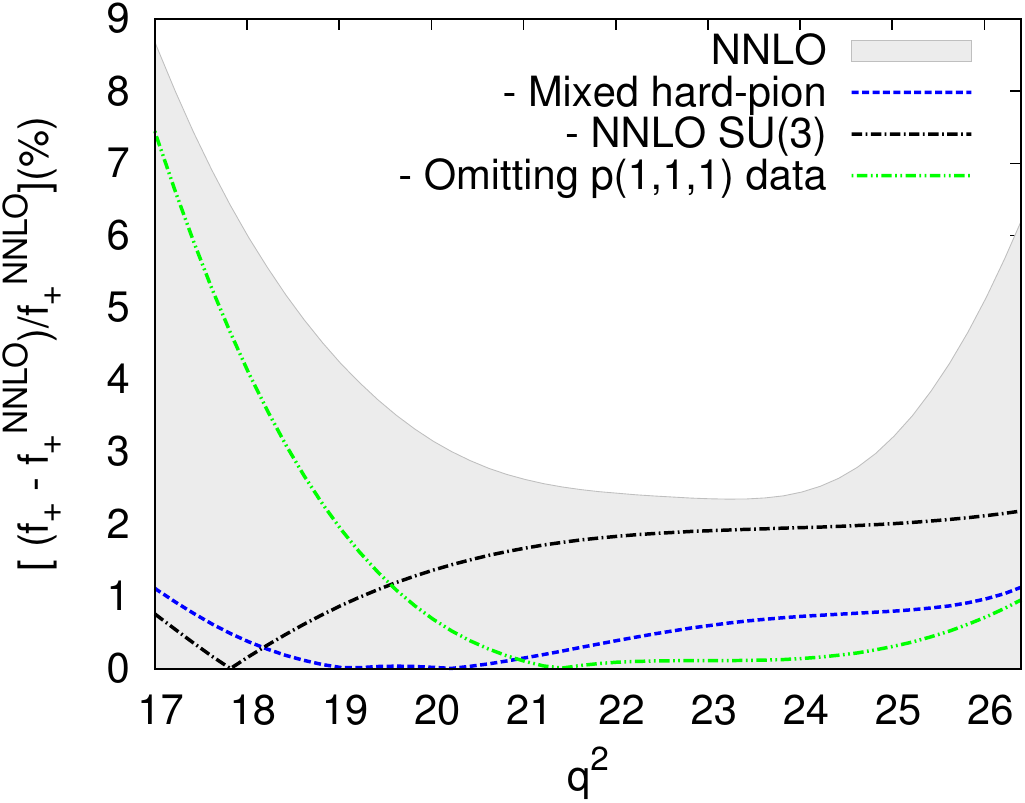}\hfill\includegraphics[scale=0.75]{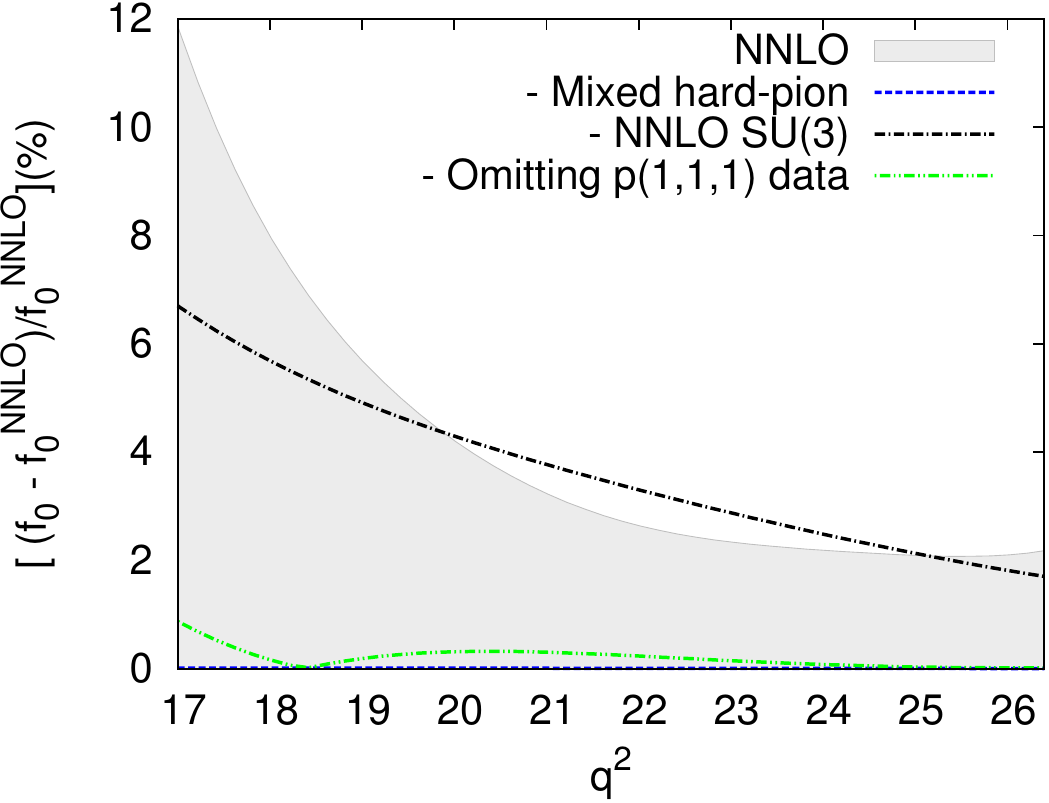}}
	
	\caption{Variations in the chiral-continuum extrapolation from different fit Ans\"atze. The shaded area shows the fit error from the preferred NNLO SU(2) fit.
		The other curves show the systematic deviations from the NNLO SU(2)
		fit under the variations discussed in the text. \label{fig:sys_chipt}}
\end{figure}

First, we study the effects of truncating the chiral expansion by adding next-to-NNLO (NNNLO) analytic terms $\delta f_{J,\text{analytic}}^\text{NNNLO}$
in our fits with coefficients constrained with the same priors
as the NNLO coefficients. The variations in $f_{+}$ due to changing the order of the $\chi$PT analytic terms
are shown in Fig.~\ref{fig:sys_NLO_NNLO_NNNLO}.
The fits of different orders are consistent in the $q^{2}$ region where most of the simulation data are located. Although the central
values and errors differ noticeably between the NLO and NNLO
fits, the central values and errors of the
NNNLO fit are very close to the NNLO fit, indicating that the chiral extrapolation has
stabilized by NNLO. As discussed earlier, the NLO coefficients are well determined by the data and we use well-motivated priors based on expectations from $\chi$PT for the NNLO and higher order terms. The fact that the error saturates with NNLO shows that the preferred fit already incorporates the uncertainty from truncating the chiral expansion, and that we do not need to add an additional systematic error. The NNLO fit error as a function of $q^2$ is shown
in Fig.~\ref{fig:sys_chipt}.

The standard soft-pion HMrS$\chi$PT fits of
$f_{\perp}$ have reasonable confidence levels, but those of $f_\parallel$ do not. Here we estimate the effect of using the hard-pion formalism
by using standard HMrS$\chi$PT for $f_\perp$ but still employing hard-pion $\chi$PT for $f_\parallel$. The resulting difference from the preferred fit is small, less than 1\% for $f_+$. The same conclusion also holds for the form factor $f_{0}$.

We use SU(2) $\chi$PT, instead of SU(3) $\chi$PT, for our central fit. To estimate the effect
of this choice, we restore the strange-quark dependence of the logarithm and analytic terms in Eq.~(\ref{eq:chiral_f}). A practical issue arises with
NNLO SU(3) $\chi$PT, where the terms proportional to the sea-quark mass, $\chi_\text{sea}$, are not well constrained by our data because the strange sea-quark mass $m'_h$ is so similar on all of our ensembles. To obtain some sensitivity to $\chi_\text{sea}$, we include data on an additional $a\approx 0.12$~fm ensemble with an unphysically small strange-quark mass, $am'_{h,\text{sea}}=am'_{l,\text{sea}}=0.005$. With the inclusion of this ensemble, we find the fit parameters for the terms involving $\chi_\text{sea}$ are better constrained. The differences between the NNLO SU(3) fits and the preferred fits are shown in Fig.~\ref{fig:sys_chipt}. For $f_+$, the difference is within the statistical error. For $f_0$, the difference lies outside the statistical error for some of the simulated $q^2$ range, but the NNLO SU(3) fit quality is poor, with a p-value of 0.01. Because SU(3) $\chi$PT does not provide a good description of our data for $f_0$, we do not take the difference between NNLO SU(2) and SU(3) fits as a systematic error.

To check how our results are affected by data with high momenta, we also perform a fit excluding
data with $\bm{p}=(2\pi/L)(1,1,1)$. As shown in Fig.
\ref{fig:sys_chipt}, the form factors $f_{+}$ and $f_0$ from the low-momentum fit agree very well with those from the preferred full-data fit for the region $q^{2}>20~\text{GeV}^2$. The systematic difference increases for
small $q^{2}$, where the highest-momentum data provide important information.

Figure \ref{fig:sys_chipt} summarizes the effects of all these variations.
Comparing the deviations between the central values of the alternate and preferred fits to the statistical error of the preferred fit, we find that the deviations are almost always smaller than the statistical error of our preferred fit. This confirms
that fit errors of our preferred fits adequately account for the systematic effects associated with these variations. We therefore do not quote any additional systematic error due to these sources.

\begin{figure}
	\center{\includegraphics[scale=0.9]{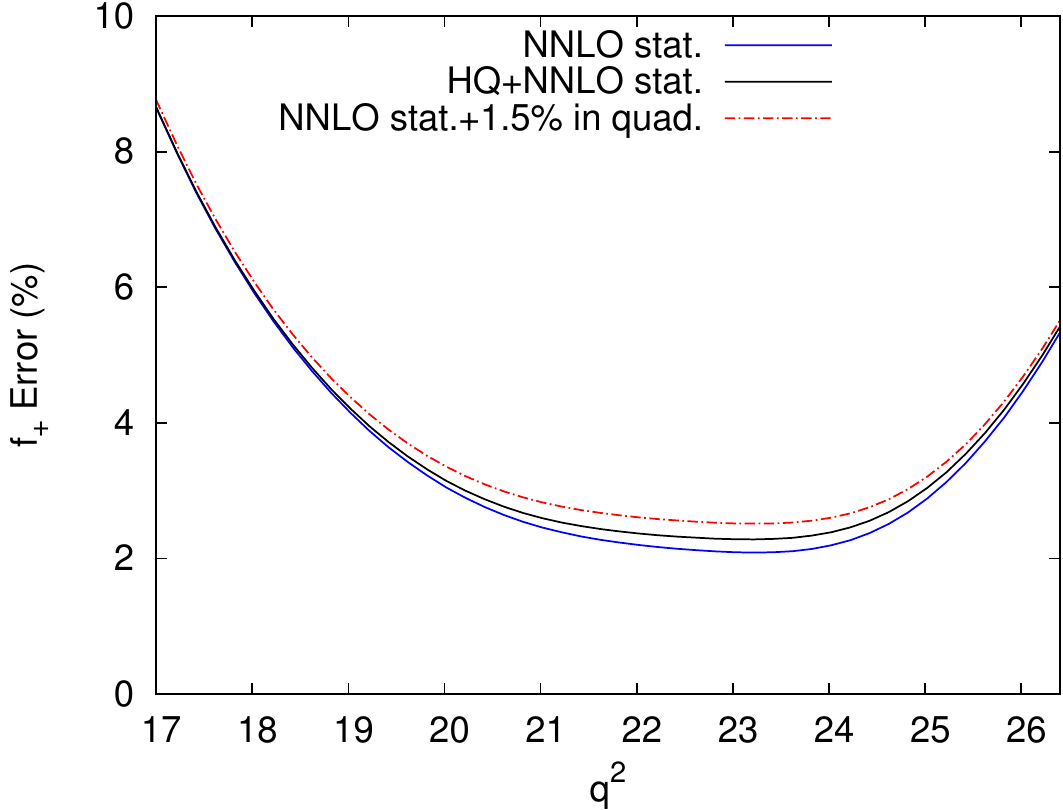}}
	
	\caption{Heavy-quark discretization effects in the chiral-continuum fit. The solid blue, solid black, and dashed red curves show the NNLO fit error, the preferred fit error, and the NNLO fit error added to a 1.5\% power-counting estimate of heavy-quark discretization errors, respectively. \label{fig:sys_HQ}}
\end{figure}

We include heavy-quark discretization effects in our chiral-continuum
extrapolation. As a consistency check, we compare our result with a
power counting estimate obtained by evaluating $\delta f{}_{J}^\text{HQ}$ in Eq.
(\ref{eq:HQ_discretization}) at the $a\approx 0.045$~fm lattice spacing, setting the coefficients $z_i=1$ and taking $\Lambda = 500$~MeV for the heavy-quark scale. 
We find $\delta f^{HQ}_J \simeq 1.5$\%. Figure \ref{fig:sys_HQ} shows that the NNLO fit error (without the heavy-quark discretization effects) added to the 1.5\% power-counting estimate in quadrature yields a similar error to that of the full fit. Thus, again, it is not necessary to add an additional error to that of the preferred chiral-continuum fit.

\subsection{Light- and bottom-quark mass uncertainties}

The effect of mistuning the $b$-quark mass in our simulation has been largely reduced via the corrections described in Sec. \ref{subsec:tuning}. Errors still arise, however, from the uncertainty in the tuned value $\kappa_b$ itself and from the procedure for shifting the form factors. From Eq.~(\ref{eq:df_f}) we estimate the relative error by
\begin{eqnarray}
\frac{\delta f}{f} & \approx & \left( \frac{\partial\ln f}{\partial\ln\bar{m}_{2}}\right) \, \frac{\delta (1/m_2)}{1/m_2} + \delta \left( \frac{\partial\ln f}{\partial\ln\bar{m}_{2}}\right)\frac{1/m_2-1/\bar{m}_2}{1/\bar{m}_2}, 
\end{eqnarray}
where $\delta(1/m_{2})$ is related to the uncertainty due to the error in $\kappa_b$
while $\delta(\frac{\partial\ln f}{\partial\ln\bar{m}_{2}})$ is the uncertainty on the normalized slope. The values of the physical $\kappa_b$ with errors are given in Table~\ref{tab:tuned_kappa}, and we can find the statistical uncertainty of the normalized slope using Table~\ref{tab:normalized_slope}. Using Eq.~(\ref{eq:df_f}), we find that the value of $\delta f/f$ on all ensembles is at most
0.6\%. We take the average value for $\delta f/f$ on all ensembles, which is 0.4\%, to be the error due to tuning $\kappa_b$, and assign the same error
to $f_{+}$ and $f_{0}$.

To obtain the physical form factors, we evaluate the result of the chiral-continuum fit at the physical light- and strange-quark masses determined from the MILC Collaboration's analysis of light pseudoscalar mesons~\cite{0903.3598}.  (Although we use SU(2) $\chi$PT, we include an analytic term proportional to $\chi_\text{sea}$ to allow for a slight shift to the physical strange sea-quark mass.)  The errors on the physical $m_l \equiv (m_u + m_d)/2$ and $m_s$ are 3.5\% and 3.0\%, respectively.  We vary the light- and strange-quark masses at which the chiral-continuum fit function is evaluated by plus and minus one standard deviation, and find that it produces differences below 0.4\% in both form factors.

\subsection{Lattice scale $r_{1}$}

We convert the lattice form factors and pion energies to physical units using the relative scale $r_{1}/a$ determined from the static-quark potential (see Table~\ref{tab:derived_params}) and the absolute scale $r_{1}=0.3117(22)$~fm \cite{1112.3051}. The statistical uncertainties on $r_1/a$ are negligible. We propagate the uncertainty in $r_1$ by shifting it $\pm1\sigma$ and repeating the chiral-continuum fit. We find shifts of at most $0.7\%$ in the range of simulated momenta. 

\begin{figure}
	\center{\includegraphics[width=0.48\textwidth]{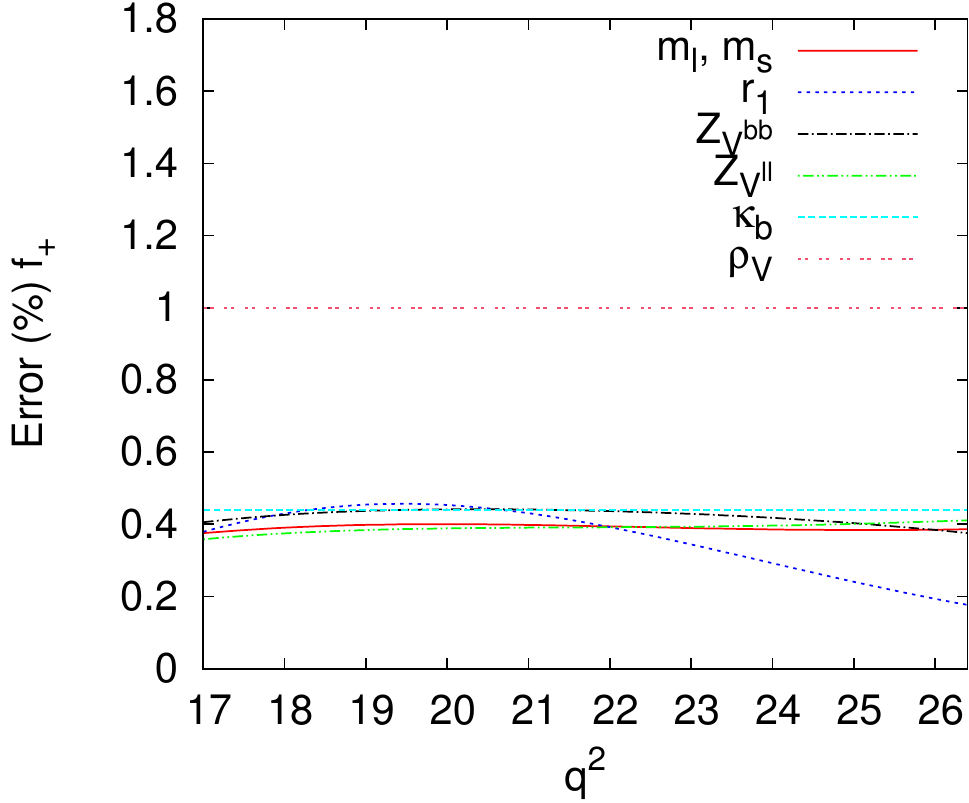}\hfill\includegraphics[width=0.48\textwidth]{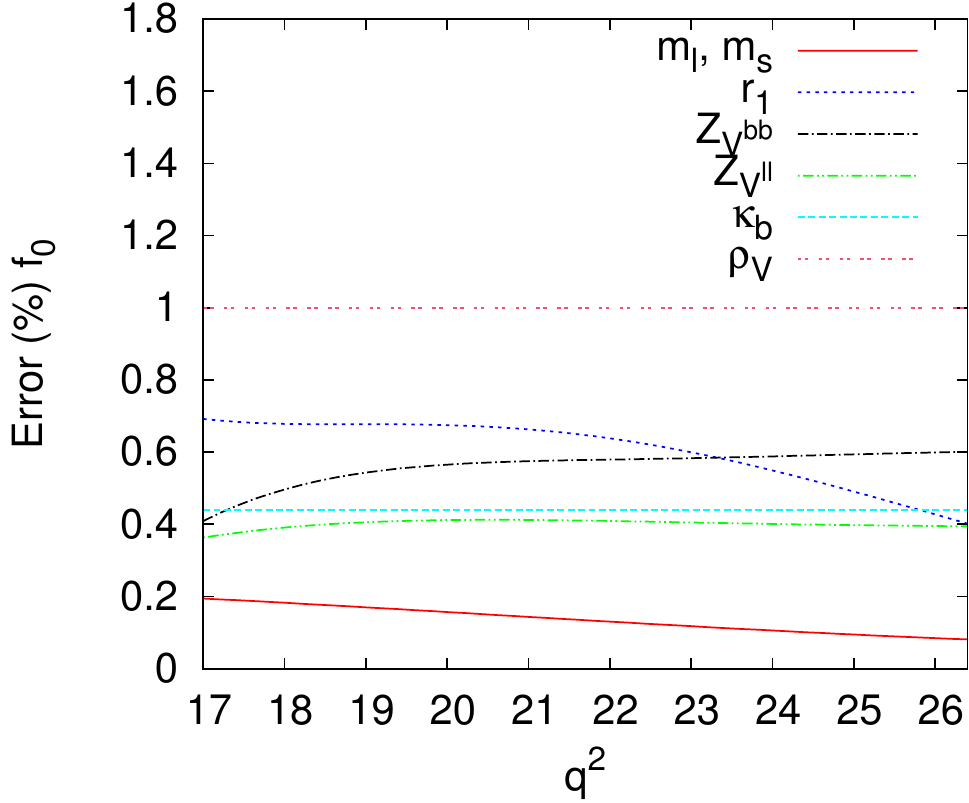}}
	
	\caption{Subdominant systematic errors over the range of simulated lattice momenta. Error estimates are described in the text. \label{fig:sys_others}}
\end{figure}

\subsection{Current renormalization}

With the mostly nonperturbative renormalization procedure that we use for the heavy-light currents, there are two sources of error. The first is due to the nonperturbatively calculated flavor diagonal factors $Z_{V^4_{bb}}$ and $Z_{V^4_{ll}}$. Their values and errors are given in Table~\ref{tab:renorm_factors}. We estimate the systematic error due to the uncertainties of $Z_{{V}^4_{bb}}$ and $Z_{{V}^4_{ll}}$ by varying their values by one sigma and looking for the maximum deviations in the form factors $f_+$ and $f_0$. The resulting deviations are small,
ranging from 0.4\% to 0.6\%.

The second source of error is due to the truncation of the perturbative expansion in the calculation of the $\rho_J$. Because the $\rho_J$ are defined from ratios of renormalization constants, their perturbative corrections are small by construction. Indeed, as seen in Table~\ref{tab:renorm_factors}, for $V^4_{bl}$ they are less than 1\% and for $V^i_{bl}$ they range between 2--3\%. For the scale-independent vector current, we observe that the one-loop corrections to $\rho_{V^4_{bl}}$ are smaller than those for
$\rho_{V^i_{bl}}$, and we use the same error estimate for both. In order to accommodate possible accidental cancellations, we take the error as
$2  \rho^{[1]}_{\rm max} \, \alpha_s^2$, where $\rho^{[1]}_{\rm max}\alpha_s$ is an upper bound of the one-loop correction to $V^\mu_{bl}$ in the range of heavy-quark mass $am_0 \leq 3$ that corresponds to the range of lattice spacings included in our analysis. The coupling is evaluated at the scale of the next-to-finest lattice spacing in our calculation, $a\approx 0.06$~fm. This procedure yields an error estimate of 1\%, which is larger than the one-loop correction to $\rho_{V^4_{bl}}$ over most of the mass range, and amounts to about 50\% of the one-loop correction to $\rho_{V^i_{bl}}$ in the mass range that corresponds to the three finest lattice spacings. This leads to an error of 1\% for both $f_+$ and $f_0$ due to the perturbative renormalization factors.

\subsection{Finite volume effects}

We estimate the size of the finite-volume effects by replacing the infinite-volume chiral logarithms with discrete sums and repeating the chiral-continuum extrapolation. The change in our preferred fit after including finite-volume corrections is very small, less than 0.01\%, which we simply neglect.

\subsection{Summary}

\begin{figure}
	\center{\includegraphics[width=0.50\textwidth, trim=40 10 40 0 ]{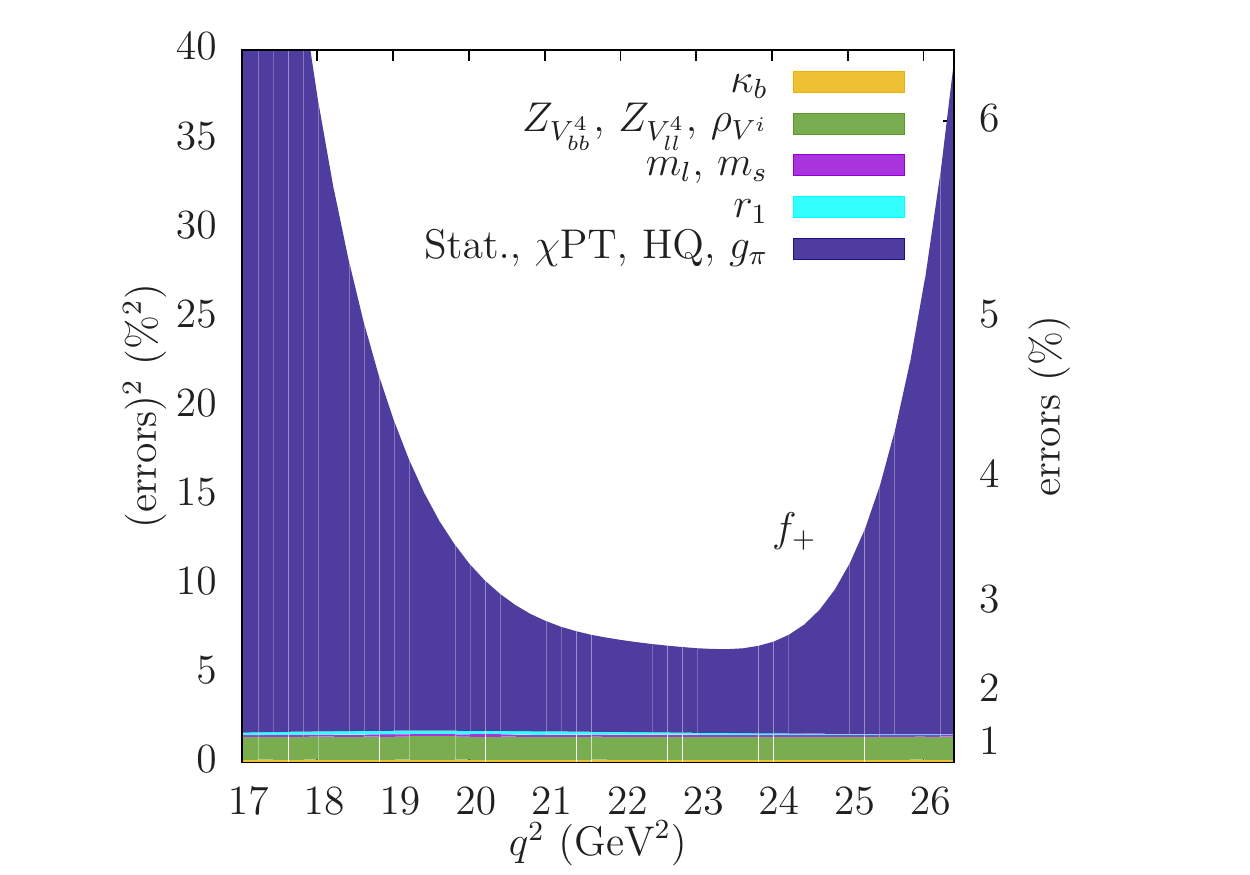}\hfill\includegraphics[width=0.50\textwidth, trim=40 10 40 0]{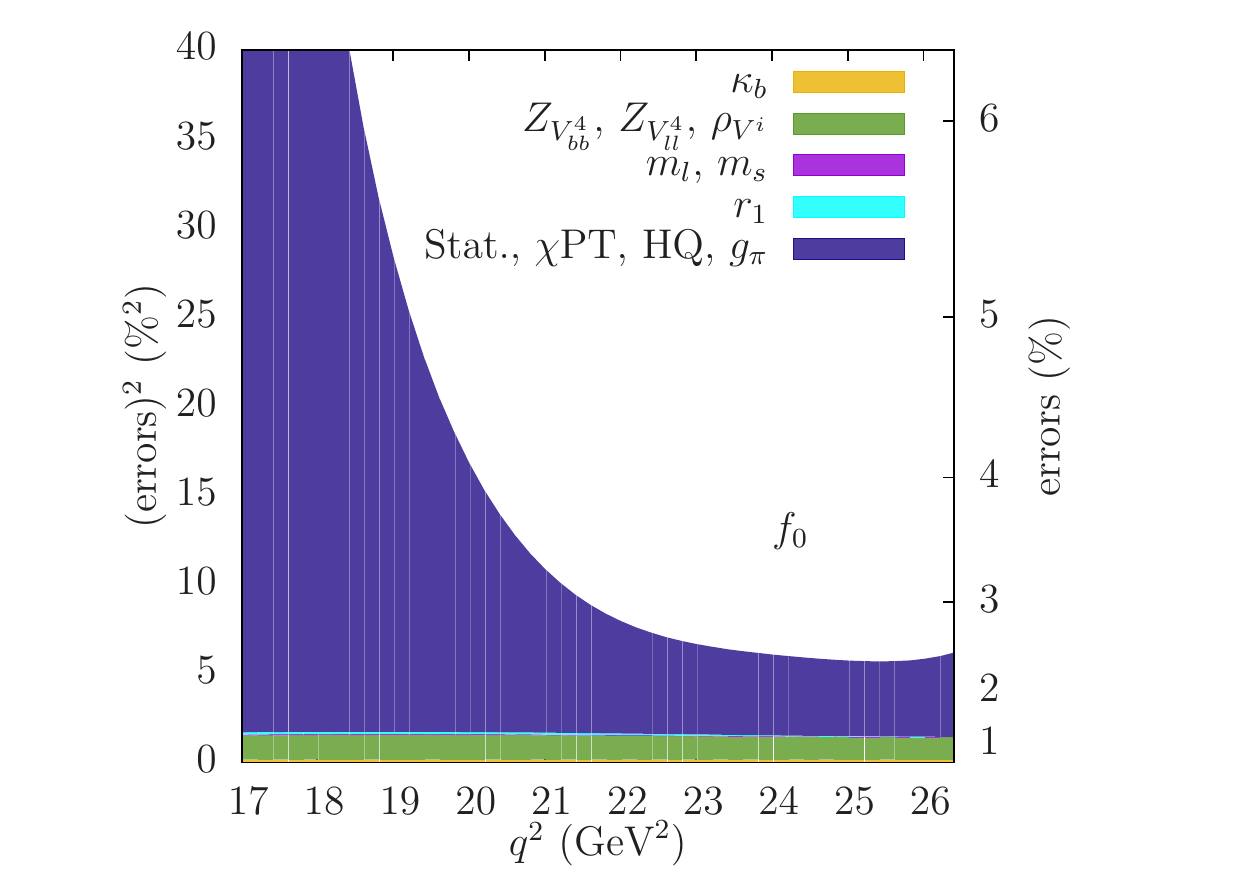}}
	
	\caption{The distribution of the errors for $f_+$ (left) and $f_0$ (right) as a function of $q^2$. The different bands in the plot show the contribution of the error source to the sum of squared errors (left y axis). The corresponding error can be read off from the right y axis.  \label{fig:errors_q2}}
\end{figure}

Figures \ref{fig:sys_others} and \ref{fig:errors_q2} visually summarize the systematic error budget for the vector and scalar form factors $f_{+}, f_0$ in the simulated lattice-QCD momentum range. By far the largest contribution to the total uncertainty is from the fit error, which includes the statistical uncertainty in addition to the chiral-continuum extrapolation and heavy-quark discretization errors. The total error on $f_+$ is smallest, about 3\%, in the region of $q^2 \approx 20$--$24$~GeV$^2$. 

The subdominant errors, such as those from heavy-quark mass tuning, the current renormalization etc., have mild
$q^{2}$ dependence, as can be seen in Fig.~\ref{fig:sys_others}. 
We therefore treat them as constant in $q^2$ when propagating them. For each source, we take the maximum estimated error in the simulated $q^2$ range; we then add these individual error estimates in quadrature to obtain an overall additional systematic error $\delta_{f}$. We
find $\delta_{f_{+}}=1.4\%$ and $\delta_{f_{0}}=1.5\%$. 

In the next section, we will use our result for $f_+$ to obtain $|V_{ub}|$ via a combined fit with experimental data to the $z$ expansion. Due to phase-space suppression, the experiments have poor access to the large-$q^2$ region. On the other hand, the
lattice-QCD form factor has a larger error than experiment at small $q^2$ due to the sizable $q^2$ extrapolation. As discussed below, the value of $|V_{ub}|$ is mostly determined in the region $q^{2}\approx 20~\text{GeV}^{2}$, which is at the low end of the $q^2$ range where the lattice-QCD form-factor error is still small. We therefore provide tabulated error budgets for the two form factors $f_{+},f_{0}$
from our calculation at the particular kinematic point $q^{2}=20\,\text{GeV}^{2}$
in Table~\ref{tab:Error-budgets-20}. The error on $f_+(20~\text{GeV}^2)$ is approximately 3.4\%, which is about one third of the error on our previously-determined form factor in Ref.~\cite{0811.3640}.

\begin{table}
	\caption{Error budgets of form factors $f_{+}$ and $f_0$ at $q^{2}=20\text{GeV}^2$.\label{tab:Error-budgets-20}}

	\begin{tabular}{lcc}
		\hline 
		\hline 
		Uncertainty  & $\delta f_{+}$  \;\;\;&\;\;\; $\delta f_{0}$ \\ 
		\hline 
		Statistical+$\chi$PT+HQ+$g_{B^{*}B\pi}$  & 3.1  & 3.8 \\ 
		Scale $r_{1}$  & 0.5  & 0.7 \\ 
		Non-perturbative $Z_{{V}^4_{bb}}$  & 0.4  & 0.6 \\ 
		Non-perturbative $Z_{{V}^4_{ll}}$  & 0.4  & 0.4 \\ 
		Perturbative $\rho$  & 1.0  & 1.0 \\ 
		Heavy-quark mass mistuning  & 0.4  & 0.4 \\ 
		Light-quark mass tuning  & 0.4  & 0.2 \\ 
		\hline 
		Total  & 3.4  & 4.1 \\
		\hline 
		\hline 
	\end{tabular}
\end{table}

\begin{figure}
	\center{\includegraphics[width=.48\linewidth]{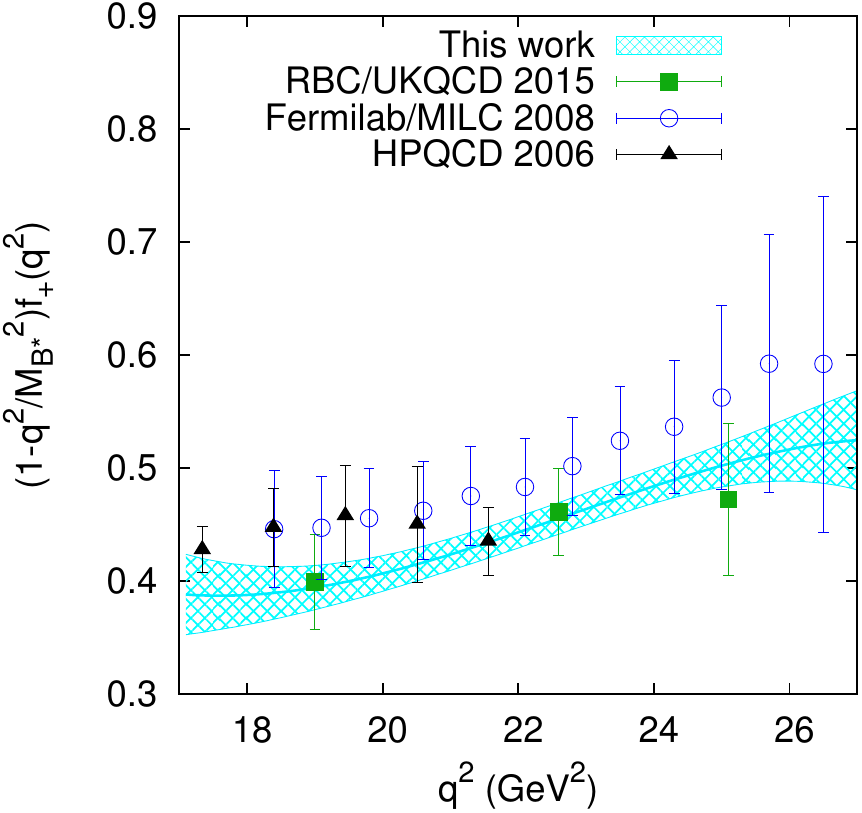}\hfill \includegraphics[width=.48\linewidth]{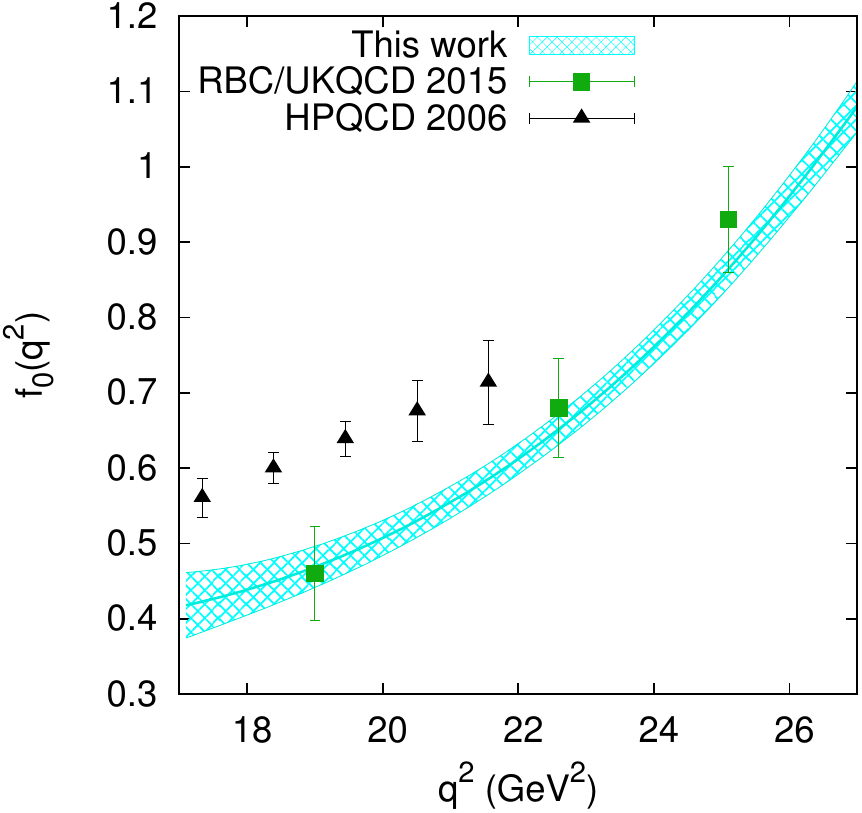}}
	
	\caption{Comparison of $f_+$ (left) and $f_0$ (right) from this work with previous lattice-QCD calculations by HPQCD \cite{hep-lat/0601021}, Fermilab/MILC\cite{0811.3640} and RBC/UKQCD \cite{Flynn:2015mha}.
		\label{fig:f+_compare}}
\end{figure}

We compare our results for $f_{+}$ and $f_{0}$ with full errors, which are obtained by adding the fit errors from the $\chi$PT fits and $\delta_f$ in quadrature, with previous lattice-QCD calculations in Fig.~\ref{fig:f+_compare}. Our result for $f_+$ agrees with previous results obtained at $q^2 \gtrsim 17$~GeV$^2$ from Refs.~\cite{hep-lat/0601021, 0811.3640,Flynn:2015mha}, but is more precise. Our result for $f_0$ is consistent with Ref.~\cite{Flynn:2015mha}, but not with Ref.~\cite{hep-lat/0601021}.

\section{\boldmath $z$ expansion and determination of $|V_{ub}|$} \label{secVI}

The chiral-continuum extrapolation described in the previous sections yields the form factors in the range $17~\text{GeV}^2 \leq q^2 \leq 26~\text{GeV}^2$. In this section, we extrapolate them to the full kinematic range using the model-independent $z$ expansion. The form factors resulting from the chiral-continuum extrapolation are functions specified by a set of parameters. One could,
in principle, incorporate the $z$ expansion with the $\chi$PT
expansion from the outset (see, {\it e.g.}, Ref.~\cite{1206.4936}). With such an approach, however, the coefficients of the $z$ expansion will have a nontrivial dependence on $m_l$ and $a$ that must be derived from the underlying chiral effective theory. Because the dependence of the coefficients on $a$ and $m_l$ is unknown, we instead carry out the extrapolation in two steps, taking the chiral-continuum extrapolated results and feeding them into the $z$ expansion. We introduce a functional
method to perform the $z$ expansion. We also apply the $z$ expansion to the experimental data and, after verifying that the fits to experiment and to lattice QCD are consistent, we carry out a combined fit to obtain $|V_{ub}|$. A byproduct of the last step is a precise determination for $f_+(q^2)$ constrained by lattice QCD at high $q^2$ and experiment at low $q^2$.

\subsection{$z$ expansions of heavy-light semileptonic form factors}

The $z$ expansion involves mapping the variable $q^{2}$
to a new variable $z$ by \cite{hep-ph/9702300}
\begin{eqnarray}
z(t,t_{0}) & = & \frac{\sqrt{t_{+}-q^2}-\sqrt{t_{+}-t_{0}}}{\sqrt{t_{+}-q^2 }+\sqrt{t_{+}-t_{0}}},\label{eq:Z(t)}
\end{eqnarray}
where $t_{\pm}=(M_{B}\pm M_{\pi})^{2}$ and $t_{0}$ is chosen for convenience below. This change of variables maps the whole complex $q^2$ plane onto
the unit disk in the $z$ plane, where the upper (lower) path along
the branch cut $[t_+,\infty)$ is mapped to the lower (upper) half of the circle enclosing
the unit disk in the complex $z$ plane. Choosing $t_{0}=(M_{B}+M_{\pi})(\sqrt{M_{B}}-\sqrt{M_{\pi}})^{2}$
centers the full kinematic range for semileptonic $B\to\pi \ell\nu$ decay around the origin $z=0$, and, moreover, restricts $z$ to $|z|<0.28$. The small, bounded interval, together with a constraint from unitarity ensures convergence of the expansion. As discussed below, we find in practice that the convergence is rapid.

The form factors $f_{+}$ and $f_0$ are analytic in $z$ except for the branch cut $[t_+,\infty)$ and poles in $[t_-,t_+]$. We can write
\begin{equation} 
P_i(z)\phi_i(z) f_i = \sum_n a_n z^n  \label{eq:BGL}
\end{equation}
where $P_i(z)$, $i=+,0$, are the Blaschke factors, which are introduced to remove the poles of $f_i$ in the region $[t_-,t_+]$, and $\phi_i(z)$ are the outer functions \cite{hep-ph/9702300, hep-ph/9412324}. We choose simple outer functions $\phi_{+,0}=1$ and employ the following formulas to expand the form factors 
\begin{eqnarray}
f_{+}(z) & = & \frac{1}{1-q^2(z)/M_{B^{*}}^{2}}\sum_{n=0}^{N_{z}-1}b^+_{j}\left[z^{n}-(-1)^{n-N_{z}}\frac{n}{N_{z}}z^{N_{z}}\right],\label{eq:f+(z)}\\
f_{0}(z) & = & \sum_{n=0}^{N_{z}}b^0_{n}z^{n}.\label{eq:f0(z)}
\end{eqnarray}
Equation (\ref{eq:f+(z)}) is known as the
Bourrely-Caprini-Lellouch (BCL) expansion \cite{0807.2722}, which is constructed to reproduce the threshold behavior at $q^2=t_+$ and the asymptotic behavior as $q^2\to\pm\infty$. Equation (\ref{eq:f0(z)}) is a simple series expansion of $f_0$ in $z$. 

The BCL coefficients in Eq.~(\ref{eq:f+(z)}) and~(\ref{eq:f0(z)}) obey the unitarity constraint \cite{0807.2722,hep-ph/9702300}
\begin{eqnarray}
\Sigma(b,N_{z})\equiv\sum_{m,n=0}^{N_{z}}B_{mn}b_{m}b_{n} & \lesssim & 1,\label{eq:unitarity}
\end{eqnarray}
where the element $B_{mn}$ satisfies $B_{nm}=B_{mn}=B_{0|m-n|}$ and depends on the choice of $t_0$~\cite{0807.2722}. We tabulate the values of $B_{0k}$
for the form factors $f_{+},f_{0}$ in Table \ref{tab:The-BCL-constants}. The inequality saturates when $N_z\to\infty$. Although we do not incorporate this constraint into our fits, we check that our results satisfy it. 

\begin{table}
	\caption{The BCL constants used to estimate $\Sigma(b,N_{z})$. \label{tab:The-BCL-constants}}

	\begin{tabular}{cccccccc}
		\hline 
		\hline 
		& $B_{00}$  & $B_{01}$  & $B_{02}$  & $B_{03}$  & $B_{04}$  & $B_{05}$  & $B_{06}$ \tabularnewline
		\hline 
		$f_{0}$  & 0.1032  & 0.0408  & $-$0.0357  & $-$0.0394  & $-$0.0195  & $-$0.0055  & $-$0.0004 \tabularnewline
		$f_{+}$  & 0.0198  & 0.0042  & $-$0.0109  & $-$0.0059  & $-$0.0002  & 0.0012  & 0.0011 \tabularnewline
		\hline
		\hline  
	\end{tabular}
\end{table}

\subsection{Functional method for the $z$ expansion}

In previous work, we have used synthetic data points generated from the $\chi$PT fit as inputs to the $z$ fit \cite{0811.3640}, but here we take a new
approach. We exploit the facts that the $\chi$PT expansion is linear in the fit parameters and that it contains only a finite number of independent functions (see Eq.~(\ref{eq:chiral_f})). We construct a covariance function $K(z_1,z_2)$, defined as the covariance of any pair of points ($z_{1},z_{2}$), using the set of functionals from the $\chi$PT expansion. Our new approach is to formulate
the $z$ expansion using the eigenfunctions of an integral operator defined from $K(z_{1},z_{2})$.

Let us start with the NLO $\chi$PT expression Eq.~(\ref{eq:chiral_f}),
as an example. Because $f_{\perp}$ and $f_{\parallel}$ are linear in their coefficients $c_{i}^{\perp}$ and $c_{i}^{\parallel}$, we can express them both in the compact form 
\begin{eqnarray}
f_{J}(m_{\ell},m_{s},a^{2},E_{\pi}) & = & C_{J}\cdot X_{J},\label{eq:f_gamma_compact}
\end{eqnarray}
where 
\begin{eqnarray}
C_{J} & \equiv & \left[\begin{array}{ccccccc}
c_{0} & c_{1} & c_{2} & c_{3} & c_{4} & c_{5} & \cdots\end{array}\right]^{J},\label{eq:C_gamma}\\
X_{J} & \equiv & f_{J}^{(0)}\left[\begin{array}{ccccccc}
(1+\delta f_{J,\text{logs}}) & \chi_\text{val} & \chi_\text{sea} & \chi_{E} & \chi_{E}^{2} & \chi_{a^{2}} & \cdots\end{array}\right]^T\label{eq:X_gamma}
\end{eqnarray}
and where $J=\perp,\parallel$ and the variables are defined in Eqs.
(\ref{eq:xval})--(\ref{eq:xE}). Any linear combination of $f_{\perp}$ and $f_{\parallel}$
can be written as 
\begin{eqnarray}
f & = & [\begin{array}{cc}
\xi & \eta\end{array}]\left[\begin{array}{c}
f_{\perp}\\
f_{\parallel}
\end{array}\right]\nonumber\\
& = & [\begin{array}{cc}
C_{\perp}^{T} & C_{\parallel}^{T}\end{array}]\left[\begin{array}{c}
\xi X_{\perp}\\
\eta X_{\parallel}
\end{array}\right],\label{eq:f_general}
\end{eqnarray}
with $\xi,\eta$ functions of $q^{2}$. The uncertainty
of the function $f$ is encoded in the uncertainty in the coefficient vector $C_{J}$.
In all these expressions, we are only interested in the terms with
$E_{\pi}$ (or $q^{2}$) dependence and, hence, $z$ dependence. We can now define the covariance function $K(z,z')$ in some valid
domain $[z_{1},z_{2}]$. Explicitly, 
\begin{eqnarray}
K(z,z') & = & Y(z)^{T}\cdot\text{Cov}\cdot Y(z'),\label{eq:K(z,z)}
\end{eqnarray}
where 
\begin{eqnarray}
Y(z) & = & \left(\begin{array}{c}
\xi(z)\, X_{\perp}(z)\\
\eta(z)\, X_{\parallel}(z)
\end{array}\right),\label{eq:Y(z)}
\end{eqnarray}
and Cov is the covariance matrix of the involved coefficients $c^J_n$
\begin{eqnarray}
\text{Cov}_{mn} & = & \langle\delta c_{m}\delta c_{n}\rangle,\label{eq:cov}
\end{eqnarray}

The covariance function $K(z,z')$ is a Mercer
kernel \cite{mercer}, and Mercer's theorem ensures that there exists
a set of orthonormal functions $\psi_{i}(z)$ defined over the domain $[z_{1},z_{2}],$
such that 
\begin{eqnarray} \label{eq:K_eigenmode}
K(z,z') & =\sum_{i}\lambda_{i} & \psi_{i}(z)\psi_{i}(z'),
\end{eqnarray}
where $\lambda_{i}$, $\psi_{i}$ are the eigenvalues and eigenfunctions
of the operator $L_{K}$ induced by the integral equation,
\begin{eqnarray}
L_{K}\psi(z) & = & \int_{z_{1}}^{z_{2}}K(z,z')\psi(z')dz'.\label{eq:L_K}
\end{eqnarray}

The form factor $f(z)$ can naturally be expanded in the basis of $\psi_{i}(z)$: we only need to project the expansions in Eqs.~(\ref{eq:f+(z)}) and~(\ref{eq:f0(z)})
onto the same basis. The process of finding the expansion coefficients
$b_{n}$ is equivalent to minimizing the following function
(in analogy to the usual $\chi^{2}$ function, replacing the sum over discrete
points with an integral over a continuous variable): 
\begin{eqnarray}
\mathcal{\chi}_\text{lat}^{2} & = & \int_{z_{1}}^{z_{2}}dz\int_{z_{1}}^{z_{2}}dz'\left[f^{\chi\text{PT}}(z)-g_{f}(b,z)\right]K^{-1}(z,z')\left[f^{\chi\text{PT}}(z')-g_{f}(b,z')\right]\nonumber \\
& = & \sum_{i=1}^{N_{\psi}}\frac{1}{\lambda_{i}}\left[\int_{z_{1}}^{z_{2}}dz[f^{\chi\text{PT}}(z)-g_{f}(b,z)]\psi_{i}(z)\right]^{2}\nonumber \\
& = & \sum_{i=1}^{N_{\psi}}\frac{1}{\lambda_{i}}\left[f_{i}^{\chi\text{PT}}-\sum_{n=0}^{N_{z}-1}b_{n}\int_{z_{1}}^{z_{2}}\theta_{n}^{f}(z)\psi_{i}(z)dz\right]^{2},\label{eq:object_L}
\end{eqnarray}
where 
\begin{eqnarray}
f^{\chi\text{PT}}(z) & = & \sum_{i}f_{i}^{\chi\text{PT}}\psi_{i}(z),\label{eq:f_i}
\end{eqnarray}
is the form factor function from the $\chi$PT fit expanded in terms of $\psi_{i}$, and 
\begin{eqnarray}
g_{f}(b,z) & = & \sum_{n=0}^{N_{z}-1}b_{n}\theta_{n}^{f}(z) \label{eq:g_f}
\end{eqnarray}
are functions rewritten from the functions defined in Eqs.~(\ref{eq:f+(z)}) and (\ref{eq:f0(z)}).
For brevity we define
\begin{eqnarray}
\theta_{n}^{+}(z) & = & \frac{1}{1-q^2(z)/M_{B^*}^2}\left[z^{n}-(-1)^{n-N_{z}}\frac{n}{N_{z}}z^{N_{z}}\right],\label{eq:theta+}\\
\theta_{n}^{0}(z) & = & z^{n}. \label{eq:theta0}
\end{eqnarray}

To summarize, we expand any form factor function $f^{\chi\text{PT}}$ obtained
from the chiral-continuum extrapolation in the basis formed by the
eigenfunctions of its covariance function $K(z,z')$. We then
project the $z$ expansion onto the same basis.
Finally, we solve for the expansion parameters $b_{n}$ by minimizing
the function $\chi^2_\text{lat}$ defined in Eq.~(\ref{eq:object_L}).

\subsection{Details on $z$ expansion of the form factors}

In addition to the fit errors from the chiral-continuum fit, we also need to propagate
the subdominant errors, which have very mild $q^2$ dependence. We treat them as constant in $q^2$ and add them in quadrature, obtaining $\delta_{f_+}=1.4\%$ and $\delta_{f_0}=1.5\%$. To include this effective subdominant error to the fit, we slightly modify the covariance function defined in Eq.~(\ref{eq:K(z,z)}) by
\begin{eqnarray}
K'(z,z') & = & Y(z)^{T}\cdot\text{Cov}'\cdot Y(z'),\label{eq:K'(z,z)}
\end{eqnarray}
where the new covariance matrix includes the subdominant error,
\begin{eqnarray}
\text{Cov}'_{mn} & = & \langle\delta c_{m}\delta c_{n}\rangle + \delta_f^2c_{m}c_{n}. \label{eq:cov'}
\end{eqnarray}

\begin{figure}
	\center{\includegraphics[scale=0.9]{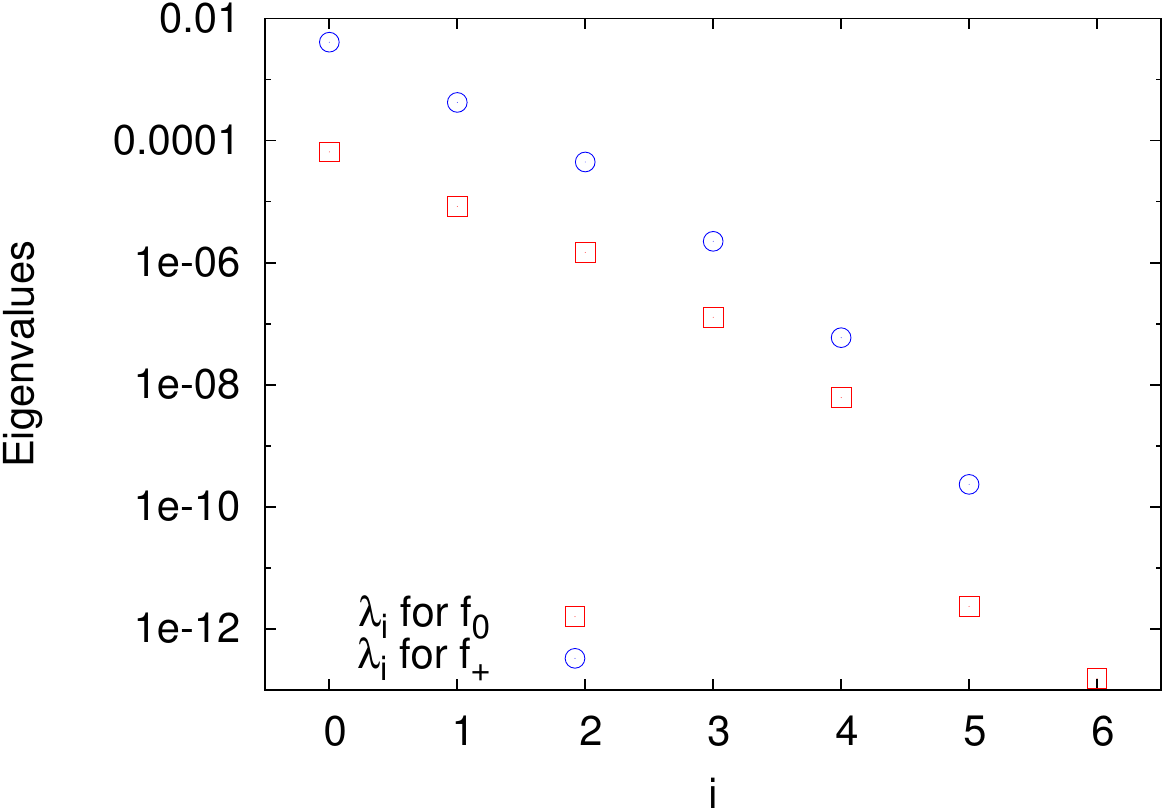}}
	
	\caption{Eigenspectrum of the kernel function $K(z,z')$ for $f_{0}$ and $f_{+}$. \label{fig:eigen_zfit}}
\end{figure}

In the function array $Y(z)$ defined in Eq.~(\ref{eq:Y(z)}),
only a relatively small number of the elements are independent functions. For example, there are 42 terms in the NNLO SU(2) $\chi$PT fit functions for $f_{\parallel, \perp}$ (including the
HQ discretization contributions). Many of them, however, are set to zero in the continuum limit or become constant once the light-quark mass is fixed at its physical value. In the end, the chiral-continuum extrapolated $f_+$ is described by only 6 independent functions. For $f_{0}$, the number of independent functions is 7.
Although we work in the functional
basis in which the covariance function $K(z,z')$ is diagonalized,
singular modes can arise because
$K(z,z')$ is built upon $\text{Cov}_{f}$, which itself may have singular modes.
Figure~\ref{fig:eigen_zfit} shows the spectra of the operator
$L_{K}$ for form factor $f_{+,0}$. The spectrum
of $f_{0}$ contains two very small eigenvalues $\lesssim10^{-12}$, 
and they are well separated from the other modes. When we discard these two modes, the fit quality of the functional $z$ fit improves from $p=0.03$  to $p=0.46$. For $f_+$, we do not need to apply any cut on the eigenvalues.

\begin{figure}
	\center{\includegraphics[width=0.49\textwidth, trim= 30 30 30 0]{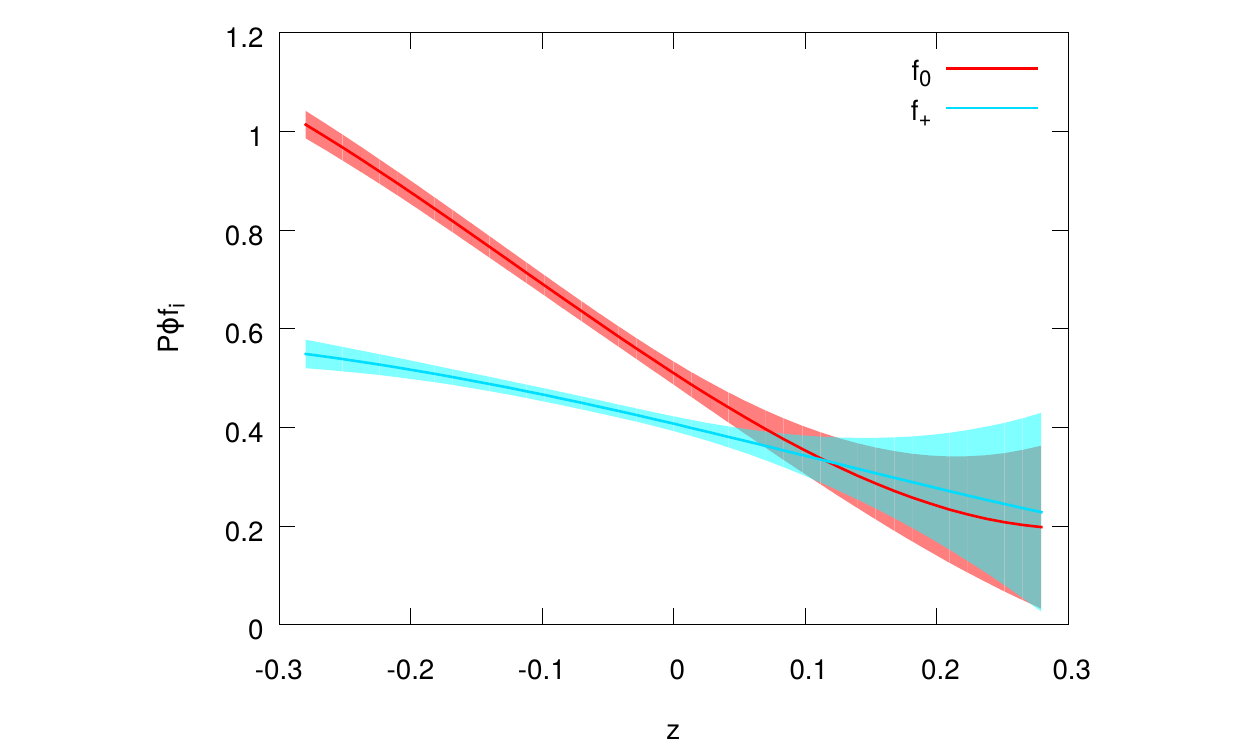}
		\;\includegraphics[width=0.49\textwidth, trim= 30 30 30 0 ]{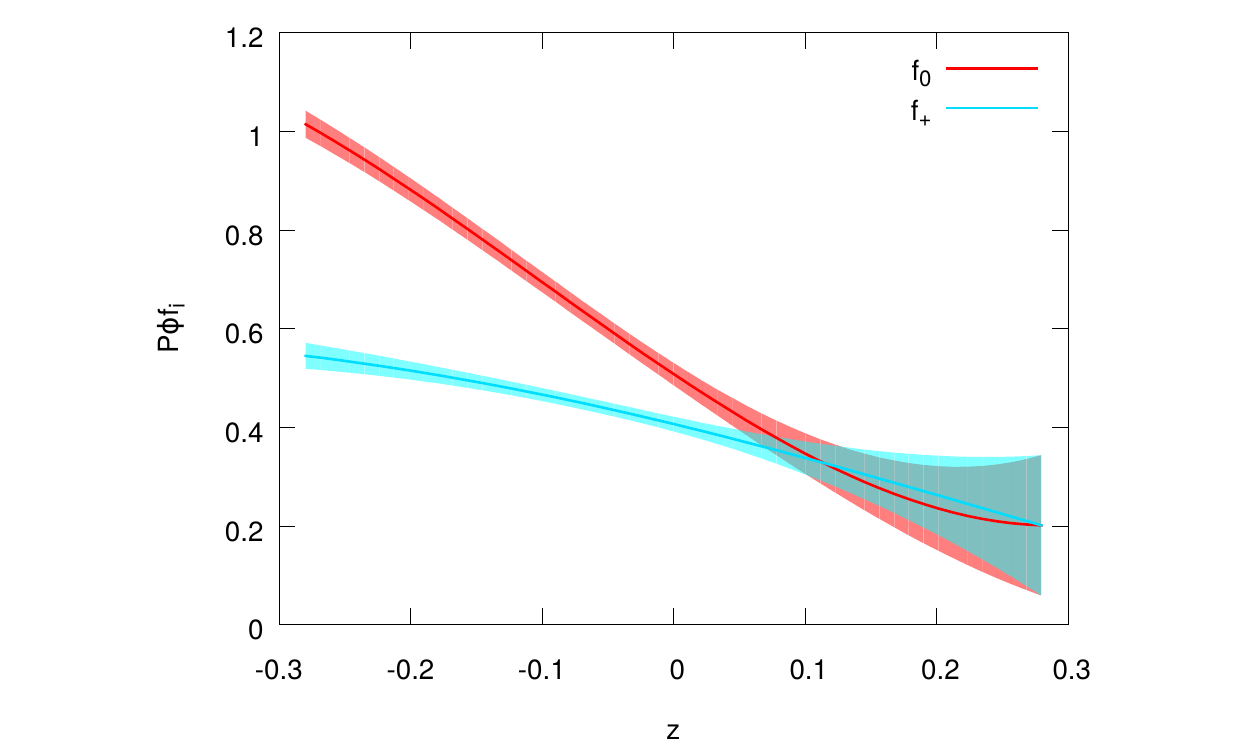}}
	\caption{The $z$ expansion of the form factors $f_+$ and $f_0$ without any constraints (left) and with the kinematic constraint. The fits use $N_z = 4$. \label{fig:kin_constraint}}
\end{figure}

We first consider separate fits of $f_+$ and $f_0$ without any constraints on the coefficients of the $z$ expansion. With $N_{z}=3$, or three free parameters $b_{0},b_{1},b_{2}$, we
obtain a low confidence level, $p=0.05$, for the fit to $f_{0}$. The analogous three-parameter fit for $f_{+}$ 
results in an acceptable confidence level, $p=0.3$. With $N_z=4$ we find good confidence levels for both form factors as well as sizable changes in the central values and errors (compared to the $N_z=3$ case). The results of unconstrained fits of $f_+$ and $f_0$ with several values of $N_z$ for $f_0$ and $f_+$ are given for comparison in Table~\ref{tab:unconstrained}. The kinematic constraint $f_+(q^2=0)=f_0(q^2=0)$ is satisfied automatically, as is shown in Fig.~\ref{fig:kin_constraint} (left). 

\begin{figure}
	\center{\includegraphics[width=0.49\textwidth, trim= 30 30 30  30]{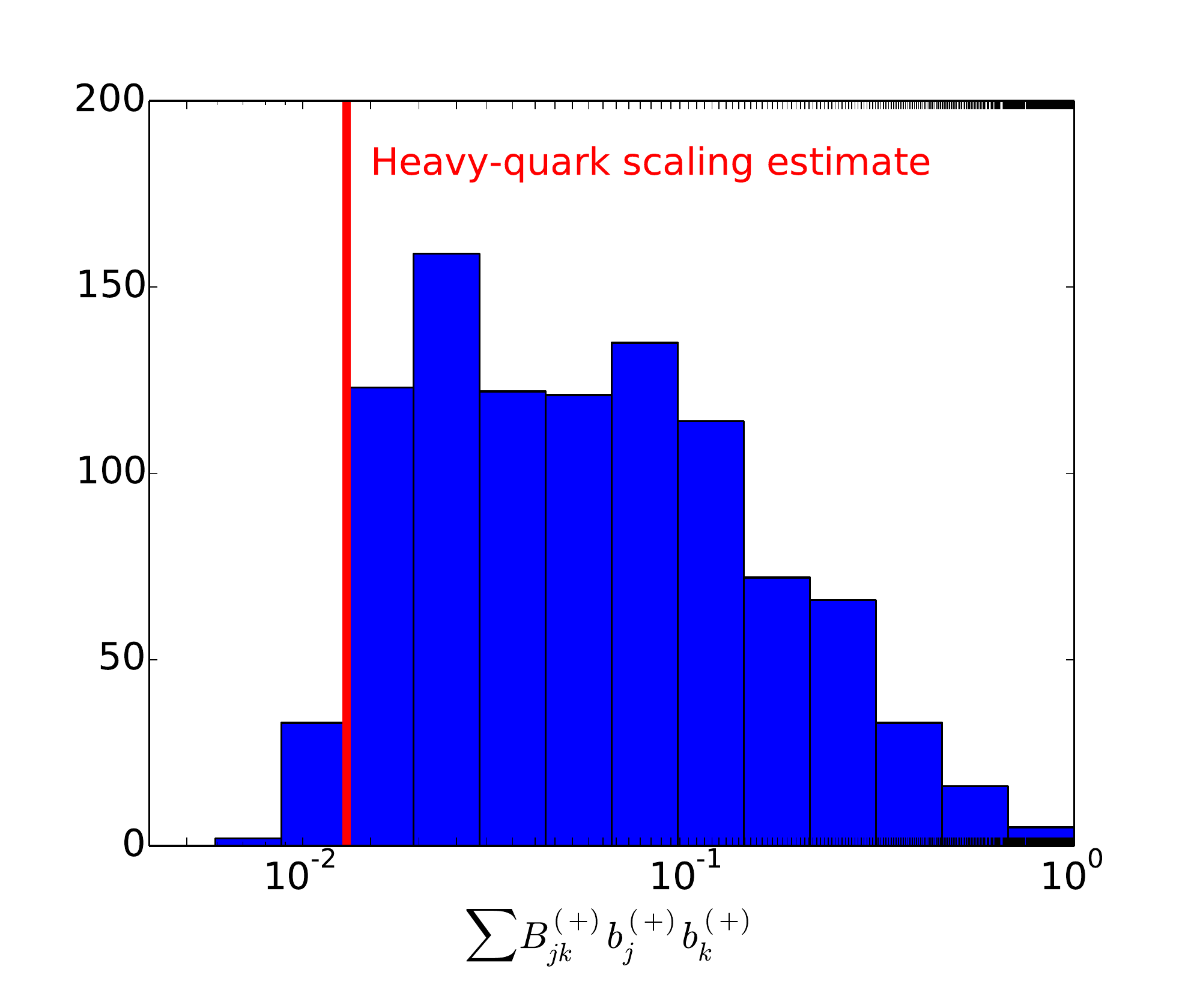}
		\hfill\includegraphics[width=0.49\textwidth, trim= 30 30 30 30 ]{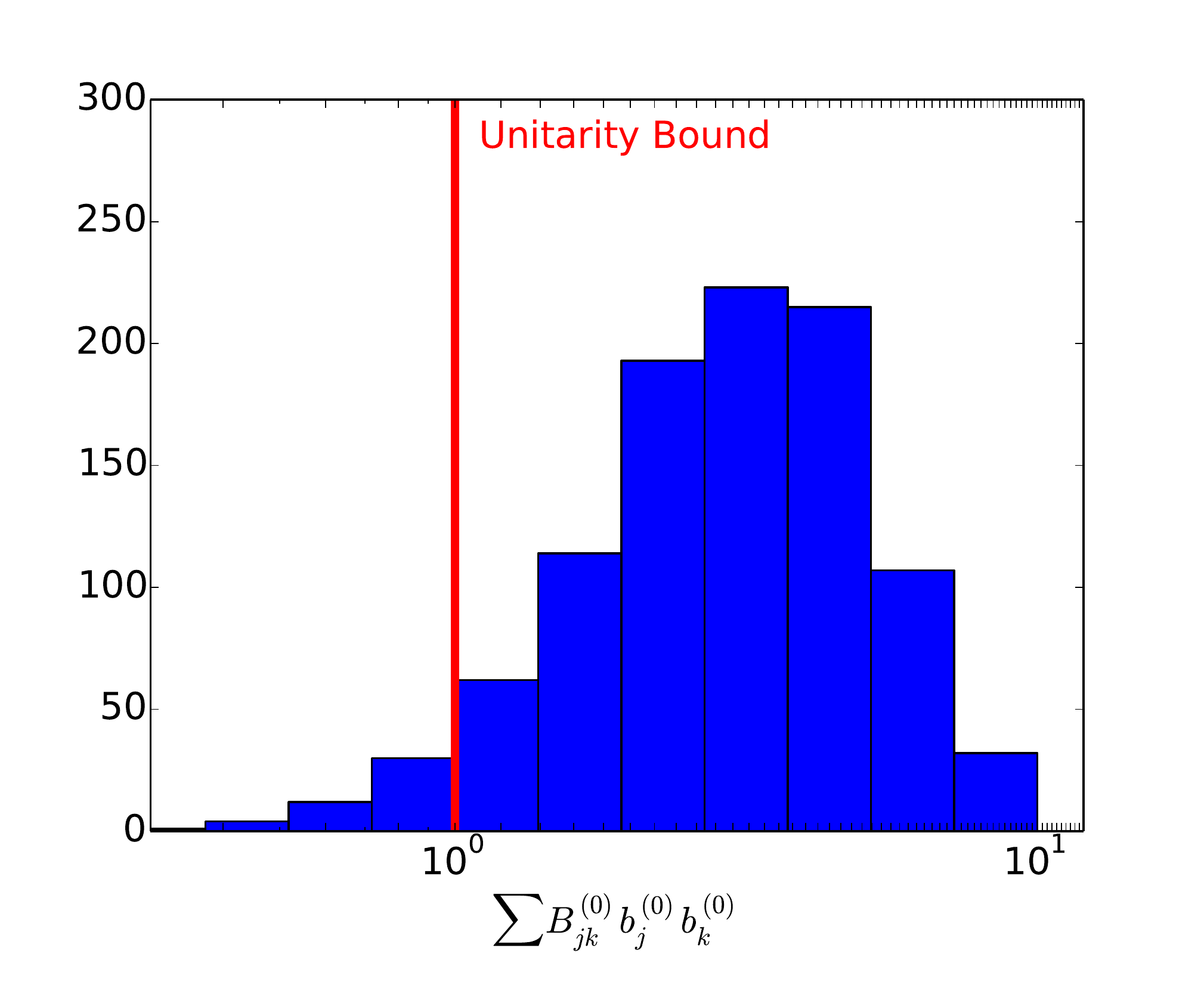}}
	\caption{Bootstrap sample distribution defined in Eq.~(\ref{eq:unitarity}) for the $z$ expansion coefficients of $f_+$ (left) and $f_0$ (right), compared with the heavy-quark scaling estimate $0.013$ (red vertical line in the left plot) and the unitarity bound (red vertical line in the right plot). The 1000 samples of fits use only the kinematic constraint and $N_z = 4$.  \label{fig:Bjksum}}
\end{figure}

The $z$-expansion coefficients $b_i$ should approximately satisfy the unitarity bound Eq.~(\ref{eq:unitarity}). Figure \ref{fig:Bjksum} (right) shows the bootstrap-sample distribution of $\Sigma(b,4)$ from the fits for $f_0$. The unitarity condition is marginally satisfied. In the case of $f_{+}$, the sum $\Sigma(b,4)\sim 0.03$ is much smaller than the unitarity bound. Reference~\cite{hep-ph/0509090} pointed out that the smallness is expected based on heavy-quark physics. The value of $\Sigma(b,4)$ should be of order $(\Lambda/m_b)^3$, which is about 0.013 with a conservative choice of $\Lambda=1$ GeV. The bootstrap-sample distribution of $\Sigma(b,4)$ from the fits for $f_+$ is shown against the heavy-quark estimate in Fig. \ref{fig:Bjksum} (left). They are consistent with each other. 

\begin{figure}
	\center{\includegraphics[scale=0.9]{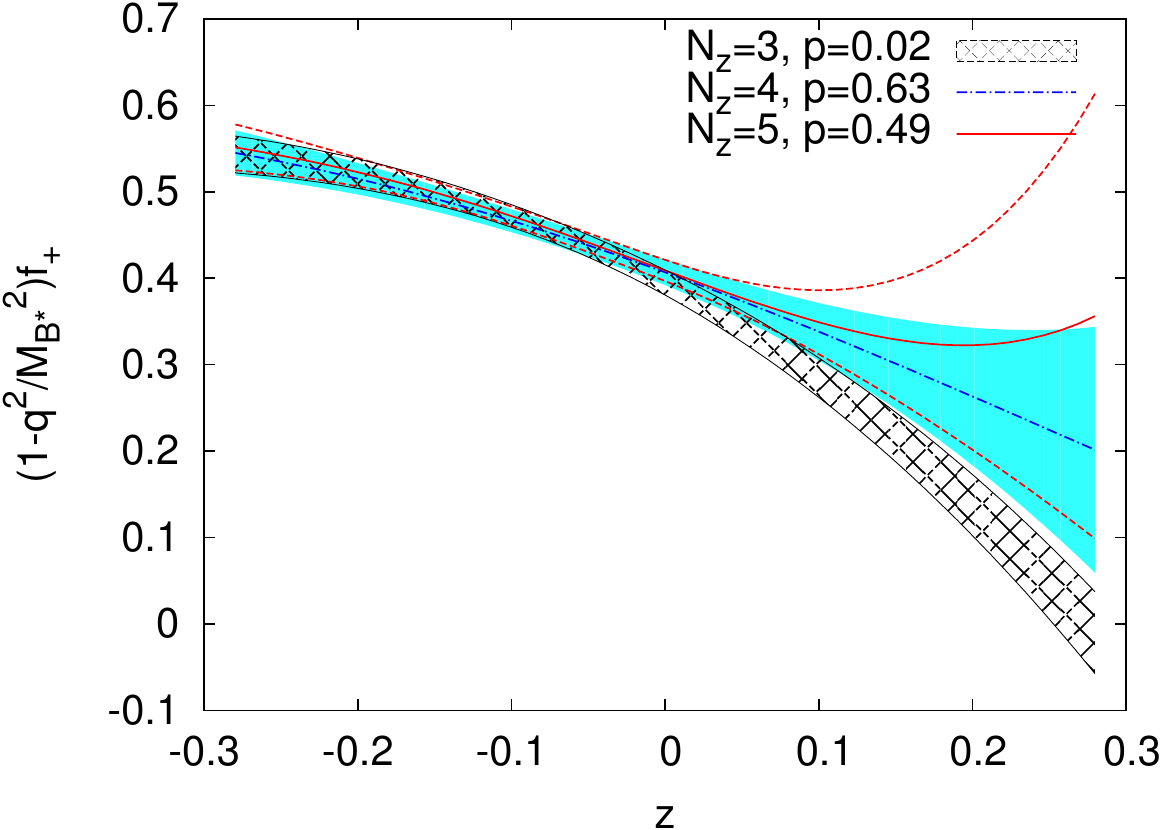}}
	
	\caption{Effects of truncating the $z$ expansion for $f_+$. The areas indicated by the hatched band, colored shaded band, and the lines are results with 1$\sigma$ errors for $N_z=3,4,5$ $z$ fits, respectively.  \label{fig:Nz}}
\end{figure}

The result of the fits of $f_{+,0}$ with the kinematic constraint are shown in Fig.~\ref{fig:kin_constraint} (right). With this constraint, we again examine how the fit varies with higher order $N_z$. We find that the fit central values do not change significantly when we change $N_z$ from 4 to 5, in contrast to the case from 3 to 4, as is shown in Fig.~\ref{fig:Nz} and Table~\ref{tab:kin_constraint}.

\begin{table}
	\caption{Results from unconstrained $z$ fits of $f_0$ and $f_+$. \label{tab:unconstrained} }
	
	\begin{tabular}{ccccccc}
		\hline 
		\hline 	
		& \multicolumn{3}{c}{$f_+$}  & \multicolumn{2}{c}{$f_0$}  \\ 
		$N_z$     & $3$        & $4$        & $ 5$   & $3$        & $4$     \\ 
		\hline
		$\chi^{2}$/dof &  1.1   &  1.0    &  1.6    &  3.1    &  0.54  \\ 
		dof  &  3   &  2  &  1  &  2  &  1  \\ 
		$p$ &  0.27   &  0.33   &  0.2      &  0.05   &  0.46    \\ 
		$\sum B_{mn}b_{m}b_{n}$ &     0.052(16) &     0.024(39)  &     2(4)  &  0.60(24)  &  2.0(11)   \\
		$f(0)$  &   0.12(6) &  0.23(20) &  0.40(34) &  $-$0.14(9)  &  0.20(17)  \\ 
		$b_{0}$   &  0.409(16)   &  0.409(12)   &  0.407(15) \;\;\;&\;\;\;  0.493(22)  &  0.510(23)    \\ 
		$b_{1}$    &  $-$0.72(13)      &  $-$0.63(20)     &  $-$0.60(22)  &  $-$2.1(2)          &  $-$1.7(2)          \\ 
		$b_{2}$   &  $-$1.0(2)   &  $-$0.3(1.3)   &  0.3(1.7)  &  $-$0.8(5)     &  1.2(9)      \\ 
		$b_{3}$  &      &  1(2)    &  4(5)    &         &  3(1)     \\ 
		$b_{4}$ &             &     &  7(7)   &             &      \\ 
		
		\hline
		\hline  
	\end{tabular}
\end{table}

\begin{table}
	
	\caption{Results of simultaneous fits of $f_+$ and $f_0$ with the kinematic constraint. The $N_z=4$ fit is our preferred result. \label{tab:kin_constraint}}
	
	\begin{tabular}{cccc}
		\hline 
		\hline 
		Fit	 \;\;\;&\;\;\; $N_{z}=3$ \;\;\;&\;\;\; $N_{z}=4$  \;\;\;&\;\;\; $N_{z}=5$ \tabularnewline		
		\hline
		$\chi^{2}$/dof
		& 2.5 &0.64& 0.73\tabularnewline
		dof 
		& 6 &4& 2\tabularnewline
		$p$ 	 & 0.02 		& 0.63 		& 0.48\tabularnewline
		$\sum B^+_{mn}b^+_mb^+_n$	
		&	0.11(2)	& 0.016(5) 		& 1.0(2.3) \tabularnewline
		$\sum B^0_{mn}b^0_mb^0_n$	
		&	0.33(8)	& 2.8(1.7)		& 8(19) \tabularnewline	
		$f(0)$		 &	0.00(4)		& 0.20(14)		& 0.36(27)  \tabularnewline	
		$b_{0}^{+}$  & 0.395(15) 	& 0.407(15) & 0.408(15)\tabularnewline	
		$b_{1}^{+}$  & $-$0.93(11) 	& $-$0.65(16) & $-$0.60(21)\tabularnewline	
		$b_{2}^{+}$  & $-$1.6(1) 	& $-$0.5(9) 	&  $-$0.2(1.4)\tabularnewline	
		$b_{3}^{+}$  &  			& 0.4(1.3) 	& 3(4)\tabularnewline	
		$b_{4}^{+}$  &  			&  			& 5(5)\tabularnewline		 		
		$b_{0}^{0}$  & 0.515(19) 	& 0.507(22) & 0.511(24) \tabularnewline	
		$b_{1}^{0}$  & $-$1.84(10) 	& $-$1.77(18) & $-$1.69(22)\tabularnewline	
		$b_{2}^{0}$  & $-$0.14(25) 	& 1.3(8) 	& 2(1)\tabularnewline		
		$b_{3}^{0}$  &  			& 4(1) 	& 7(5)\tabularnewline	
		$b_{4}^{0}$  &  			&  			& 3(9)\tabularnewline	
		\hline 
		\hline
	\end{tabular}
	
\end{table}

\begin{table}
	\caption{ Central values, errors, and correlation matrix of the coefficients of $f_+$ and $f_0$ from the $N_z = 4$ lattice-only $z$-fit with the kinematic constraint.  \label{tab:fpf0fT_correlation}}
	\begin{tabular}{ccccccccc}
		\hline 
		\hline 
		& $b_{0}^{+}$ & $b_{1}^{+}$ & $b_{2}^{+}$ & $b_{3}^{+}$ & $b_{0}^{0}$ & $b_{1}^{0}$ & $b_{2}^{0}$ & $b_{3}^{0}$ \\ 
		\hline 
		& 0.407(15) & $-$0.65(16) & $-$0.46(88) & 0.4(1.3) & 0.507(22) & $-$1.77(18) & 1.27(81) & 4.2(1.4) \\
		$b_{0}^{+}$ & 1 & 0.451 & 0.161 & 0.102 & 0.331 & 0.346 & 0.292 & 0.216 \\ 
		$b_{1}^{+}$ &  & 1 & 0.757 & 0.665 & 0.430 & 0.817 & 0.854 & 0.699 \\ 
		$b_{2}^{+}$ &  &  & 1 & 0.988 & 0.482 & 0.847 & 0.951 & 0.795 \\ 
		$b_{3}^{+}$ &  &  &  & 1 & 0.484 & 0.833 & 0.913 & 0.714 \\ 
		$b_{0}^{0}$ &  &  &  &  & 1 & 0.447 & 0.359 & 0.189 \\ 
		$b_{1}^{0}$ &  &  &  &  &  & 1 & 0.827 & 0.500 \\ 
		$b_{2}^{0}$ &  &  &  &  &  &  & 1 & 0.838 \\ 
		$b_{3}^{0}$ &  &  &  &  &  &  &  & 1 \\
		\hline 
		\hline
	\end{tabular}
	
\end{table}

We perform several additional checks to confirm the stability of our results against various fit choices. In our preferred fit, we set the integral range in Eq.~(\ref{eq:object_L}) to be $z=[-0.25, 0.01]$ (or equivalently $q^2=[19.8, 26.0]$). The results, however, do not change noticeably if we extend the integral range to $z=[-0.249,0.069]$ which covers the full range of simulated lattice momenta. This is because the statistical fluctuations and correlations of the form-factor functions are largely decided by the region $-0.1\lesssim z \lesssim 0$, where the $\chi$PT fit results are the most precise. 

We also try removing the smallest eigenvalue from the covariance function $K(z,z')$ for $f_+$; we find that the resulting central values are essentially unaffected. Finally, we also try the fit using, instead of the BCL formula, the Boyd-Grinstein-Lebed (BGL) formula, which uses more complicated outer functions \cite{hep-ph/9412324}. We find that the resulting form factors are within one standard deviation of the BCL result. 

\begin{figure}
	\center{\includegraphics[width=0.85\linewidth]{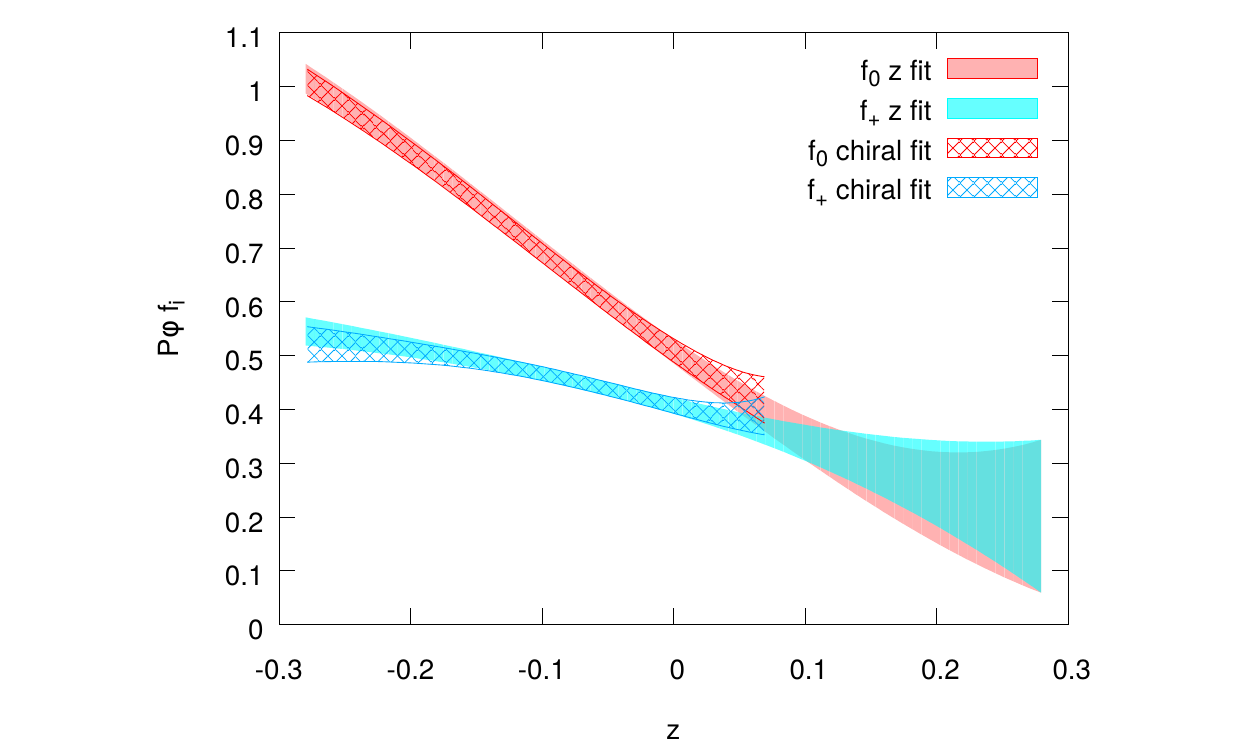}}
	
	\caption{$z$-fit results for the form factors $f_{0}$ and $f_{+}$ as functions of $z$. 
		\label{fig:zfit_all}}
\end{figure}

To summarize, we obtain our preferred result from a simultaneous fit to $f_+$ and $f_0$ with $N_z=4$ and with imposing the kinematic constraint $f_+(q^2=0)=f_0(q^2=0)$. The results for the two form factors $f_{+}$ and $f_{0}$ are plotted in Fig.~\ref{fig:zfit_all}. In this plot, the form factors obtained from the $\chi$PT fit are overlaid on the results of the $z$ fit. The $z$ fit faithfully reproduces the $\chi$PT fits in the region where $\chi$PT is reliable (indicated by the ranges of the hatched bands). 
The $z$ coefficients with errors from our preferred fit and their correlation matrix are provided in Table~\ref{tab:fpf0fT_correlation}. This information is sufficient to reproduce the lattice form-factor results over the full kinematic range. 

\begin{figure}
	\center{\includegraphics[width=.48\linewidth, trim= 80 10 80  10 ]{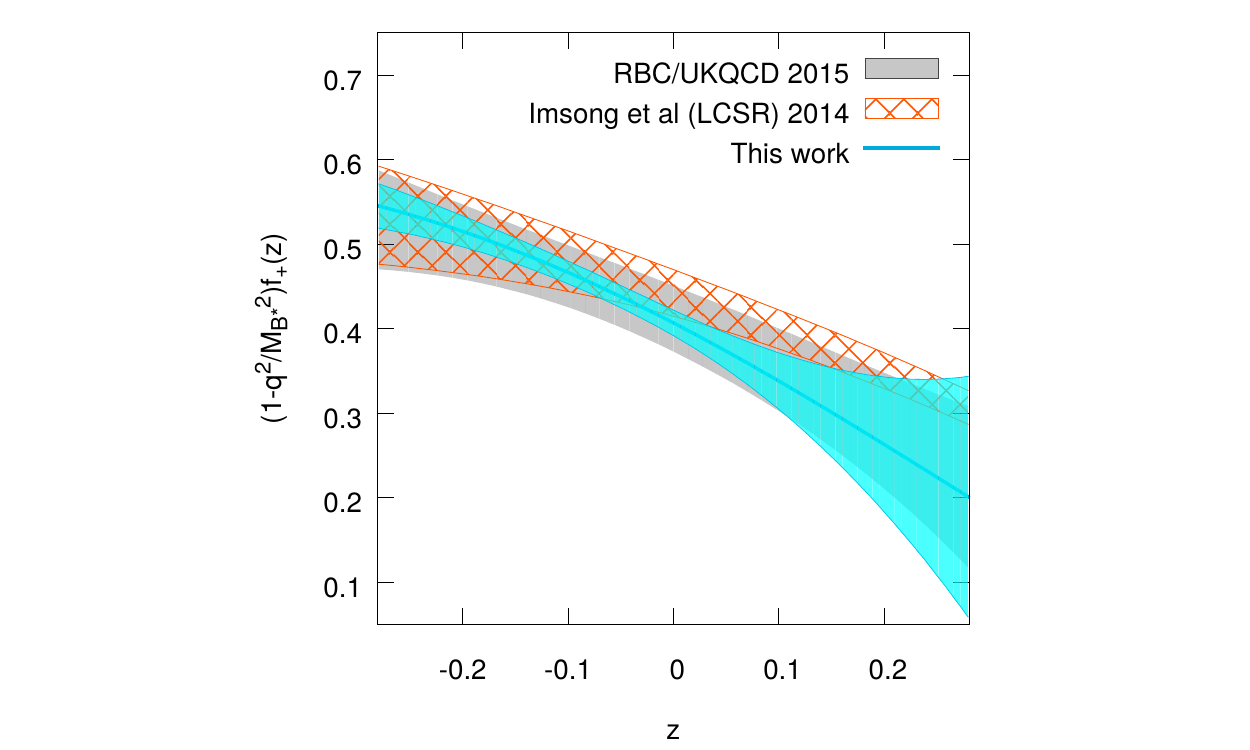}\hfill \includegraphics[width=.48\linewidth, trim= 80 10 80  10]{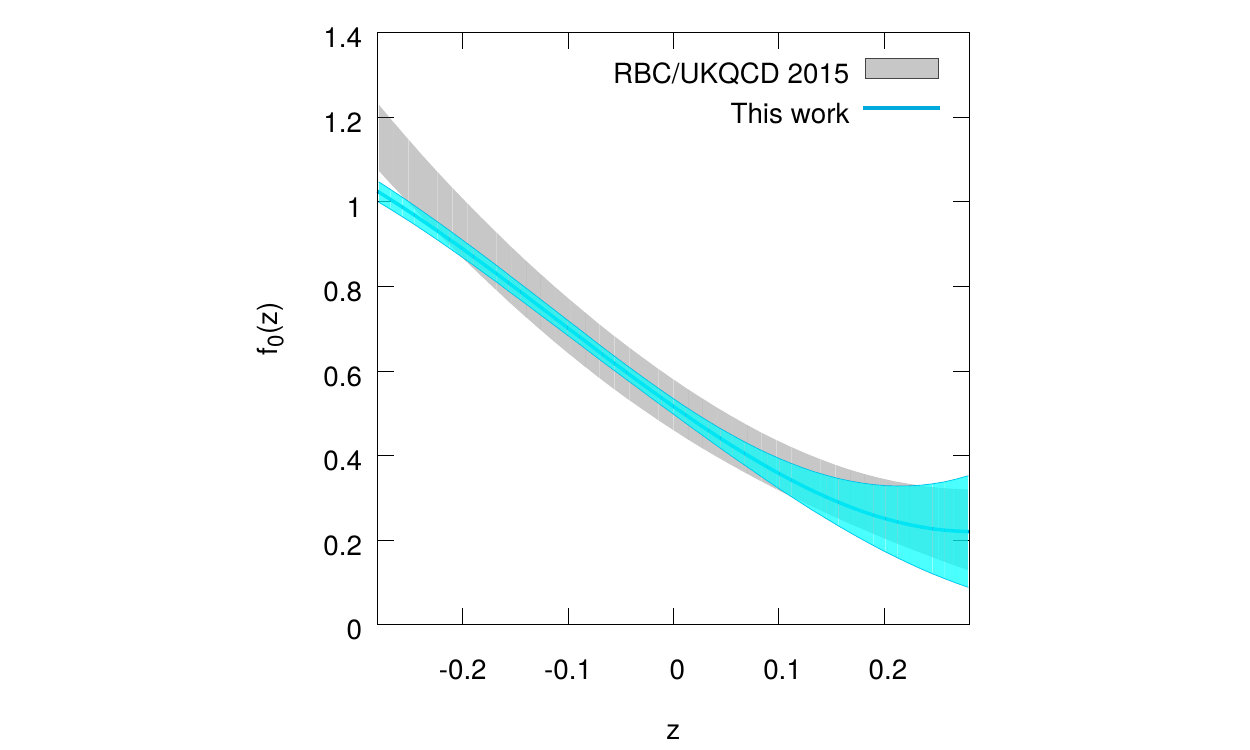}}
	
	\caption{Comparison of $f_+$ (left) and $f_0$ (right) from the $z$-expansion fit results of this work with recent theoretical calculations using light-cone sum rules (LCSR) \cite{Imsong:2014oqa} and lattice QCD \cite{Flynn:2015mha}.
		\label{fig:zfit_compare}}
\end{figure}

Figure~\ref{fig:zfit_compare} shows a comparison of our results with other theoretical calculations of the form factors \cite{Imsong:2014oqa, Flynn:2015mha}. While our results are consistent with the previous results, ours are significantly more precise in the region of $z \leq 0.1$. 

\begin{figure}
	\center{\includegraphics[width=0.61\linewidth, trim= 40 22 40 0]{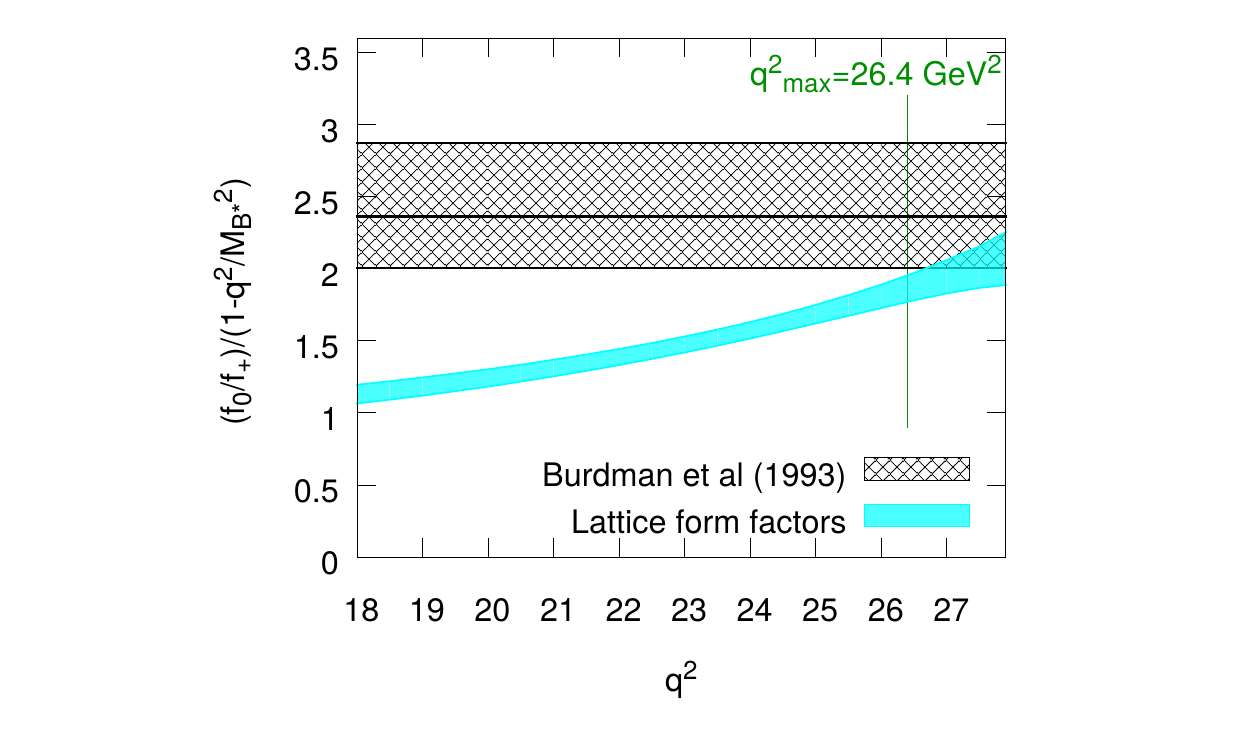}}
	
	\caption{ The lattice form-factor ratio $(f_0/f_+)/(1-q^2/M^2_{B^*})$ versus $q^2$ (cyan shaded band) compared with the prediction in the soft-pion limit from heavy-quark symmetry and $\chi$PT~\cite{Burdman:1993es} (hatched band). The width of the hatched band reflects only the uncertainty from $g_{B^*B\pi}=0.45(8)$ and not other theoretical errors.
		\label{fig:f0_over_fplus}}
\end{figure}

Finally, it is interesting to compare the lattice form factors with theoretical expectations from heavy-quark symmetry.  In the soft-pion limit, the vector and scalar form factors $f_+$ and $f_0$ are related as~\cite{Burdman:1993es}
\begin{equation}
\lim_{q^2 \to M_B^2} \frac{f_0(q^2)}{f_+(q^2)} = \left(\frac{f_{B}}{f_{B^*}}\right) \frac{1-q^2/M_{B^*}^2}{g_{B^*B\pi}}
\label{eq:f0f+lowrecoil}
\end{equation}
up to corrections of ${\mathcal O}(1/m_b^2)$. This expression updates the leading-order result of Ref.~\cite{Wise:1992hn} to include the $1/m_b$ correction, which turns out to be simply the additional multiplicative factor $(f_{B^*}/f_B)^{-1}$ in the soft-pion limit. In Fig.~\ref{fig:f0_over_fplus} we 
plot the ratio of $(f_0/f_+)/(1-q^2/M_{B^*}^2)$ obtained using the coefficients of our preferred $z$-expansion in Table~\ref{tab:fpf0fT_correlation}. We also show the theoretical expectation from Eq.~(\ref{eq:f0f+lowrecoil}), taking the HPQCD Collaboration's recent three-flavor lattice-QCD result for the decay-constant ratio $f_{B^*}/f_B = 0.941(26)$~\cite{Colquhoun:2015oha}, and using the same value of $g_{B^*B\pi} = 0.45(8)$ as in our chiral-continuum extrapolation. The large width of the expected band is due to the generous range taken for $g_{B^*B\pi}$.  Higher-order 
corrections in the heavy-quark expansion are expected to be small. Taking a conservative value for $\Lambda = 500$~MeV and $m_b = 4.2$~GeV, one would estimate $(\Lambda/m_b)^2$ corrections to be about 1\%.  The difference of $f_{B^*}/f_B$ from one also provides a measure of $\Lambda/m_b \sim 6\%$, which would indicate that $(\Lambda/m_b)^2$ corrections may even be below the percent level. The lattice form factors agree with the theoretical expectation for $q^2 \gtrsim 27~\rm{GeV}^2$.

\subsection{\boldmath Determination of $|V_{ub}|$}

We now combine our lattice form factors with experimental data for $B\to\pi\ell\nu$ to obtain $|V_{ub}|$. The Standard-Model partial branching fraction is $\tau_B d\Gamma/dq^2$, where $d\Gamma /dq^2$ is defined in Eq.~(\ref{eq:partialdecayrate}).
The contribution from $f_0$ is negligible due to the small lepton mass. Given $f_{+}(q^{2})$, the branching fraction in the $i$th $q^{2}$
bin $[q_{i}^{2},q_{i+1}^{2}]$ is 
\begin{eqnarray}
\Delta\mathcal{ B}_{i}^{\text{fit}} & = & C_{B}^{2}|V_{ub}|^{2}\int_{q_{i}^{2}}^{q_{i+1}^{2}}|\bm{p}_\pi(q^{2})|^{3}|f_{+}(q^{2})|^{2}dq^{2},\label{eq:D}
\end{eqnarray}
where $C_{B}^{2}=(\tau_{B}G_{F}^{2})/(24\pi^{3})$ is a constant. For the combined lattice plus experiment $z$ fit, we define a $\chi^2$ for the experimental measurements $\Delta\mathcal{B}_i^\text{exp}$ as 
\begin{eqnarray}
\chi_{\text{exp}}^{2} & = & \sum_{i,j}(\Delta\mathcal{B}_{i}^{\text{exp}}-\Delta\mathcal{B}_{i}^{\text{fit}})\text{Cov}_{ij}^{\text{exp}}(\Delta\mathcal{B}_{j}^{\text{exp}}-\Delta\mathcal{B}_{j}^{\text{fit}}), \label{eq:x_exp}
\end{eqnarray}
where $\Delta \mathcal{B}_i^\text{exp}$ is the experimentally-measured branching fraction in the $i$th $q^2$ bin ($i$ is a shorthand notation for each bin in each experiment 
included in the fit) and $\text{Cov}^{\text{exp}}$ is the experimental covariance matrix, including the statistical and all systematic errors.

We use the experimental results compiled by the Heavy Flavor Averaging Group (HFAG) \cite{1207.1158}: BaBar untagged 6-bin analysis (2011) \cite{1005.3288}, Belle
untagged 13-bin analysis (2011) \cite{1012.0090}, BaBar untagged 12-bin analysis (2012) \cite{1208.1253} and Belle
tagged analysis with 13 bins for the $B^{0}$ and 7 bins for the $B^-$ mode (2013) \cite{1306.2781}. For convenience in the fit, we assume isospin symmetry to convert the Belle tagged $B^-$ data to the $B^0$ mode via 
\begin{equation}
\Delta\mathcal{B}(B^{0}\to\pi^{+}\ell^{-}\nu)_{\text{Belle},B^{-}}  =2\frac{\tau_{B^{0}}}{\tau_{B^{-}}}  \Delta\mathcal{B}(B^{-}\to\pi^{0}\ell^{-}\nu),\label{eq:isospin}
\end{equation}
where $\tau_{B^{0}}=1.519(7)$~{\rm ps} and $\tau_{B^{-}}=1.641(8)~{\rm ps}$ are from the
PDG \cite{PDG2014}.

We omit systematic correlations between the BaBar
and Belle analyses, because they do not share any major systematic errors. The BaBar 6-bin and 12-bin data have
very small overlaps in the selection of samples, so the statistical errors
can be considered approximately uncorrelated. There is some systematic
correlation between the two analyses, which is, however, supposed
to be insignificant \cite{JDingfelder}. The Belle untagged and tagged data
are also largely uncorrelated because
the dominant source of systematic errors in these two measurements
are very different. In summary, we take the four experimental analyses as independent
measurements. 

On the other hand, there are systematic correlations between
the two isospin modes of the Belle tagged data, which we estimate
as follows. Let $\Delta\mathcal{B}_{i}^{-}$ and $\Delta\mathcal{B}_{\alpha}^{0}$
be the branching fractions in the $i$th and $\alpha$th bin of the
charged and neutral decay modes, respectively. Let $\sigma_{x}^{-},\sigma_{x}^{0}$
be the systematic uncertainties of the two modes from
source $x$ and $r_{x}^{-0}$ be the correlation between them. Then
we estimate the off-block-diagonal elements of the systematic error covariance matrix by
\begin{eqnarray}
S{}_{i\alpha} & = & \sum_{x\in\text{all sys.}}r_{x}^{-0}\left(\sigma_{x}^{-}\sigma_{x}^{0}\Delta\mathcal{B}_{i}^{-}\Delta\mathcal{B}_{\alpha}^{0}\right),
\end{eqnarray}
where the sum is over all sources of systematic errors. That said, only a few of the systematic errors contribute noticeably to the sum and the biggest source of error, the tag
calibration, dominates. From the correlation matrices, we construct the total
covariance matrices of each isospin decay mode by adding the statistical
matrices and the systematic matrices. We then take the direct sum of
the covariance matrices of the $B^-$ and $B^0$ modes block-diagonally and add the off-block-diagonal elements $S_{i\alpha}$ so that we can fit them simultaneously.

\begin{figure}
	\center{\includegraphics[scale=0.6]{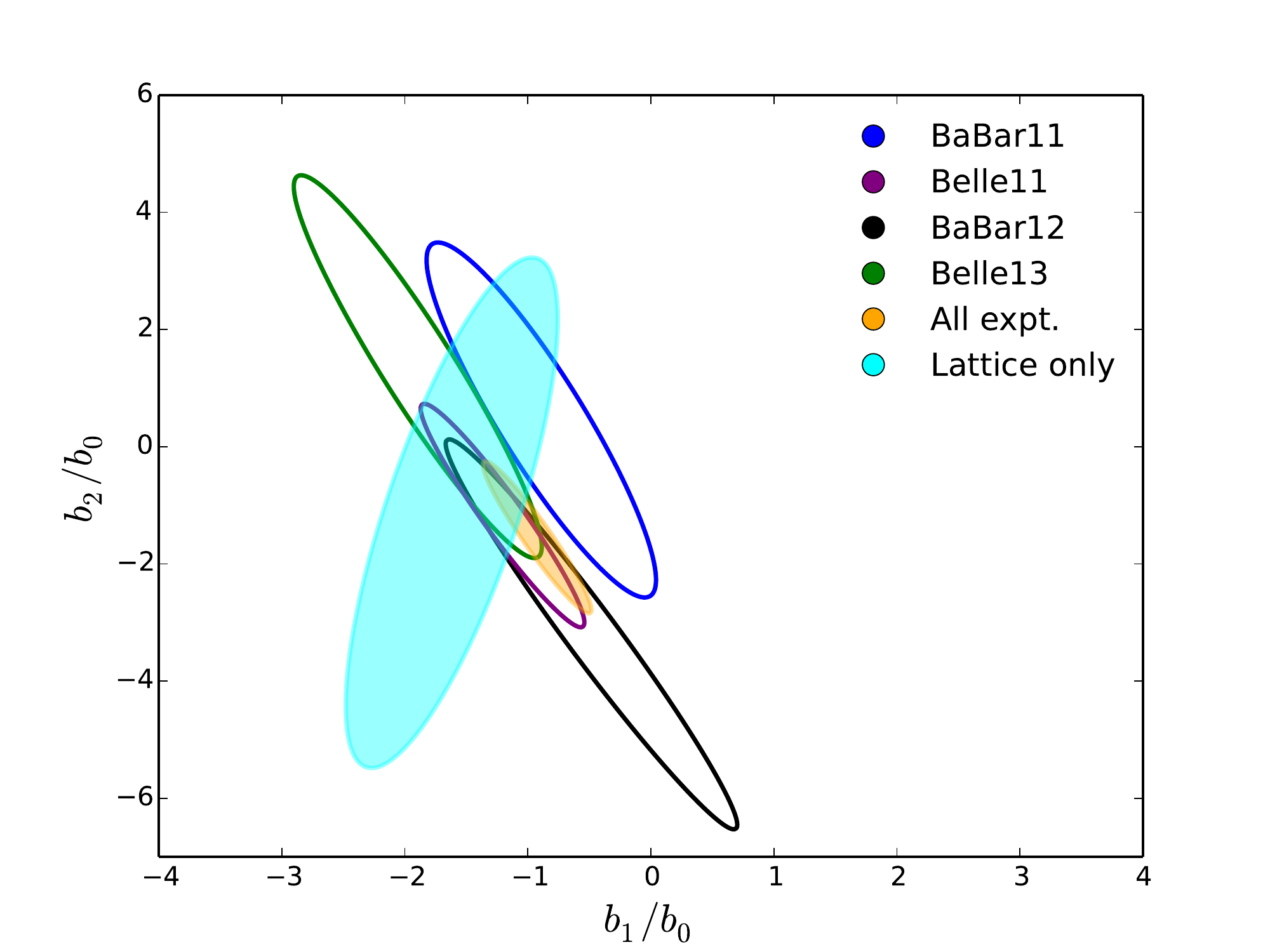}}
	
	\caption{ The contour plot of the slope $b_1/b_0$ and curvature $b_2/b_0$ of the form factor $f_+$. The open ellipses are the 1-$\sigma$ contour of the slope and curvature constructed from the 3-parameter $z$ fit to individual experimental data. The gold filled ellipse is from the combined fit of all experimental data. The cyan filled ellipse is from the 4-parameter $z$ fit to lattice form factors.
		\label{fig:shape_compare}}
\end{figure}

We first fit the $z$ expansion to the experimental data only and without any constraints on the coefficients. We use the BCL formula with
three parameters, $N_{z}=3$, where the normalization is $|V_{ub}|b_{0}$
. The result is shown in Table~\ref{tab:exp-fits}. 
To check the consistency in the shape among the experimental data sets, we also fit each experimental data set separately. The individual fits all have acceptable confidence levels and $p$ values, but the combination of all four data sets gives a rather poor fit that is not improved by going to higher order in $z$, $e.g.$, $N_z=4$. The poor fit stems from the BaBar11 measurement, which is only marginally consistent with the other three. Figure \ref{fig:shape_compare} compares the shapes (slopes $b_1/b_0$ and curvatures $b_2/b_0$) of the separate and combined experimental fits with the lattice-only fit. The lattice form factor shape is consistent with all of the experimental results. 

\begin{table}
	\caption{The results of fits to experimental data only.\label{tab:exp-fits} }
	\begin{tabular}{ccccccc}
		
		\hline 
		\hline 
		Fit & $\chi^{2}$/dof &\;\; dof\;\; & $p$ & $b_{1}/b_{0}$ & $b_{2}/b_{0}$ & $b_0|V_{ub}|\times 10^{-3}$\tabularnewline
		\hline 
		All exp. & 1.5 &48 & 0.02 & $-$0.93(22) & $-$1.54(65)		&	1.53(4)\tabularnewline
		BaBar11 \cite{1005.3288}& 2 &3    & 0.12  & $-$0.89(47) &  0.5(1.5)  & 	1.36(7)\tabularnewline
		BaBar12 \cite{1208.1253}& 1.2 &9 & 0.31 & $-$0.48(59) & $-$3.2(1.7)	&   1.54(9)\tabularnewline
		Belle11 \cite{1012.0090}& 1.1 &10 & 0.36 & $-$1.21(33) & $-$1.18(95)	&	1.63(7)\tabularnewline
		Belle13 \cite{1306.2781}& 1.2 &17 & 0.23 & $-$1.89(50) & 1.4(1.6)	&	1.56(8)\tabularnewline
		\hline 
		\hline 
	\end{tabular}
	
\end{table}

To perform a combined fit to the lattice and experimental data, we define the total
chi squared function,
\begin{eqnarray}
\chi^{2} & = & \mathcal{\chi}_\text{lat}^{2}+\chi_\text{BaBar11}^{2}+\chi_\text{Belle11}^{2}+\chi_\text{BaBar12}^{2}+\chi_\text{Belle13}^{2},\label{eq:L_all}
\end{eqnarray}
where the lattice and experimental chi squared functions are defined in Eqs.~(\ref{eq:object_L})
and (\ref{eq:x_exp}), respectively. The fit is performed to these five independent
data sets with common shape parameters $b_m$ and overall normalization $|V_{ub}|$ by minimizing Eq.~(\ref{eq:L_all}). Table~\ref{tab:lat+exp} summarizes the various fit results. Due to the tension between the experimental data sets, the $p$ value of
the fit to the lattice result and all experiments is only 0.02. Table~\ref{tab:chi_details} shows the contributions to the total $\chi^2$ from each data set of the combined fit. By far the largest contribution to $\chi^2/{\rm dof}$ is from the BaBar 6-bin data set, similar to what we find for the experiment-only fits presented in Table~\ref{tab:exp-fits}.  

In the combined fit to lattice form factors and experimental data, the kinematic constraint between $f_+$ and $f_0$ at $q^2=0$ is unimportant for the determination of $|V_{ub}|$. This is because the experimental data constrain the shape at low $q^2$. Removing the kinematic constraint from the combined fit and fitting only with the vector form factor $f_+$ changes neither the coefficients of the $z$ expansion nor the value of $|V_{ub}|$. We also try varying the number of parameters $b_m$ in the $z$ expansion ($N_z$). The results are shown in Table~\ref{tab:Nz_combined}. Compared to our preferred fit with $N_z=4$, the fit using $N_z=3$ gives a very low $p$~value and a shift of about $1\sigma$ in both the form factor and  $|V_{ub}|$, while the fit result using $N_z=5$ nearly coincides with that of the $N_z=4$ fit and the values of $|V_{ub}|$ are almost identical.

\begin{table}
	\caption{  Results of the combined lattice+experiment fits with $N_z=4;$. % The errors shown here are inflated by a factor $(\overline{\chi^2})^{1/2}$.
		\label{tab:lat+exp}}
	
	\begin{tabular}{ccccccccc}
		\hline 
		\hline 
		Fit  \;&\; $\chi^{2}$/{\rm dof} \;&\; dof \;& $p$~value  & $b^{+}_{0}$  & $b^{+}_{1}$  & $b^{+}_{2}$  & $b^{+}_{3}$  &  $|V_{ub}|$($\times10^{3}$)\tabularnewline
		\hline 
		Lattice+exp.(all) & 1.4 & 54  & 0.02  & 0.419(13)  & $-$0.495(55)  & $-$0.43(14)  & 0.22(31)  & 3.72(16)\tabularnewline
		Lattice+BaBar11 \cite{1005.3288}& 1.1 &9  & 0.38  & 0.414(14)  & $-$0.488(73)  & $-$0.24(22)  & 1.33(44)  & 3.36(21)\tabularnewline
		Lattice+BaBar12 \cite{1208.1253}& 1.1 &15  & 0.34  & 0.415(14)  & $-$0.551(72)  & $-$0.45(18)  & 0.27(41)  & 3.97(22)\tabularnewline
		Lattice+Belle11 \cite{1012.0090}& 0.9 &16  & 0.55 & 0.412(13)  & $-$0.574(65)  & $-$0.40(16)  & 0.38(36)  & 4.03(21)\tabularnewline
		Lattice+Belle13 \cite{1306.2781}& 1.0 &23 & 0.42 & 0.406(14) & $-$0.623(73) & $-$0.13(22) & 0.92(45) & 3.81(25)\tabularnewline
		\hline
		\hline  
	\end{tabular}
\end{table}

\begin{table}
	\caption{The contribution to the total $\chi^2$ from each data set of the combined fit. The column ``\#~data'' gives the number of independent functions (for lattice QCD) or the number of bins (for experiment). The total $\chi^2/(\#~\text{data})$ agrees with the $\chi^2/\text{dof}$ in Table~\ref{tab:lat+exp}, once the constraint and number of fit parameters have been taken into account. \label{tab:chi_details}}
	\begin{tabular}{cccc}
		\hline 
		\hline
		data set \;&\; \#~data \;&\; $\chi^{2}$ \;&\; $\chi^2$/\#~{\rm data}\tabularnewline
		\hline 
		Lattice & 11 & 4.8 & 0.44\tabularnewline
		BaBar11 \cite{1005.3288}& 6 & 20.9 & 3.5\tabularnewline
		BaBar12 \cite{1208.1253}& 12 & 15.1 & 1.3 \tabularnewline
		Belle11 \cite{1012.0090}& 13 & 13.8 & 1.1\tabularnewline
		Belle13 \cite{1306.2781}& 20 & 23.5& 1.2 \tabularnewline
		\hline
		Total & 62 & 78.2& 1.26\tabularnewline
		\hline 
		\hline
	\end{tabular}

\end{table}

\begin{table} [t]
	\caption{Combined lattice+experiments $z$ fits with $N_z=3,4$ and 5. 
		\label{tab:Nz_combined} }
	
	\begin{tabular}{cccccccccc}
		\hline 
		\hline 
		$N_z$ \; &\; $\chi^{2}$/{\rm dof} \;&\; dof  \;&\; $p$~value  & $b^{+}_{0}$  & $b^{+}_{1}$  & $b^{+}_{2}$  & $b^{+}_{3}$  & $b^{+}_{4}$ & $|V_{ub}|$		\tabularnewline
		\hline 
		$3$  		& 2.5 & 56 & 0.0  & 0.425(12)  & $-$0.424(31)  	& $-$0.59(9)  &   	&  & 3.63(11)			\tabularnewline
		$4$ 		& 1.4 & 54  & 0.02  & 0.419(13)  & $-$0.495(55)  	& $-$0.43(14)  & 0.22(31)  	&    & 3.72(16)	\tabularnewline
		$5$		& 1.5 & 52 & 0.01  	& 0.418(13)  & $-$0.491(56)  	& $-$0.31(30)  & 0.01(55)  	& $-$0.6(1.9)	& 3.72(16)\tabularnewline
		\hline
		\hline  
	\end{tabular}
\end{table}

\begin{figure}
	\center{\includegraphics[width=0.48\linewidth, trim= 40 8 40 2]{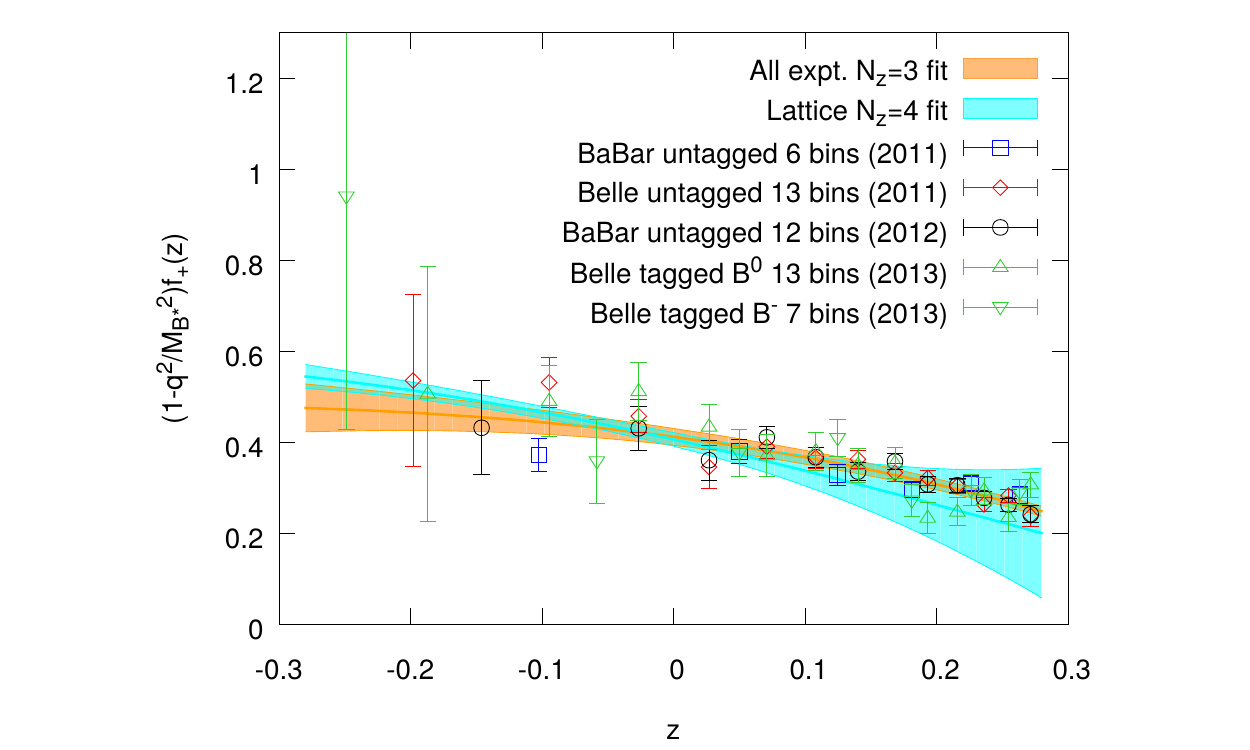}\hfill\includegraphics[width=0.51\linewidth, trim= 40 22 40 0]{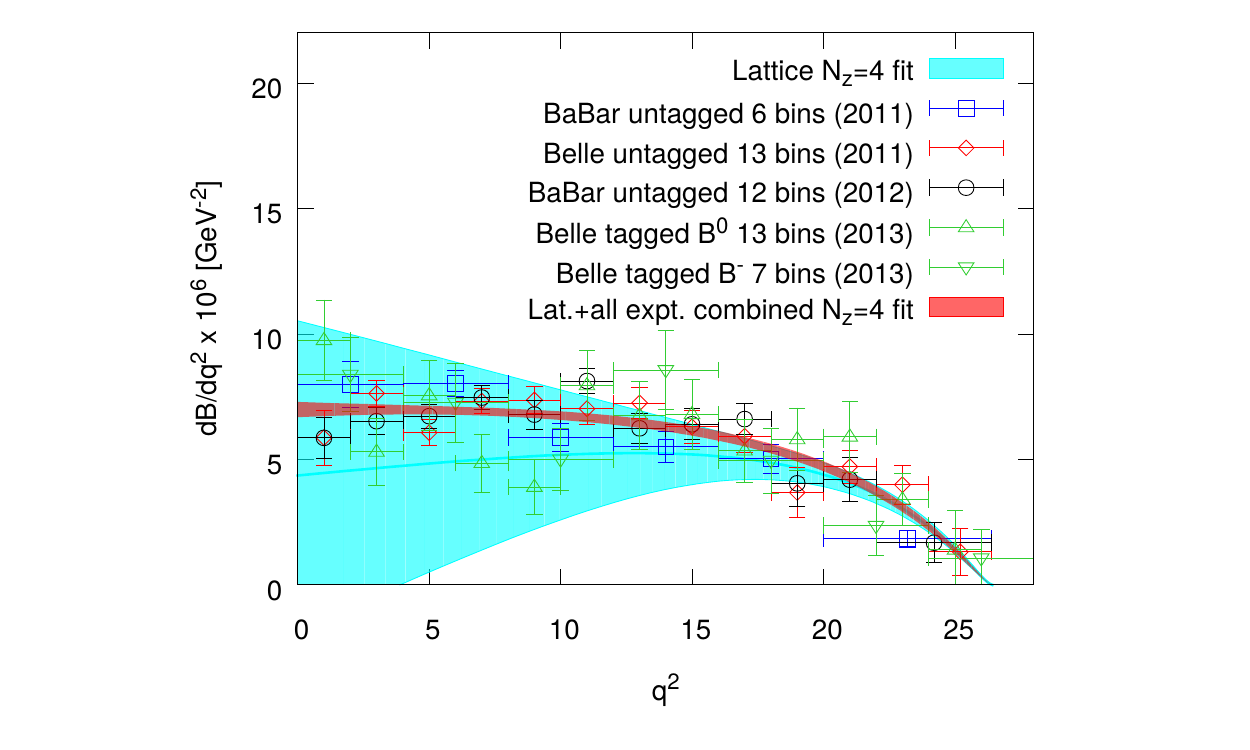}}
	
	\caption{ Left: comparison of vector form factor $f_{+}(z)$ from $z$ expansion fits to: only the lattice-QCD data (cyan band) and only experimental data including all four measurements (gold band). Right: the similar plot for the partial branching fraction $dB/dq^2$. The fits including lattice results use $N_z=4$, while the experiment-only fit uses $N_z=3$. The experimental
		data points and the experiment-only $z$-fit result in the left plot have been converted from $\left(\Delta\mathcal{B}/\Delta q^{2}\right)^{1/2}$ to $f_+$ using $|V_{ub}|$ from the combined fit. The lattice-only fit result(cyan band) and the combined-fit result (red band) in the right plot is converted from the form factor with the same $|V_{ub}|$. 
		\label{fig:lat_exp}}
\end{figure}

The experimental data are plotted in Fig.~\ref{fig:lat_exp} (left) along with the $z$ fits to the lattice data and to all experimental data. The lattice form factor and experimental measurements provide complementary information and, when combined, yield an accurate description of the form factor over the full-$q^{2}$
range and hence a precise determination of $|V_{ub}|$. The plot shows
that the experimental data dominate the determination of the form-factor shape in the large-$z$ (small-$q^{2}$)
region while the lattice-QCD form factor dominates the small-$z$ (large-$q^{2}$) region. In the intermediate region around $q^{2}\sim20\, \text{GeV}^{2}$
($z\sim0$), the lattice-QCD and experimental uncertainties are similar in size.
This region is decisive in determining $|V_{ub}|$ and, hence, can be used to estimate the separate contributions from lattice and experimental data to the $|V_{ub}|$ uncertainty. At $q^2=20~\text{GeV}^2$, the error on the lattice-QCD form factor $f_+$ is about 3.4\% (see Table~\ref{tab:Error-budgets-20}) and the error on $f_+|V_{ub}|$ from the experiment-only fit is 2.8\% at the same momentum. Adding these two errors in quadrature gives a total uncertainty of $4.4\%$, which is consistent with the error on $|V_{ub}|$ obtained from the full fit, $4.3\%$. Another estimate of the individual error contribution to $|V_{ub}|$ can be obtained from the uncertainty on the fit parameters from the separate lattice-QCD and experiment fits. From the fit to all experimental data in Table~\ref{tab:exp-fits}, the normalization is $|V_{ub}|b_0=(1.53\pm0.04)\times 10^{-3}$. Similarly, the lattice-only $z$ fit gives the normalization $b_0=0.407\pm 0.015$. Assuming no correlation, one would obtain $|V_{ub}|=(3.76\pm0.17)\times 10^{-3}$, which is close to what we obtain from the combined fit. 

\section{Results and conclusion} \label{secVII}

Our final result for $|V_{ub}|$, obtained from our preferred $z$ fit combining our lattice-QCD calculation of the $B\to\pi\ell\nu$ form factor with experimental measurements of the corresponding decay rate, is
\begin{eqnarray}
|V_{ub}| & = & (3.72 \pm 0.16) \times 10^{-3}.
\end{eqnarray}
The error includes all experimental and lattice-QCD uncertainties. The contribution from lattice QCD to the total error is now comparable to that from experiment. The error reported here, following HFAG~\cite{1207.1158}, does not apply the PDG prescription for discrepant data; that prescription~\cite{PDG2014} would scale the error by a factor of $\sqrt{\chi^2/{\rm dof}} = 1.2$. As can be seen from Table~\ref{tab:chi_details} and Fig.~\ref{fig:shape_compare}, the low fit quality is due to the tension between the BaBar11 data set and the others. An inspection of all the experimental data in Fig.~\ref{fig:lat_exp} shows that the point near $z=-0.1$ in the BaBar11 data set is lower than the others and a bit more precise than one might have anticipated, but does not suggest that this or any of the data sets have any systematic problems.

\begin{figure}
	\center{\includegraphics[width=0.8\textwidth]{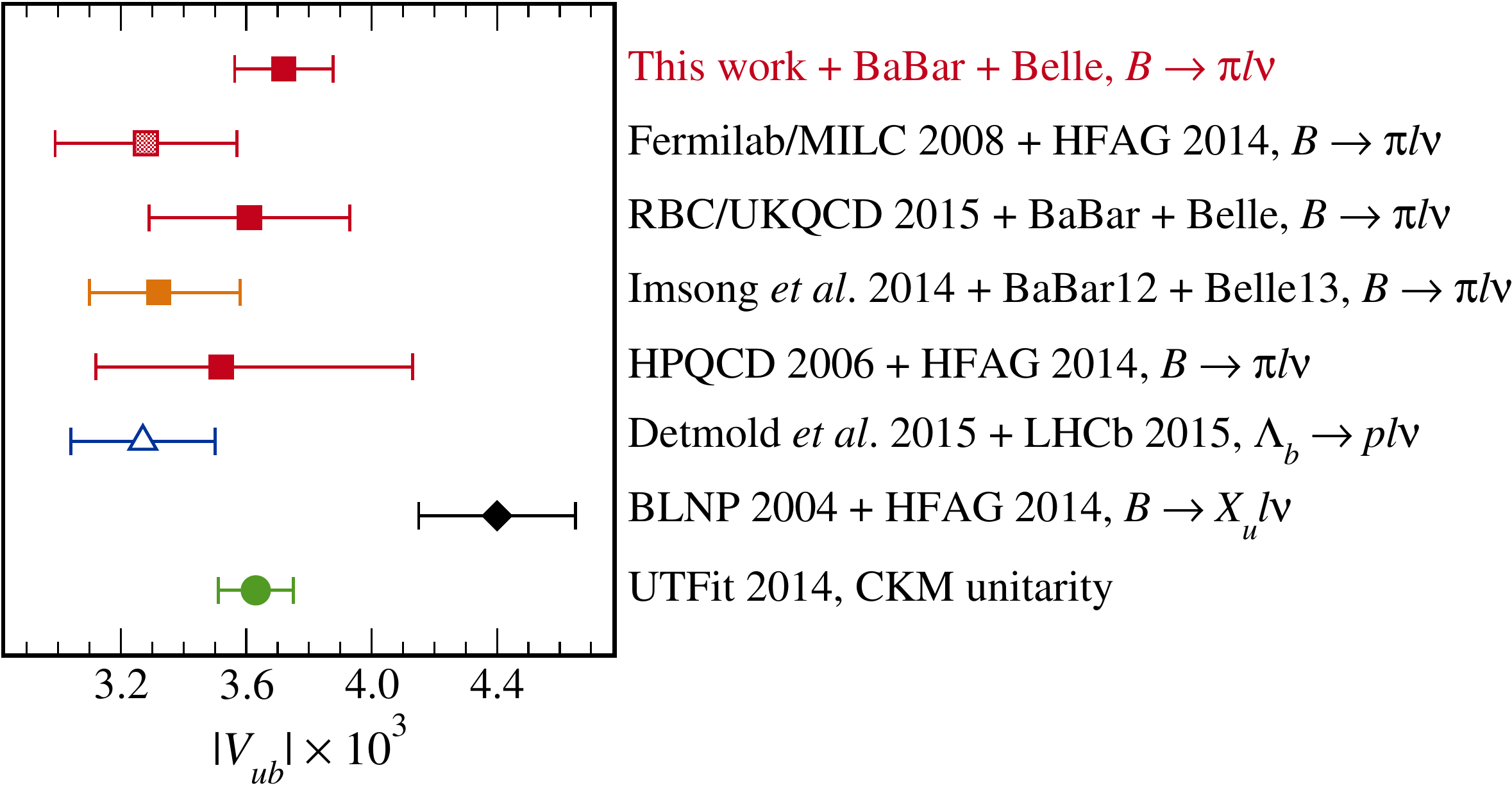}}
	\caption{  			Determinations of $|V_{ub}|$. The squares are obtained from $B \to \pi \ell \nu$ decay using theoretical form factors from this analysis, our earlier work~\cite{0811.3640} (now superseded, but with updated experimental input  from HFAG 2014 \cite{1207.1158}), a three-flavor lattice calculation by RBC/UKQCD~\cite{Flynn:2015mha}, light-cone sum rules (orange square)~\cite{Imsong:2014oqa}, and HPQCD~\cite{hep-lat/0601021} (using the $q^2>16$~GeV$^2$ experimental data only). The blue upward-pointing triangle is obtained from $\Lambda_b \to p \ell \nu$ decay using lattice-QCD form factors from Ref.~\cite{Detmold:2015aaa} and experimental data from LHCb~\cite{Sutcliffe}. The black diamond shows the inclusive determination using $B\to X_u\ell\nu$ decays \cite{1207.1158} with the theoretical approach of Ref.~\cite{Bosch:2004bt}. Also shown is the expectation from CKM unitarity~\cite{Bona:2006ah} (green filled circle). For the exclusive determinations from $B \to \pi \ell \nu$ decay (squares), all four experimental results~\cite{1005.3288,1012.0090,1208.1253,1306.2781} are used except in the LCSR $z$-fit where only the more recent BaBar \cite{1208.1253} and Belle \cite{1306.2781} data are used. 
		\label{fig:vub}}
\end{figure}

We compare our determination of $|V_{ub}|$ with other results in Fig.~\ref{fig:vub}. In particular, our result is consistent with the recent determination from HFAG using our collaboration's 2008 form-factor determination~\cite{0811.3640} obtained from a small subset of the gauge-field ensembles used in this work. The difference in the central values is due to a small shift in the central values for the form factor $f_+$ of this analysis compared to our previous analysis \cite{0811.3640}. As shown in Fig.~\ref{fig:f+_compare} (left), the form factor $f_+$ from this analysis is consistent within errors with the previous analysis, but shifted slightly downward and with an error smaller by roughly a factor of three. The two analyses have very little statistical and systematic correlation. Our result is also compatible with Standard-Model expectations from CKM unitarity \cite{Laiho:2009eu,Bona:2006ah}. Although our determination of $|V_{ub}|$ is higher than that in Ref.~\cite{0811.3640}, and thus closer to the determination from inclusive $B\to X_u$ semileptonic decays~\cite{1207.1158}, the inclusive-exclusive disagreement is still greater than 2$\sigma$. 

\begin{figure}
	\center{\includegraphics[width=0.48\textwidth, trim = 60 10 60 0]{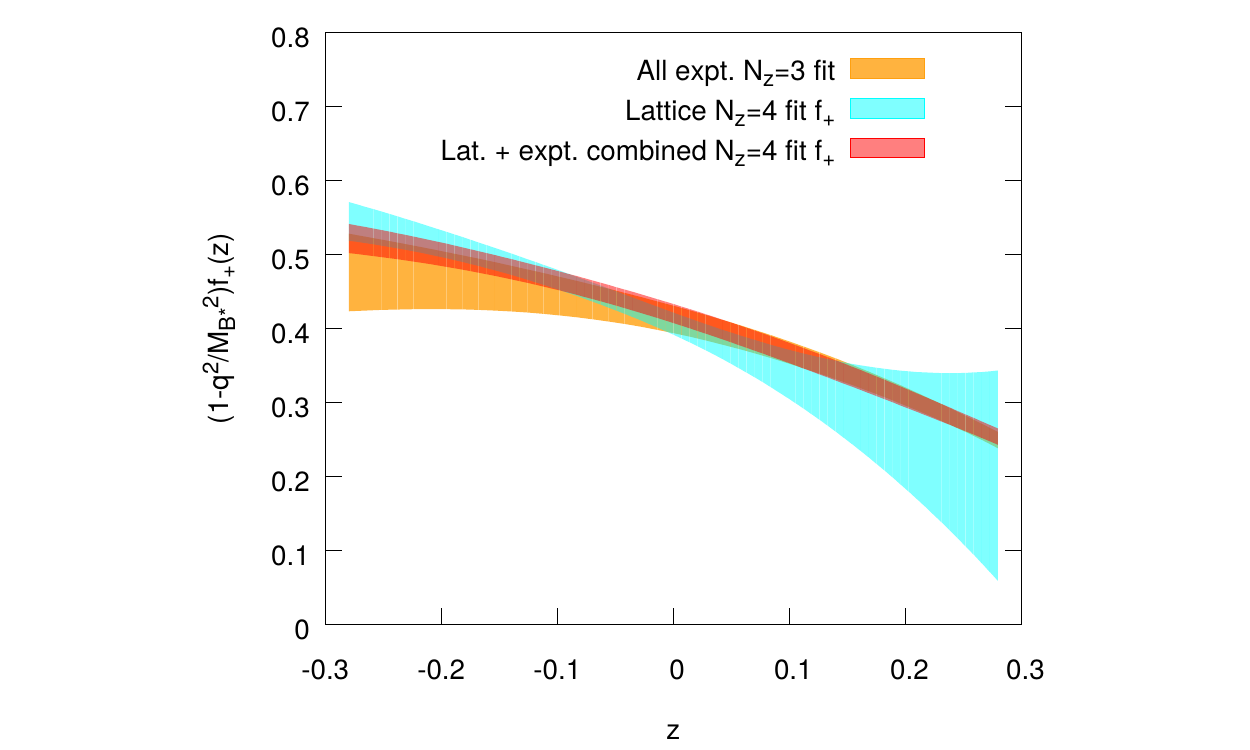}\hfill \includegraphics[width=0.48\textwidth, trim = 60 10 60 30]{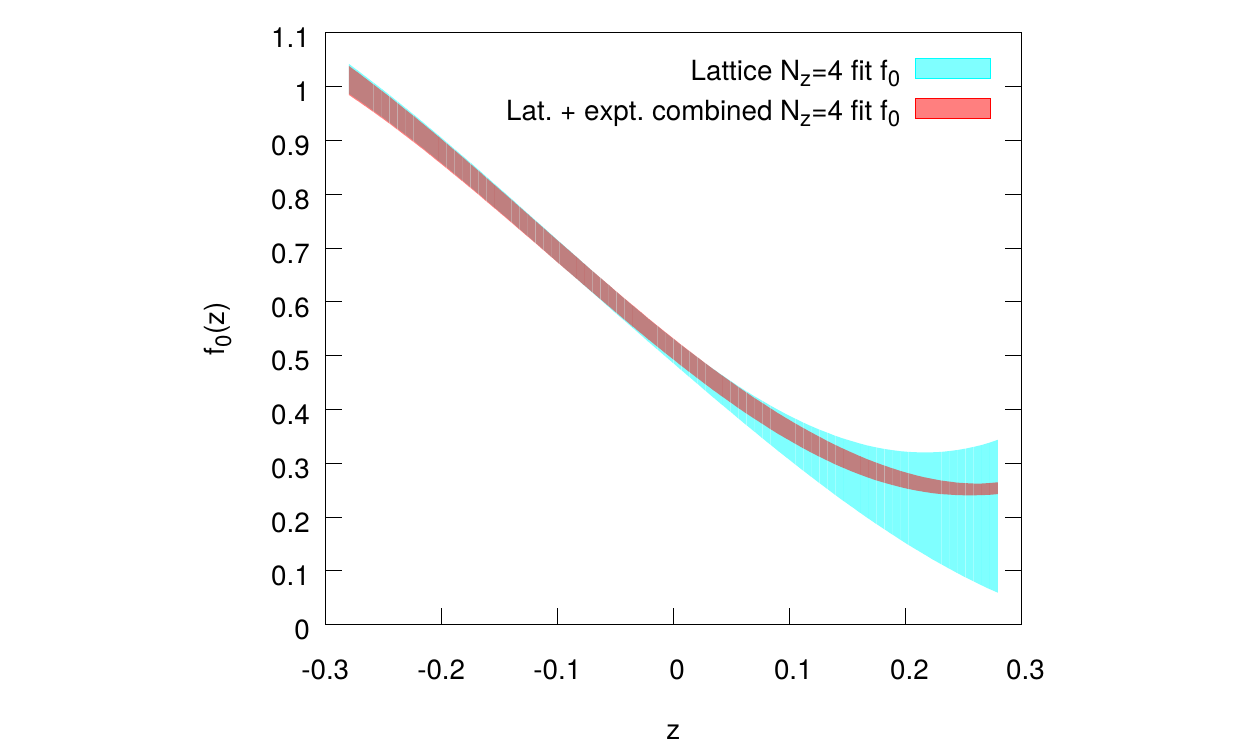}}
	\caption{ The form factors $f_{+}(z)$ (left) and $f_0(z)$ (right) from the $z$ fits to the lattice data (cyan band), to all experiments (gold band), and to the lattice data plus experiments (red band). The fits including lattice results use $N_z=4$ and the kinematic constraint, while the experiment-only $z$ fit uses $N_z=3$. The experiment-only $z$-fit result has been converted from $\left(\Delta\mathcal{B}/\Delta q^{2}\right)^{1/2}$ to $f_+$ using $|V_{ub}|$ from the combined fit.  
		\label{fig:formfactorsfinal}}
\end{figure}

A byproduct of the combined lattice and experiment fit is a more precise determination of the vector and scalar form factors than from the lattice-QCD calculation alone. Both form factors $f_+$ and $f_0$ are well determined from lattice QCD in the high $q^2$ region, and $f_+$ is strongly constrained by experiment in the low $q^2$ region. This information is then transferred to $f_0$ via the kinematic constraint $f_0(0)=f_+(0)$. The resulting form factors are shown in Fig.~\ref{fig:formfactorsfinal}. The corresponding $z$-expansion coefficients and their correlations are given in Table~\ref{tab:fpf0_correlation2}. These represent the present best knowledge of the $B\to\pi\ell\nu$ form factors, and can be used in other phenomenological applications or to test other nonperturbative QCD calculations.

Future improvements in the determination of the $B\to\pi$ semileptonic form factor $f_+$ will further reduce the uncertainty on $|V_{ub}|$. If the uncertainty of $f_+^{B\to\pi\ell\nu}$ at $q^2\sim 20$ GeV$^2$ can be reduced further from $3.4$\% to $1.5$\%, we would expect a precision of 3\% in $|V_{ub}|$, using the current experimental input. With the anticipated improvement in the experimental rate measurement from Belle II, this error would be reduced further. The reduction of uncertainty in $f_+^{B\to\pi\ell\nu}$ is expected with the newly-available MILC gauge ensembles that are being generated using the highly improved staggered quark (HISQ) action \cite{Bazavov:2012xda}. 
The new HISQ ensembles have statistics similar to the asqtad ensembles, but with much smaller light-quark discretization effects. Further, the HISQ ensembles simulated at the physical light-quark masses will remove the need for a chiral extrapolation, thereby eliminating a significant source of uncertainty in this work. These ensembles have already helped to determine the form factor $f_+^{K\to \pi\ell\nu}(0)$ \cite{Bazavov:2013maa} and the leptonic decay constants $f_{D_{(s)}}$ and $f_K$ \cite{Bazavov:2014wgs}, and hence the relevant CKM matrix elements $|V_{us}|$, $|V_{cd}|$ and $|V_{cs}|$, with high precision. All of these improvements will further refine and reduce the uncertainties in $|V_{ub}|$, and may also help to resolve the inclusive/exclusive puzzle.

\begin{acknowledgments}
	We thank Jochen Dingfelder for the helpful information about the experimental measurements and HFAG averaging procedure. D.D. thanks Peter Lepage for sharing his lsqfit code (github.com/gplepage/lsqfit) which is extensively used in the fitting procedures of the analysis. We also thank Heechang Na for valuable discussions. Computations for this work were carried out with resources provided by the USQCD
	Collaboration, the Argonne Leadership Computing Facility, the National Energy Research
	Scientic Computing Center, and the Los Alamos National Laboratory, which are funded
	by the Office of Science of the United States Department of Energy; and with resources provided by the National Institute for Computational Science, the Pittsburgh Supercomputer
	Center, the San Diego Supercomputer Center, and the Texas Advanced Computing Center,
	which are funded through the National Science Foundation's Teragrid/XSEDE Program.
	This work was supported in part by the U.S. Department of Energy under Grants No.~DE-
	FG02-91ER40628~(C.B., J.K.), No.~DE-FC02-12ER41879~(C.D., J.F., L.L.), No.~DE-SC0010120~(S.G.), No.~DE-FG02-91ER40661~(S.G., R.Z.), No.~DE-FC02-06ER41443~(R.Z.), No.~DE-FG02-13ER42001~(D.D., A.X.K.), No.~DE-FG02-13ER41976~(D.T.), No.~DE-SC0010114 (Y.M.); by the National Science
	Foundation under Grants No.~PHY-1067881, No.~PHY-10034278~(C.D., L.L., S-W.Q.), No.~PHY-1417805~(J.L., D.D.), No.~PHY-1212389~(R.Z.), No.~PHY-1316748~(R.S.); by the URA Visiting Scholars' program~(C.M.B., D.D., A.X.K., Y.L.); by the MINECO (Spain) under Grants
	FPA2010-16696, FPA2006-05294, and
	Ram\'{o}n y Cajal program (E.G.); by the Junta de Andaluc\'{i}a (Spain) under Grants FQM-101 and FQM-6552 (E.G.); by European Commission
	hunder Grant No. PCIG10-GA-2011-303781 (E.G.); by the German Excellence Initiative and the European Union Seventh Framework Programme under grant agreement No.~291763 as well as the European Union's Marie Curie COFUND program (A.S.K.); and by the Basic Science Research Program of the National Research Foundation of Korea (NRF) funded by the Ministry of Education (No. 2014027937) and the Creative Research Initiatives Program (No. 2014001852) of the NRF grant funded by the Korean government (MEST) (J.A.B.). This manuscript has been co-authored by an employee of Brookhaven Science
	Associates, LLC, under Contract No. DE-AC02-98CH10886 with the U.S. Department of
	Energy. Fermilab is operated by Fermi Research Alliance, LLC, under Contract No. DE-AC02-07CH11359 with the U.S. Department of Energy.
\end{acknowledgments}

\appendix

\begin{table}
	\caption{ Central values, errors, and correlation matrix of the coefficients of $f_+$ and $f_0$ from the $N_z = 4$ $z$-fit combining lattice and all four experiments. \label{tab:fpf0_correlation2}}
	\begin{tabular}{cccccccccc}
		\hline 
		\hline 
		&$|V_{ub}|\times 10^3$ & $b_{0}^{+}$ & $b_{1}^{+}$ & $b_{2}^{+}$ & $b_{3}^{+}$ & $b_{0}^{0}$ & $b_{1}^{0}$ & $b_{2}^{0}$ & $b_{3}^{0}$ \\ 
		\hline 
		& 3.72(16) & 0.419(13)  & $-$0.495(54)  & $-$0.43(13)  & 0.22(31) & 0.510(19) & $-$1.700(82) &1.53(19) & 4.52(83) \\
		$|V_{ub}|$ & 1 &   $-$0.870 &   $-$0.400 &   $$0.453 &   0.428 &   $-$0.175 &   $-$0.201 &   $-$0.119 &   $-$0.009 \\
		$b_{0}^{+}$  & &   1 &   0.140 &   $-$0.455 &   $-$0.342 &   0.224 &   $$0.174 &   $$0.047 &   $-$0.033 \\
		$b_{1}^{+}$  & &        &   1 &   $-$0.789 &   $-$0.874 &   $-$0.068 &   0.142 &   0.025 &   $-$0.007 \\
		$b_{2}^{+}$  & &        &        &   1 &   0.879 &   $-$0.051 &   $-$0.253 &   0.098 &   0.234 \\
		$b_{3}^{+}$  & &        &        &        &   1 &   $$0.076 &   0.038 &   0.018 &   $-$0.200 \\
		$b_{0}^{0}$  & &        &        &        &        &   1 &   $-$0.043 &   $-$0.604 &   $-$0.388 \\
		$b_{1}^{0}$  & &        &        &        &        &        &   1 &   $-$0.408 &   $-$0.758 \\
		$b_{2}^{0}$  & &        &        &        &        &        &        &   1 &   0.457 \\
		$b_{3}^{0}$  & &        &        &        &        &        &        &        &   1 \\
		\hline 
		\hline
	\end{tabular}
	
\end{table}

\bibliographystyle{unsrt}

%%%%% CLEAR DOUBLE PAGE!
\newpage{\pagestyle{empty}\cleardoublepage}

\end{document}